\documentclass[jcp, floatfix, nobibnotes, reprint, superscriptaddress]{revtex4-1}
\usepackage{docs}%
\usepackage{siunitx}
\usepackage{amsmath}
\usepackage{booktabs}
\DeclareSIUnit[number-unit-product = {\,}]
\cal{cal}
\DeclareSIUnit\kcal{\kilo\cal}
\DeclareSIUnit[number-unit-product = {\,}]
\Btu{Btu}
\DeclareSIUnit[number-unit-product = {\,}]
\Fahr{\degree F}
\DeclareSIUnit[number-unit-product = {\,}]
\usepackage{bm}%
\usepackage[colorlinks=true,linkcolor=black]{hyperref}%
\usepackage{graphicx}
\usepackage[utf8]{inputenc}
\usepackage{tabularx} % provides a column type called "X" that should satisfy your professed need to have several equal-width columns

\usepackage{dcolumn}
\usepackage{epstopdf}
\usepackage{afterpage}
\usepackage{setspace}
\usepackage{multirow}
\usepackage{dsfont}
\usepackage{sidecap}
\usepackage{epstopdf}
\usepackage{braket}
\usepackage{siunitx} % align numbers in tables w.r.t. decimal point
\usepackage{color, colortbl}

\definecolor{Gray}{gray}{0.9}

\AppendGraphicsExtensions{.gif}

\DeclareMathAlphabet\mathbfcal{OMS}{cmsy}{b}{n}

\setcitestyle{super}
\begin{document}
\setstretch{1.0}

\title[]{Kernel based quantum machine learning at record rate:  \\
Many-body distribution functionals as compact representations}

%Many Body Distribution Functional (Density) Representations for Quantum Machine Learning}

\author{Danish Khan}

\affiliation{Department of Chemistry, University of Toronto, St. George Campus, Toronto, ON, Canada}
\affiliation{Vector Institute for Artificial Intelligence, Toronto, ON, M5S 1M1, Canada}

\author{Stefan Heinen}
\affiliation{Vector Institute for Artificial Intelligence, Toronto, ON, M5S 1M1, Canada}

\author{O. Anatole von Lilienfeld}
\email{anatole.vonlilienfeld@utoronto.ca}
\affiliation{Vector Institute for Artificial Intelligence, Toronto, ON, M5S 1M1, Canada}
\affiliation{Departments of Chemistry, Materials Science and Engineering, and Physics, University of Toronto, St. George Campus, Toronto, ON, Canada}
\affiliation{Machine Learning Group, Technische Universit\"at Berlin and Institute for the Foundations of Learning and Data, 10587 Berlin, Germany}
\begin{abstract}

The feature vector mapping used to represent chemical systems
is a  key factor governing the superior data-efficiency of
kernel based quantum machine learning (QML) models applicable throughout chemical compound space. 
%For kernel ML models, a key factor underpinning both their accuracy and computational cost is the form of the feature vector mapping used to represent the chemical system.
%In principle, a feature vector of the order $4N$ for any $N$ atom system should be sufficient. 
Unfortunately, the most accurate representations require a high dimensional feature mapping, thereby imposing a considerable computational burden on model training and use.
We introduce compact yet accurate, linear scaling QML representations based on atomic Gaussian many-body distribution functionals (MBDF), and their derivatives. % and their frequencies.
%Our representation significantly reduces the feature vector dimensionality and, by extension, the computational cost of the resulting models. 
%Herein we tackle this problem by using functionals of atomic Gaussian distributions to form efficient atomic representations that significantly reduce this dimensionality.
Weighted density functions (DF) of MBDF values are used as global representations which are constant in size, i.e.~invariant with respect to the number of atoms. 
%We further generate a density function of this atomic representation that does not scale with the number of atoms to test how well this information can be compressed into a compact form.
We report predictive performance and training data efficiency that is competitive with state of the art for two diverse datasets of organic molecules, QM9 and QMugs.
Generalization capability has been investigated for atomization energies, HOMO-LUMO eigenvalues and gap, internal energies at 0 K, zero point vibrational energies, dipole moment norm, static isotropic polarizability, and heat capacity as encoded in QM9. 
MBDF based QM9 performance lowers the optimal Pareto front spanned between sampling and training cost to compute node minutes,~effectively sampling chemical compound space with chemical accuracy at a sampling rate of $\sim 48$ molecules per core second.
%We also show that when using non-compact representations, the train/test time of kernel models is restricted by the kernel evaluation step which depends on the representation size.
%Furthermore, since the train/test time of kernel models is restricted by the kernel evaluation step, it depends on the representation size.
%We show that using our compact representation this bottleneck can be removed at $\sim$ 100k training set size for both small and large molecules.
\end{abstract}
\maketitle
\section{Introduction}
Modern data-driven statistical Machine Learning (ML) models have emerged as powerful tools over the past decade for inferring quantum mechanical observables throughout chemical compound space, without explicitly solving electronic Schrödinger equations\cite{CM, QMLessayAnatole, felix_google}.
Similar success was obtained for ML based interatomic potentials and force-fields\cite{Rabitz1996,NN_Tucker2006,Behler-Parrinello_NN,GAP,PES_NN,ANI-1}
as well as electronic structure modeling throughout Chemical Compound Space (CCS)\cite{ML4Kieron2012,ML4Graeme2012}. 
For an entire set of extensive in-depth reviews on these and other related ML applications, we refer the reader to the recent special issue in Chemical Reviews\cite{ceriotti2021editorial, ceriotti_clementi_anatole_jcp2021}.
%ML models are surrogate which all have in common that they exploit similarities between query and training instances.
%Given sufficient training data, Quantum Machine Learning (QML) models\cite{QMLessayAnatole}
%applicable across CCS have been shown to reach the accuracy of popular quantum chemistry methods such as density functional theory (DFT) for many molecular properties while incurring orders of magnitude less computational cost\cite{felix_google}.
Various aspects in the development of ML model architecture and training protocols have proven to be essential for data-efficiency. In particular, the molecular representation is known to have a strong impact on the performance of similarity based ML models, such as kernel ridge regression (KRR)\cite{FourierDesc., desc_role_Scheffler, physics-inspired-reps-ceriotti}.  
This is not surprising as the representation controls the information about the systems, how its weighed and the consistency of these Quantum Machine Learning (QML) models with ab-\textit{initio} methods~\cite{Ramakrishnan_vonLilienfeld_2015}.
These representations are non-linear mappings of the atomistic systems to a suitable Hilbert Space where a statistical regression model can easily be applied. 
The Hilbert Space constraint applies due to the requirement of measuring similarity in terms of an inner product\cite{Vapnik1998}.
These mappings should have some desirable features of which the most important are i) uniqueness such that systems with different properties necessarily possess different representations\cite{FourierDesc.}and 
ii) invariance with respect to transformations that leave the target property invariant, such as global translations, rotations, and atomic index permutations of the same chemical elements. Other desirable features include
iii) an analytical and continuous form of the representation function,
iv) differentiability with respect to nuclear coordinates, charges, number of electrons, and number of atoms,
v) as general as the Hamiltonian, 
vi) computationally efficient evaluation,
vii) compact or even constant size, 
limiting  the computational cost for larger systems\cite{const_size}. 
\\
Due to their critical role, many representations have been introduced and investigated within the context of atomistic simulation\cite{acsf,OM,mbtr,wacsf, ace,pip,wst,mtp,nice,gaussian_moments,kresse2020,snap,soap2018}. 
For recent comprehensive reviews, the reader is referred to Refs.~\citenum{reps_review_Rupp,physics-inspired-reps-ceriotti}.
These representations can either describe the molecule as a whole (global) or each atom (local or atomic) separately. 
For the sake of brevity we restricted most of our comparative benchmarks within this study to  the following representations which are commonly used to train QML models throughout CCS:
Faber-Christensen-Huang-Lilienfeld (FCHL19) representation\cite{fchl19,fchl18}, 
Smooth Overlap of Atomic Positions (SOAP)\cite{soap},
spectrum of London and Axilrod–Teller–Muto potentials (SLATM)\cite{amons_slatm}, 
atom index sorted Coulomb matrix (CM)\cite{CM}, 
and its vectorized form, Bag of Bonds (BOB)\cite{bob}. Other representations/models tested are mentioned in the data and code section.
\\
While these representations satisfy most of the aforementioned qualities, seen the immense size of CCS, a more compact and scalable representation would still be desirable. 
Formally, the number of degrees of freedom of any material or molecule would prescribe usage of a $4M$ dimensional feature vector (3 spatial coordinates and one nuclear charge coordinate for each of the atoms). 
However, all aforementioned representations when using optimal hyperparameters require a higher dimensional feature vector mapping in order to be accurate and training-data-efficient at regression tasks, and some (e.g. CM or BOB) even scale quadratically with $M$.
While verbosity facilitates the inclusion of invariances, 
the $4M$ degrees of freedom suggest that the same performance can be obtained using more compact representations.
This is especially an issue for kernel based ML models, where the size of the representation directly affects the distance/kernel evaluation time\cite{Vapnik1998, fchl19}. 
Although the scaling for kernel inversion\cite{gpr_rasmussen} is larger ($\propto \mathcal{O}(n^{3})$ for Cholesky solvers), 
for highly data-efficient (i.e. efficient in training data) QML models it is the kernel generation and evaluation that consumes the most compute time as demonstrated later.
The kernel evaluation pre-factor becomes even worse when using atomic (or local) representations in conjunction with a suitable local kernel\cite{local_kernels_ceriotti}.
Obvious solutions by reducing finite local cutoffs within the representation come at the expense of reducing the predictive power, or, conversely, increasing training data needs. 
As such, a computationally efficient yet accurate solution is desirable as was shown in the discretization of the FCHL18 representation\cite{fchl18, fchl19}.
Other solutions to this problem such as sparse kernel models\cite{sparse_GP_bartok} and the recently introduced conjugate gradients based iterative approach for training kernel based models\cite{conjugate_gradient_kernels} would also be well complemented by a compact molecular representation.
\\\\
Herein, we propose a methodology for generating representations that minimize feature size.
We use functionals of many-body distributions (MBDFs) and their derivatives to encode any local chemical environment of any atom in an integrated compact fashion.
The representations thus generated preserve the system's various symmetries (translation, rotation, atom index invariance), and can be tailored to the physical property of interest through scaling functions. 
MBDFs are easily extendable to include higher order many-body interactions with minimal increase in size. 
In the current formulation, while including three-body interactions, MBDF scales as $5M$. 
We further tackle the issue of storing this information in a manner that remains invariant to the number of atoms in a molecule. 
We do this by generating a discretized density (DF) function of MBDF values, and using it as a global molecular representation.
Using two diverse datasets the performance of MDBF is tested against aforementioned SOAP and FCHL19, which are commonly used and state-of-the-art, as well as SLATM, BOB, CM representations and a few other QML models
mentioned later.
Lastly, we explore the bottleneck cross-over from kernel evaluation to inversion. 
%MBDF based QML models, kernel inversion becomes the new computational bottleneck with respect to model evaluation efficiency. 
%explore why such a compact representation could be attractive given the aforementioned, pre-existing library of well performing and elegant molecular representations. 

\section{Theory and Discussion}
\subsection{Many-Body Distribution Functionals}
We begin the discussion of our local representation using distribution functions over the internal coordinates which can be constructed using the atomic coordinates.

An analytical and continuous distribution over the pair-wise internal coordinate defined as the inter-atomic distances (pair correlation function), is easily built using Gaussian probability density functions (PDFs) centered at each inter-atomic distance with respect to an atom $i$:
\begin{align}
    \rho_{i}(r,R_{ij})=\frac{1}{\sqrt{2\pi \sigma_{r}^2}}\sum_{j\neq i}^{M}Z_{j}\exp{\left(-{\frac{(r-R_{ij})^{2}}{\sigma_{r}}}\right)}
    \label{eq:rdf}
\end{align}
where $\rho_{i}(r,R_{ij})$ is the normalized distribution with atom $i$ as the origin, $\sigma_{r}$ is the Gaussian length-scale (or variance) parameter, $M$ denotes the total number of atoms in the system (or within a radial cutoff if employed), $R_{ij}$ denotes the inter-atomic distance between atoms $i$ and $j$ and $Z_{j}$ is the nuclear charge. 
Scaling by nuclear charges defines elemental identities, but could also be done in other ways e.g. having different length-scale parameters $\sigma_{r}$ for each unique chemical element, or multiple dimensions such as period and group specifications~\cite{Elpasolite_2016}. 
In a similar fashion, a continuous distribution (triplet correlation function) over the 3-body internal coordinate, inter-atomic angles  $\theta$, is defined as:
\begin{align}
    \rho_{i}(\theta,\theta_{jik})=\frac{1}{\sqrt{2\pi \sigma_{\theta}^2} }\sum_{j\neq i}^{M}\sum_{k\neq j}^{M}Z_{j}Z_{k}~\exp{\left(-\frac{(\theta-\theta_{jik})^{2}}{\sigma_{\theta}}\right)}, \nonumber
    \\
    \theta_{jik} = \cos^{-1}\frac{(\mathbf{R}_{i}-\mathbf{R}_{j})^{T}(\mathbf{R}_{i}-\mathbf{R}_{j})}{R_{ij}^{2}}
    \label{eq:adf}
\end{align}
where $\theta_{jik}$ is the inter-atomic angle centered on atom $i$.
This can be generalized to define a continuous distribution, or correlation function, over any $m$-body internal coordinate $\tau$.
\\
%Such internal coordinate distribution functions have previously been used as descriptors of atomic environments in Atom Centered Symmetry Functions\cite{acsf}, their weighted variants\cite{wacsf} and other similar methods\cite{fchl18, fchl19, amons_slatm}.
Such analytical distribution functions of the internal coordinates have been used as descriptors of atomic environments in Atom Centered Symmetry Functions\cite{acsf,Behler-Parrinello_NN} (ACSF), their weighted variants wACSFs\cite{wacsf} and other similar methods. 
Instead of using these distribution functions as atomic descriptors, we define the MBDF representation as functionals of these $m$-body distributions which leads to a more compact descriptor and allows inclusion of higher order terms with minimal change in size (a single scalar for each $m$-body term).
%Instead of using these distribution functions as atomic descriptors, we define the MBDF representation as functionals of these $m$-body distributions which leads to a single number corresponding to each distribution function resulting in a compact representation. 
With the 2- and 3-body distributions defined above, each atom can then be represented by the two zeroth-order functionals:
\begin{align}
    F_{0}^{(2)}[i] =\int_{0}^{r_{c}} dr~g_{0}(r,R_{ij})~\rho_{i}(r,R_{ij})
    \label{eq:F02}
\end{align}
\begin{align}
    F_{0}^{(3)}[i] =\int_{0}^{\pi}d\theta~ g_{0}(\theta,R_{ij},R_{jk},R_{ik})~\rho_{i}(\theta,\theta_{jik}) 
    \label{eq:F03}
\end{align}
where $r_{c}$ denotes the radial cut-off distance, and $g_{0}(r,R_{ij})$ and $g_{0}(\theta,R_{ij},R_{jk},R_{kj})$ denote 2- and 3-body weighting functions. 
Note that when the weighting functions $g_{0}(r,R_{ij})$ and $g_{0}(\theta,R_{ij},R_{jk},R_{kj})$ correspond to suitable 2 and 3-body potentials, the functionals $F_{0}^{(2)}$, $F_{0}^{(3)}$ become the average of the corresponding 2, 3-body inter-atomic interactions weighted by the pair and triplet correlation functions $\rho_{i}(r,R_{ij})$, $\rho_{i}(\theta,\theta_{jik})$, respectively. 
These functionals then form a coarse approximation to the average 2, 3-body interactions experienced by a chemical species. Furthermore, we exploit the advantage of using the infinitely differentiable Gaussian PDFs to define higher order functionals such as:

\begin{align}
    F_{1}^{(2)}[i] =\int_{0}^{r_{c}} dr~g_{1}(r,R_{ij})~\frac{\partial \rho_{i}}{\partial r }(r,R_{ij})
    \label{eq:F12}
\end{align}
\begin{align}
    F_{1}^{(3)}[i] =\int_{0}^{\pi} d\theta~g_{1}(\theta,R_{ij},R_{jk},R_{ik})~\frac{\partial \rho_{i}}{\partial \theta }(\theta,\theta_{jik}) 
\end{align}

with potentially different weighting functions $g_{1}(r)$, $g_{1}(\theta,R_{ij},R_{jk},R_{kj})$. 
The derivative functionals are useful since the functional of any arbitrary distribution is not unique. 
These functionals also encode the change in the $m$-body distribution in an atom's local neighborhood and have not been used in previous works involving internal coordinate distribution functions. 
For any $n$-th derivative of the 2-body distribution we can define the functional:
\begin{align}
    F_{n}^{(2)}[i] =\int_{0}^{r_{c}}dr~g_{n}(r) ~\partial_{r}^{n}\rho_{i}(r,R_{ij}), \nonumber
    \\
    \partial_{r}^{n}\rho_{i}(r) = \frac{\partial^{n} \rho_{i}}{\partial r^{n}}(r,R_{ij})
    \label{eq:Fn2}
\end{align}
where $g_{n}(r)$ is, again, a suitable radial weighting function.
Generalizing this to all internal coordinates, a functional $F_{n}^{(m)}[i]$ can be defined over the $n$-th derivative of any $m$-body distribution function centered at atom $i$:
\begin{multline}
    F_{n}^{(m)}[i] =\int_{0}^{\tau_{c}} d\tau~g_{n}(\tau) ~\partial_{\tau}^{n}\rho_{i}(\tau, \tau_{i_{1} i_{2}..i_{m}}),
    \\
    \partial_{\tau}^{n}\rho_{i}(\tau, \tau_{i_{1} i_{2}..i_{m}}) = \sum_{i_{1} < i_{2}.. < i_{m}}^{M} H_{n}(\tau - \tau_{i_{1} i_{2}..i_{m}})
    \\
    \times \mathcal{N}(\tau_{i_{1} i_{2}..i_{m}},\sigma_{\tau}^2) \prod_{j=i_{1}}^{i_{m}} Z_{j} 
    \label{eq:Fnm}
\end{multline}
where $\tau$ denotes the $m$-body internal coordinate, $\rho_{i}$ is the $m$-body distribution function w.r.t atom $i$, $g_{n}$ is the weighting function for the $n$-th derivative of $\rho_{i}$, $H_{n}$ denotes the Hermite polynomial of degree $n$ and $\mathcal{N}(\tau_{i_{1} i_{2}..i_{m}},\sigma_{\tau}^2)$ denotes the normalized Gaussian distribution. 
The Hermite polynomials arise due to the use of Gaussian PDFs and allow convenient computation of $n$ derivatives of the distribution function at any point $\tau$.
\\

We note here that an alternative way to describe a (bounded) distribution in a compact form is through a moment expansion of the form:
\begin{align}
    G^{(m)}[i] = \int_{0}^{\tau_{c}}d\tau ~ (\tau - \tau_{i_{1} i_{2}..i_{m}})^{m} g_{m}(\tau) \rho_{i}(\tau, \tau_{i_{1} i_{2}..i_{m}})
    \label{eq:Fnm_moments}
\end{align}
where $G^{(m)}[i]$ denotes the $m$-th moment of the distribution centered at atom $i$. 
The set of $m$ moments $G^{(m)}[i]$ would then form the local representation of the atom $i$. 
These moments can also be evaluated by placing a set of Gaussians (or any basis functions) on each atom $i$, and then evaluating the moments of the atomic density $\mathbf{\rho}_{i}(\mathbf{r})$ w.r.t each atomic position within a radial cutoff $\mathbf{r}_{c}$:
\begin{multline}
    G^{(m)}[i] = \int_{0}^{\mathbf{r}_{c}}d\mathbf{r}~|\mathbf{r}|^{m}g_{m}(|\mathbf{r}|) \rho_{i}(\mathbf{r}),
    \\
     \rho_{i}(\mathbf{r}) = \frac{1}{\sqrt{2\pi \sigma^2}} \sum_{j}\exp{\left(-\frac{||\mathbf{r}-\mathbf{R}_{ij}||_{2}^{2}}{2\sigma^{2}}\right)}
    \label{eq:moments_general}
\end{multline}
where $|.|$ is any metric.
This form has the advantage of being independent of many-body orders and the computational cost of evaluating these moments scales solely with the number of atoms within the cutoff radius $\mathbf{r}_{c}$.
The integral can be simplified in spherical polar coordinates by expanding the density $\rho_{i}(\mathbf{r})$ in a basis set composed of spherical harmonics $Y_{l}^{m'}$ and orthogonal radial functions $u_{n}$ (which is a common practice\cite{soap}):
\begin{multline}
    G^{(m)}[i] = \int_{0}^{r_{c}}\int_{0}^{2\pi}\int_{0}^{\pi}dr ~ d\theta ~ d\phi  ~ r^{m+2} ~ g_{m}(r) ~ \sin(\phi)
    \\
    \times \sum_{nlm'} c_{nlm'}^{i} ~ u_{n}(r) ~ Y_{l}^{m'} (\theta, \phi),
    \\
     c_{nlm'}^{i} = \braket{ \rho_{i} (\mathbf{r}) | u_{n}(r)~Y_{l}^{m'} (\theta, \phi)}
    \label{eq:moments_spherical}
\end{multline}
Throughout our work we use the derivative formalism from eq. \eqref{eq:Fnm} since our numerical results indicated superior performance than the moments expansion in the internal coordinates (eq. \eqref{eq:Fnm_moments}).
However, the moment expansion in eq. \eqref{eq:moments_general}, being independent of many-body terms, offers a promising alternative  and could be the subject of a future work.
\\

We have tested multiple weighting functions to identify the best combination. 
In particular, since these functionals correspond to correlation function averages of $m$-body interactions, we have tested the Harmonic, Morse\cite{morse}, Lennard-Jones\cite{lennard-jones2} potentials and simple decaying functions (power laws, exponential, gaussian decays and their combinations) for 2-body terms. 
For 3-body terms we have tested Cosine Harmonic,  Axilrod-Teller\cite{axilrod_teller}, Stillinger-Weber\cite{stillinger-weber}  potentials and a scaled Fourier series.
Through trial and error combined with cross-validated hyper-parameter optimization, we have identified the following 5 suitable functionals corresponding to 2 and 3-body distributions:
\begin{align}
    F_{0}^{(2)}[i]=&
    \int_{0}^{r_{c}}
    dr~ \left[e^{-\eta r}-\frac{\sqrt{2}\pi}{10(r+1)^{3}}\right]~\rho_{i}(r,R_{ij})
    \label{eq:F02}\\
    F_{1}^{(2)}[i]
    =&\frac{\sqrt{2}\pi}{10}
    \int_{0}^{r_{c}}dr~
    \frac{\partial_{r}\rho_{i}(r,R_{ij})}{(r+1)^{6}}
    \label{eq:F12}\\
    %F_{1}^{(2)}[i]
    %=&\frac{\sqrt{2}\pi}{10}
    %\int_{0}^{r_{c}}dr~
    %\frac{\sum_{j \neq i}^{M} Z_{j} H_{1}(r - R_{ij})}%{(r+1)^{6}} \mathcal{N}(R_{ij},\sigma_{r}^2)
    %\label{eq:F12}\\
    F_{2}^{(2)}[i]
    =&\int_{0}^{r_{c}}
    dr~e^{-\alpha r} ~  \partial_{r}^{2}\rho_{i}(r,R_{ij})
    \label{eq:F22}\\
    F_{0}^{(3)}[i]
    =&\int_{0}^{\pi}
    d\theta~\frac{\sum_{n=0}^{3} a_{n}\cos(n\theta)}{(R_{ij}R_{jk}R_{ik})^{4}}~\rho_{i}(\theta,\theta_{jik}) 
    \label{eq:F03}\\
    \begin{split}
    F_{1}^{(3)}[i]
    =&\int_{0}^{\pi}d\theta~
    \left[\frac{1+\cos(\theta)\cos(\theta_{kji})\cos(\theta_{ikj})}{(R_{ij}R_{jk}R_{ik})^{4}}\right]\\
    &\times \partial_{\theta}\rho_{i}(\theta,\theta_{jik})
    \end{split}
\label{eq:F13}
\end{align}
where $\eta$, $\alpha$, $a_{n}$ and the various power laws are all hyperparameters of the representation. 
Note the scaling functions of MBDF contributions $F_0^{(2)}, F_1^{(2)},F_0^{(3)},F_1^{(3)}$ being respectively reminiscent of Buckingham type-potential, softened London dispersion potential, Fourier series scaled by Lennard-Jones repulsion and Axilrod-Teller-Muto potential scaled by Lennard-Jones repulsion. 
The specific reason as to why this particular selection of weighting functions has proven advantageous will be subject of future research. 

\begin{figure}[h!]
          \centering          
          \includegraphics[width=\columnwidth]{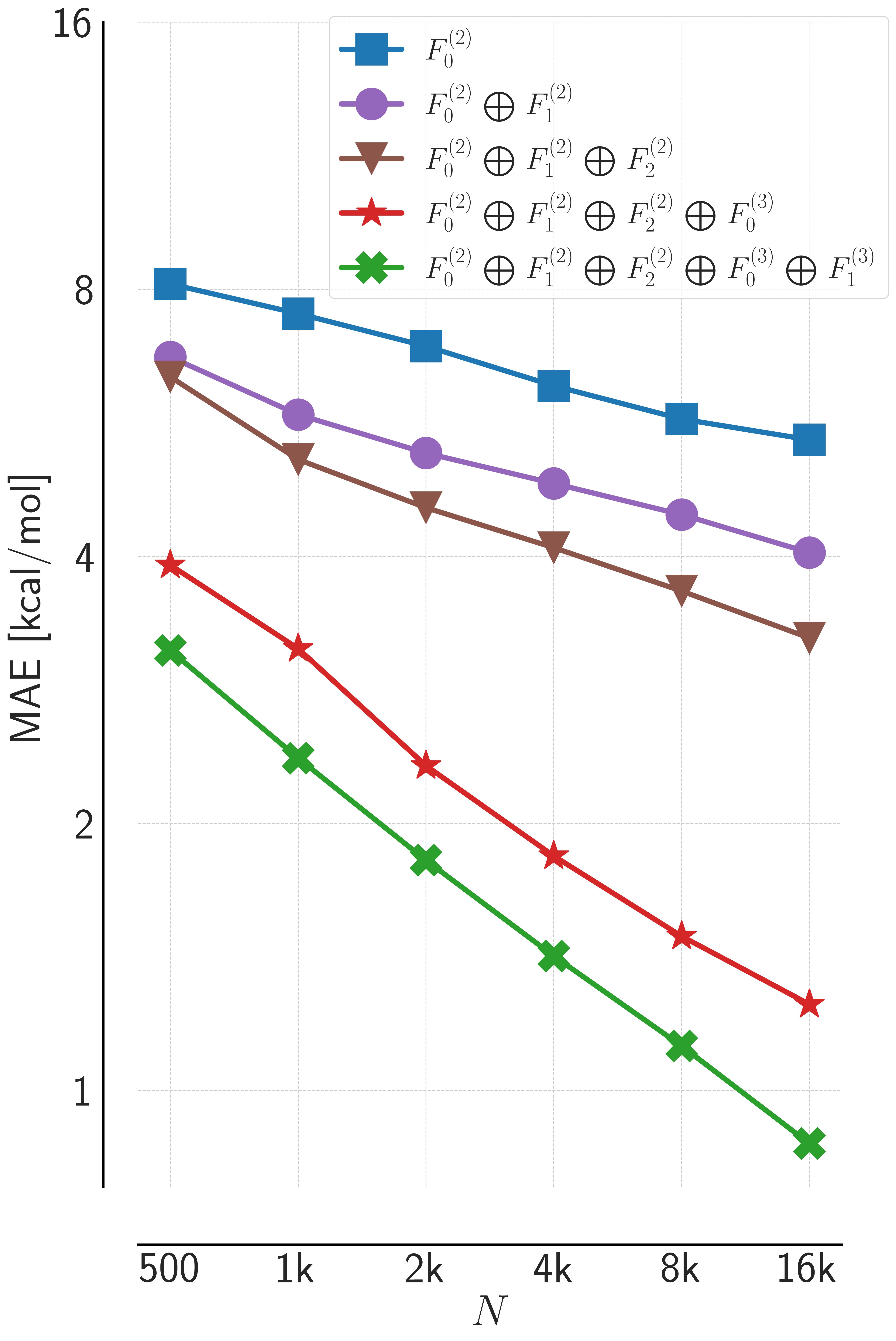}
          \caption{MBDF based QML learning curves using concatenated increasingly higher order and body functionals (Eqs.~(\ref{eq:F02}-\ref{eq:F13})). Mean absolute error (MAE) of predicting atomization energies of the QM9~\cite{qm9} dataset is shown as a function of training set size $N$.                    }
     \label{fig:fig2}
 \end{figure}

Figure \ref{fig:fig2} shows the effect of each functional on the learning capacity of MBDF for the task of predicting atomization energies of the QM9\cite{qm9} dataset.
It is apparent that both the derivative and many-body terms improve the learning capacity, albeit by different magnitudes. 

Throughout our testing on the QM7\cite{qm7}, QM9\cite{qm9} and QMugs\cite{QMugs} datasets, we have found these 2, 3-body functionals to be sufficient at discriminating between all of the molecular structures. 
Cases where the 2-body information does not suffice include homometric pairs, 
as already discussed many years ago~\cite{FourierDesc.}.
And even 3- and 4-body information does not suffice for some cases, as recently discussed in Ref.~\citenum{incompleteness_ceriotti}. 
We note here that, whenever necessary, arbitrarily higher order derivative and many-body information could also be included in MBDFs at minimal increase in size, i.e.~one additional term per order (See eq. \eqref{eq:Fnm}).
In particular, we believe that the inclusion of the 4-body term as a functional of the dihedrals could further improve the learning capacity for conformational isomers. 
Inclusion of 4-body information has been shown to result in further improvements of learning curves~\cite{communication_bing}.
%It would also allow the representation to discriminate between molecular conformers only differing in torsional angles, a feature that is lacked by most other representations (with finite cutoffs) discussed earlier \cite{incompleteness_ceriotti}.
Also note that the size of MBDF is invariant to the cutoffs used which can be raised to arbitrarily higher values while employing a suitable long-range functional (hence increasing the farsightedness of the representation) without affecting the kernel evaluation cost. 
We further note that other weighting functions and response terms could also be useful for QML models of physical observables such as dipole moments, vibrational frequencies, heat capacities etc.
%Due to the central role of the energy in terms of the Hamiltonian, and for the sake of brevity, we have restricted our function space search to the weighting functions mentioned earlier in this proof-of-principle study. 

\subsection{Density of functionals}
MBDF is a local representation and its size  scales linearly with the number of atoms in the system.
%This scaling can be eliminated by transforming
In order to eliminate this scaling we can transform to the frequency space of MBDF functional values.
The frequencies can be evaluated by normalizing the functional values to lie within an arbitrary range and then using, for e.g., kernel density estimation\cite{parzen_density}. 
For a finite set of MBDF functional values $\{X_{i}\}_{i=1}^{5M}$ over the range $[a,b]$, a "smooth histogram" of their frequencies can be constructed by placing a set of kernel functions $K$ at each point $X_{i}$:
\begin{align}
    f(u) = \frac{1}{5M}~\sum_{i=1}^{5M}~K(u,X_{i})
    \label{eq:kde}
\end{align}
where $f(u)$ gives the density of the samples at any point $u ~\epsilon ~[a,b]$. 
If the set $\{X_{i}\}_{i=1}^{5M}$ are the MBDF functional values for any molecule, their distribution density can be evaluated using eq. \eqref{eq:kde}.
The density $f(u)$ can then be used as a global molecular representation whose size is independent of the number of MBDF functional values, and number of atoms by extension. The (dis-)similarity between two molecules $A$ and $B$ can be evaluated as, e.g., the l2-distance:
\begin{align}
    d(A,B)^2 = \int_{-c}^{c} du~|f_{A}(u)-f_{B}(u)|^{2}
    \label{eq:dist_kde}
\end{align}
where $[-c,c]$ is the normaliztion range chosen as $[-10,10]$ in our work. The form of the density function used in our work is:
\begin{align}
    f_{A}(u) = \frac{1}{5M}~\sum_{i=1}^{5M}~\frac{\sqrt{|X_{i}|}}{\sqrt{2\pi}\sigma_{b}}~\exp{\left(-\frac{(u-X_{i})^2}{2\sigma_{b}^{2}}\right)}
    \label{eq:kde2}
\end{align}
where $X_{i}$ are MBDF functionals for molecule $A$, $\sigma_{b}$ is the variance of the Gaussian function and is a hyperparameter (also called bandwidth). Comparing with eq. \eqref{eq:kde}, the function $K$ is defined as: 
\begin{align}
    K(u,X_{i}) := \frac{\sqrt{|X_{i}|}}{\sqrt{2\pi}\sigma_{b}}~\exp{\left(-\frac{(u-X_{i})^2}{2\sigma_{b}^{2}}\right)}
    \label{eq:ku}
\end{align}
Note that this is a divergence\cite{amari2000methods} but not a kernel function because it is asymmetric: 
\begin{align}
    K(x,y) = &\frac{\sqrt{|y|}}{\sqrt{2\pi}\sigma_{b}}~\exp{\left(-\frac{(x-y)^2}{2\sigma_{b}^{2}}\right)}
    \nonumber
    \\
    \begin{split}
     \neq &\frac{\sqrt{|x|}}{\sqrt{2\pi}\sigma_{b}}~\exp{\left(-\frac{(x-y)^2}{2\sigma_{b}^{2}}\right)} = K(y,x)
    \end{split}
    \label{eq:div}
\end{align}
This function is used because it weights the MBDF functional frequencies by the functional value itself resulting in the distance measurement (eq. \eqref{eq:dist_kde}) being weighted by the difference in functional values.
Another advantage of this function is that it eliminates the frequency of null values (or "ghost atoms") within the MBDF representation which might be present due to the procedure of zero-padding\cite{CM}. 
\\
In our work we generate a separate density function $f(u)$ for each of the 5 MBDF functionals in eq. (11-15), and for each unique chemical element present in the dataset.
These are then concatenated to form the global representation of the molecule.
Alternatively, it could be done by using multivariate Gaussian functions for the density estimation. Let $\mathbf{x}_{i}$ denote the 5-dimensional vector of MBDF functional values (eq. 11-15) for any atom $i$ in molecule $A$. Then the multivariate density function $f_{A}(\mathbf{u})$ of this molecule can be evaluated as :
\begin{align}
    f_{A}(\mathbf{u}) = \frac{1}{M}~\sum_{i=1}^{M}~K(\mathbf{u},\mathbf{x}_{i})
    \label{eq:vector_kde}
\end{align}
\begin{align}
    f_{A}(\mathbf{u}) = \frac{1}{M}\sum_{i=1}^{M}\frac{(\mathbf{x}_{i}^{T}\mathbf{x}_{i})^{1/4}}{\sqrt{2\pi}\sigma_{b}}\exp{\left(-\frac{(\mathbf{u}-\mathbf{x}_{i})^{T}(\mathbf{u}-\mathbf{x}_{i})}{2\sigma_{b}^{2}}\right)}
    \label{eq:vector_kde2}
\end{align}
The l2-distance between molecules $A$ and $B$ then takes the form:
\begin{multline}
    d(A,B)^2 = \int d\mathbf{u}~(f_{A}(\mathbf{u})-f_{B}(\mathbf{u}))^{T}(f_{A}(\mathbf{u})-f_{B}(\mathbf{u}))
    \label{eq:dist_vector_kde}
\end{multline}
where the integral is over the normalization region.
Since the former method generates a more compact representation we have chosen to work with it. The abbreviation DF (Density of functionals) will be used throughout for this global representation.\\

\subsection{Numerical analysis}
\begin{figure}[htb]
          \centering
          \includegraphics[width=\columnwidth]{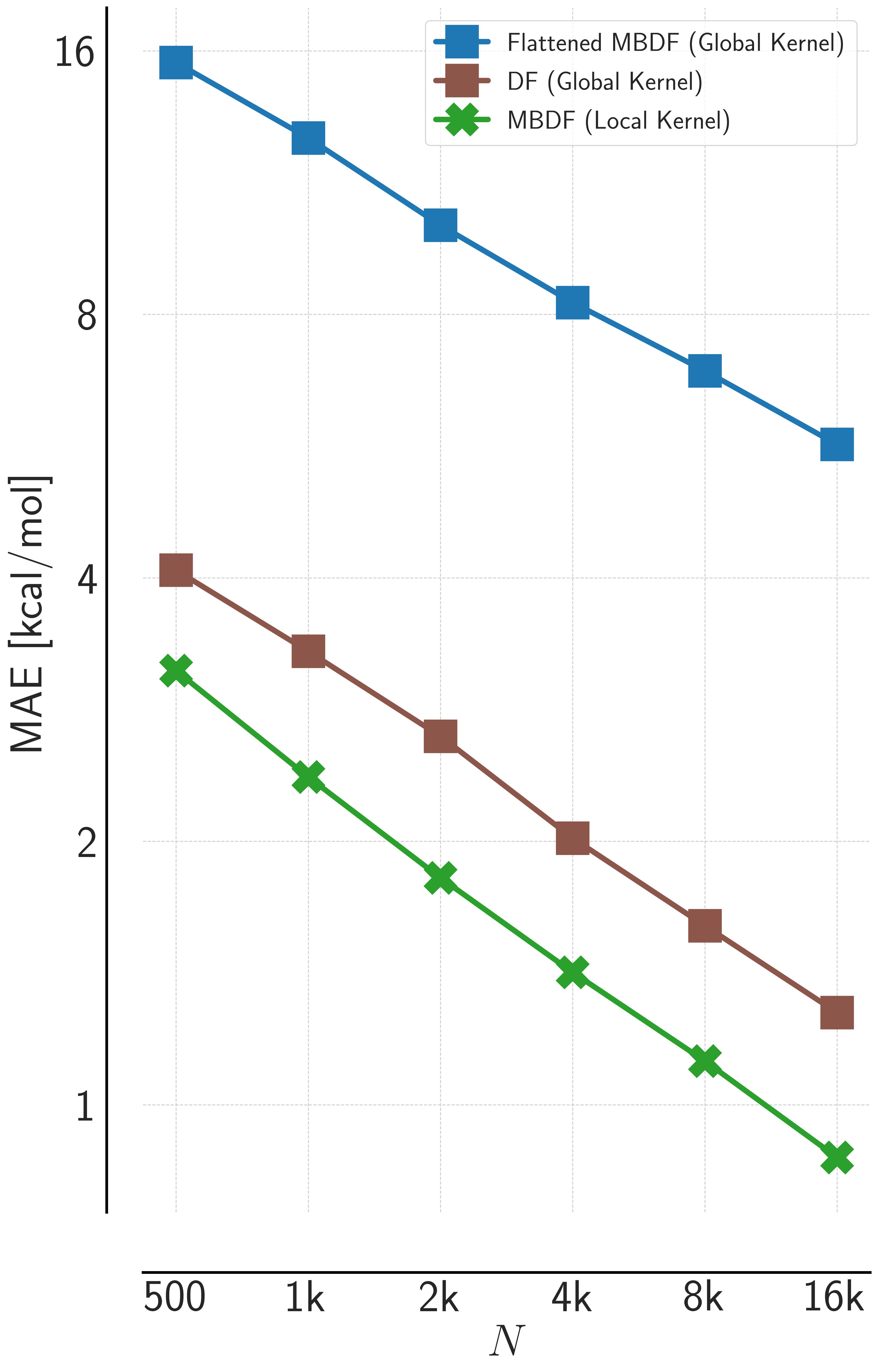}
     \caption{QML learning curves for MBDF with local kernel, DF (global), and flattened MBDF (global and sorted by row norm). Mean absolute error (MAE) of predicting atomization energies of the
            QM9~\cite{qm9} data-set as a function of training set size N.
          }
    \label{fig:flat_mbdf}
 \end{figure}

The DF method allows generating a representation that does not scale with the number of atoms in the system. 
However, in order to use it as a feature vector the density functions have to be discretized.
Through convergence testing we have set the grid spacing to 0.2 throughout our work.
However, we note that this grid spacing could be changed for a different data-set in order to achieve the desirable accuracy vs computational cost trade-off. 
Furthermore, DF corresponds to a flattened feature vector which can be used with global kernels (or other ML methods), and which exhibits superior performance when compared to a straightforward concatenation of all MBDF rows (see Figure \ref{fig:flat_mbdf}).
The flattened MBDF representation is generated by sorting the MBDF matrix of each molecule by the row norm, and then flattening the matrix by concatenation of the rows to vectorize it\cite{CM}.
 
Figure 3 shows molecular fingerprints generated using the 1 and 5 functional DF representations for three diverse and relevant organic  molecules (glucose, uric acid, and testosterone) on the same grid. 
For each molecule, a distinct fingerprint is obtained, with peak-positions depending  on the local chemical environment of each atom. Consequently, peaks of atoms with chemically similar environments are located closer to each other.
Peak heights encode both number and type (because of the density estimate being weighted) of chemical environments [See Eq.~\ref{eq:ku}]. 
Figure 3 demonstrates that for molecules with increasing size, corresponding DF based fingerprints will grow in magnitude, not in size.
In the SI, we also show how DF fingerprints distinguish conformational isomers, as exemplified for the chair and boat conformations of cyclohexane.

\begin{figure*}[htb]
          \centering
          \includegraphics[width=\linewidth]{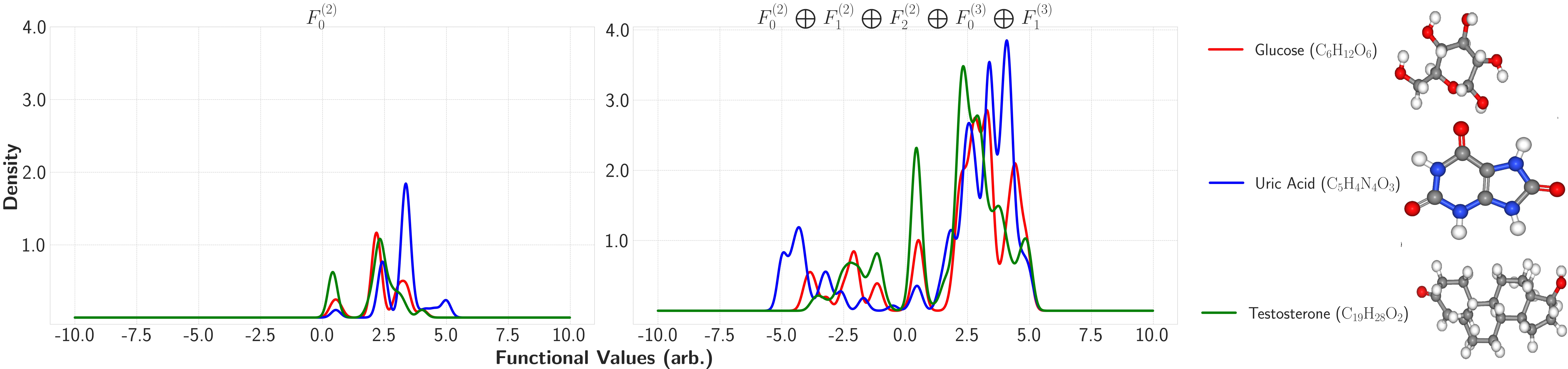}\label{fig:fig3a}
          \caption{Two-body (left) and three-body (right) DF versions (Eqs.~\ref{eq:F02}, \ref{eq:F03}) representations for Glucose, Uric Acid, and Testosterone. 
          }
 \end{figure*}

\section{Methods and Data}
\subsection{Kernel ridge regression}
The ML method that we focus on, and use throughout this work, is the supervised learning method called Kernel ridge regression\cite{Vapnik1998, gpr_rasmussen} (KRR). This method has been covered extensively earlier\cite{CM,fchl18,fchl19,felix_google,QMLessayAnatole} so we skip the details here.

The kernel functions\cite{Deisenroth2020,anatolebook} we use in our work along with global representations are the Gaussian kernel,
\begin{align}
    k(\mathbf{x}_I, \mathbf{x}_J) = \exp{\left( -\frac{\vert \vert \mathbf{x}_I - \mathbf{x}_J \vert \vert^{2}_{2}}{2 \sigma^2} \right)}
    \label{eq:kernel_gaussian}
\end{align}
and Laplacian kernel,
\begin{align}
    k(\mathbf{x}_I, \mathbf{x}_J) = \exp{\left( -\frac{\vert \vert \mathbf{x}_I - \mathbf{x}_J \vert \vert_{1}}{ \sigma} \right)}
    \label{eq:kernel_laplacian}
\end{align}
where $\mathbf{x}_I$ denotes the representation vector of molecule $I$.
\\
The kernel function used for the local representations FCHL19, SOAP and MBDF is a summation of atomic kernels:
\begin{align}
    k(\mathbf{M}_I, \mathbf{M}_J) = \sum_{a\epsilon i}\sum_{b\epsilon j} k^{l}(\mathbf{x}_{Ia}, \mathbf{x}_{Jb})
    \label{eq:kernel_local}
\end{align}
with the local Gaussian kernel:
\begin{align}
    k^{l}(\mathbf{x}_{Ia}, \mathbf{x}_{Jb}) = \mathbf{\delta}_{Z_{a},Z_{b}}  \exp{\left( -\frac{\vert \vert \mathbf{x}_{Ia} - \mathbf{x}_{Jb} \vert \vert^{2}_{2}}{2 \sigma^2} \right)}~
    \label{eq:local_gaussian}
\end{align}
the local Laplacian kernel:
\begin{align}
    k^{l}(\mathbf{x}_{Ia}, \mathbf{x}_{Jb}) =  \mathbf{\delta}_{Z_{a},Z_{b}}  \exp{\left( -\frac{\vert \vert \mathbf{x}_{Ia} - \mathbf{x}_{Jb} \vert \vert_{1}}{ \sigma} \right)}~
    \label{eq:local_laplacian}
\end{align}
or the local exponential kernel with the Euclidean norm:
\begin{align}
    k^{l}(\mathbf{x}_{Ia}, \mathbf{x}_{Jb}) =  \mathbf{\delta}_{Z_{a},Z_{b}}  \exp{\left( -\frac{\vert \vert \mathbf{x}_{Ia} - \mathbf{x}_{Jb} \vert \vert_{2}}{ \sigma} \right)}~
    \label{eq:local_exponential}
\end{align}
where $\mathbf{M}_I$ denotes the representation matrix of molecule $I$, $\mathbf{x}_{Ia}$ denotes the representation vector of atom $a$ within molecule $I$ and $\mathbf{\delta}_{Z_{a},Z_{b}} $ denotes a Kronecker Delta over the nuclear charges $Z_{a},Z_{b}$ which restricts the similarity measurement between atoms of the same chemical element\cite{fchl19}.
Other kernel functions will also be tested in the future\cite{wasserstein,metric_learning,local_kernels_ceriotti}.
\\
Throughout this study, we evaluate performance of ML methods through learning curves for the task of predicting physical properties of molecular systems. Learning curves quantify the model prediction error $\varepsilon$ (often measured as mean absolute error (MAE)) against the number of training samples $N$ and are key to understand the efficiency of ML models. 
It is generally known\cite{Vapnik1998,vapnik1994learningcurves,StatError_Muller1996} that they are linear on a log-log scale,
\begin{align}
    \log{ \left( \varepsilon \right) }   \approx I +  S \log{(N)}~
    \label{eq:log_learning}
\end{align}
where $I$ is the initial error and $S$ is the slope indicating model improvement given more training data. 
We also note that according to the central limit theorem the distribution of the errors $\epsilon$ approaches the normal distribution with standard deviation $\frac{\sigma}{\sqrt{N}}$, and mean 0 as $N\xrightarrow{} \infty$. Hence eq.~\eqref{eq:log_learning} becomes :
\begin{align}
    \log{ \left( \varepsilon \right) }   \approx \log(\sigma) -  \frac{1}{2} \log{(N)}
    \label{eq:log_learning_slope_half}
\end{align}
%Therefore, the expected slope $S$ of such a learning curve corresponds to $-\frac{1}{2}$.

\subsection{Hyperparameter Optimization} 
The current form of the representations has been optimized for Kernel based learning models. 
It depends on the weighting functions used and a number of hyperparameters which include variances ($\sigma_r, \sigma_\theta$) of the Gaussian PDFs, weighting function hyperparameters mentioned in eq. \eqref{eq:F02}-\eqref{eq:F13}, and bandwidth $\sigma_{b}$ for the DF representation. 
The hyperparameter optimization was done on a random subset of two thousand molecules from the QM7 dataset\cite{qm7}, and then kept fixed for all other data-sets. 
We note that further improvements might be possible if they had been
optimized simultaneously on all data-sets. 
The weighting functions $g_{n}(\tau)$ in eq. \eqref{eq:Fnm} 
were chosen by straightforward screening of the functions mentioned earlier. 
The optimization minimized the atomization energy prediction errors on the QM7 subset using Gaussian Process (GP) based Bayesian Optimization (BOpt)\cite{gpr_rasmussen}.
Starting with a Gaussian prior, the method fits a posterior distribution over the objective function using successive function evaluations. 
The posterior distribution is then used to construct an efficient acquisition function which can be optimized using, for instance, a quasi-Newton method to determine the next query point. 
Table 1 shows the optimized hyperparameter values used throughout this work. 
\begin{table}
\begin{center}
\begin{tabular}{ |c|c| } 
 \hline
 Parameter & Value  \\ 
 \hline
 $\sigma_{r}$ & 1  \\ 
 $\sigma_{\theta}$ & 2  \\
 $\eta$ & 10.8 \\
 $\alpha$ & 1.5 \\
 $a_{0}, a_{1}, a_{2}, a_{3}$ & 3, 100, -200, -164 \\
 $\sigma_{b}$ & 0.07 \\
 $r_{c}$ (Å) & 6 \\
 \hline
\end{tabular}
\caption{MBDF and DF hyperparameters after optimization on atomization energies of QM7~\cite{qm7} subset.}
\end{center}
\end{table}

We used the scikit-optimize\cite{scikit-optimize} implementation of GP based BOpt using the default Mat\'ern kernel with unit variance and the limited memory BFGS optimizer\cite{lbfgs} for the acquisition function. 
In order to enable comparison to other representations (such as FCHL or SOAP) that rely on distance cutoffs, we have chosen to set inter-atomic distance cutoff $r_{c}$ for MBDF  to 6 Å  throughout this work.
We note that larger cutoffs for MBDF would not change its size. 

All MBDF functionals were evaluated using the trapezoidal numerical integration method.
The grid spacing for discretizing DF densities has been set to 0.2 throughout our work as noted earlier.
The bandwidth $\sigma_{b}$ = 0.07 was found to work well on the QM7 subset however it is recommended to be screened once (in the range $\sim$ [0.01,1]) along with the grid-spacing when using with other datasets.
The Numpy\cite{numpy} and Numba\cite{numba} libraries are used in the representation generation code.
\subsection{Data and Code}
The QM9 dataset\cite{qm9} consists of $\sim$134k small organic molecules with up to 9 heavy atoms (C, N, O, F). The calculations were performed at the B3LYP/6-31G(2df,p)\cite{b3,lyp,631g2dfp} level of theory.\\
QMugs is a dataset containing $\sim$665k biologically and pharmacologically relevant drug-like molecules.
It consists of larger molecules than QM9 with up to 100 heavy atoms (C, N, O, F, P, S, Cl, Br, or I) per molecule.
The training and predictions were performed on the DFT ($\omega$B97X-D/def2-SVP)\cite{wb97xd,def2} values reported in the dataset.
The QMugs subsets we used for Figure \ref{fig:qmugs} were drawn at random and consist of 20k molecules. 
Throughout, we used zero-padding for all representations studied in order to accommodate training and test molecules smaller than the maximum present in the data. 

In order to keep the FCHL19 and SOAP kernel evaluations computationally tractable, 
we have (a) 
restricted ourselves to a maximum 100 atoms per QMugs-molecule,
and (b)
reduced the default hyperparameters of the FCHL19 and SOAP representations to nRs2 = 12, nRs3 = 10, $r_{cut}$ = 6$\mathrm{\AA}$ and $n_{max}$ = $l_{max}$ = 3, $\sigma$ = 0.1, $r_{cut}$ = 6$\mathrm{\AA}$, respectively. 
For consistency, we used the same parameters for all other results reported in this article.
These two versions of the representations with the reduced basis sets are denoted as FCHL19* and SOAP* in all reported figures.
Note that the latter choice of hyperparameters negligibly deteriorates the predictive accuracy for QM9 (as assessed below and when comparing to the published prediction errors on QM9 for FCHL19 and SOAP2013).
For FCHL19 and SOAP based prediction errors reported here within for QMugs, it could still be that the accuracy could improve further if these parameters were optimized. 

\begin{figure*}[htb]
          \centering           
          \includegraphics[width=\columnwidth]{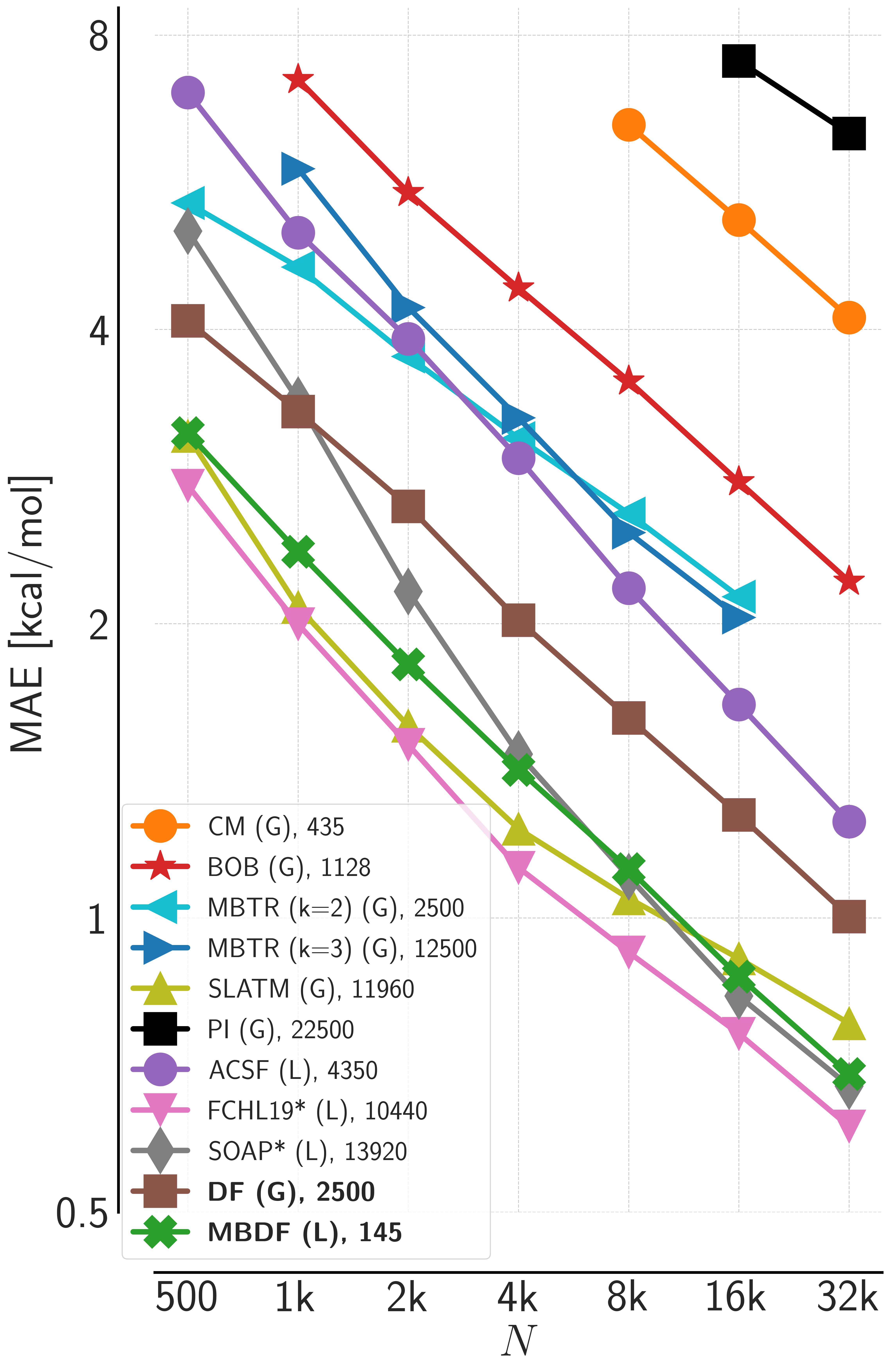}
          \includegraphics[width=\columnwidth]{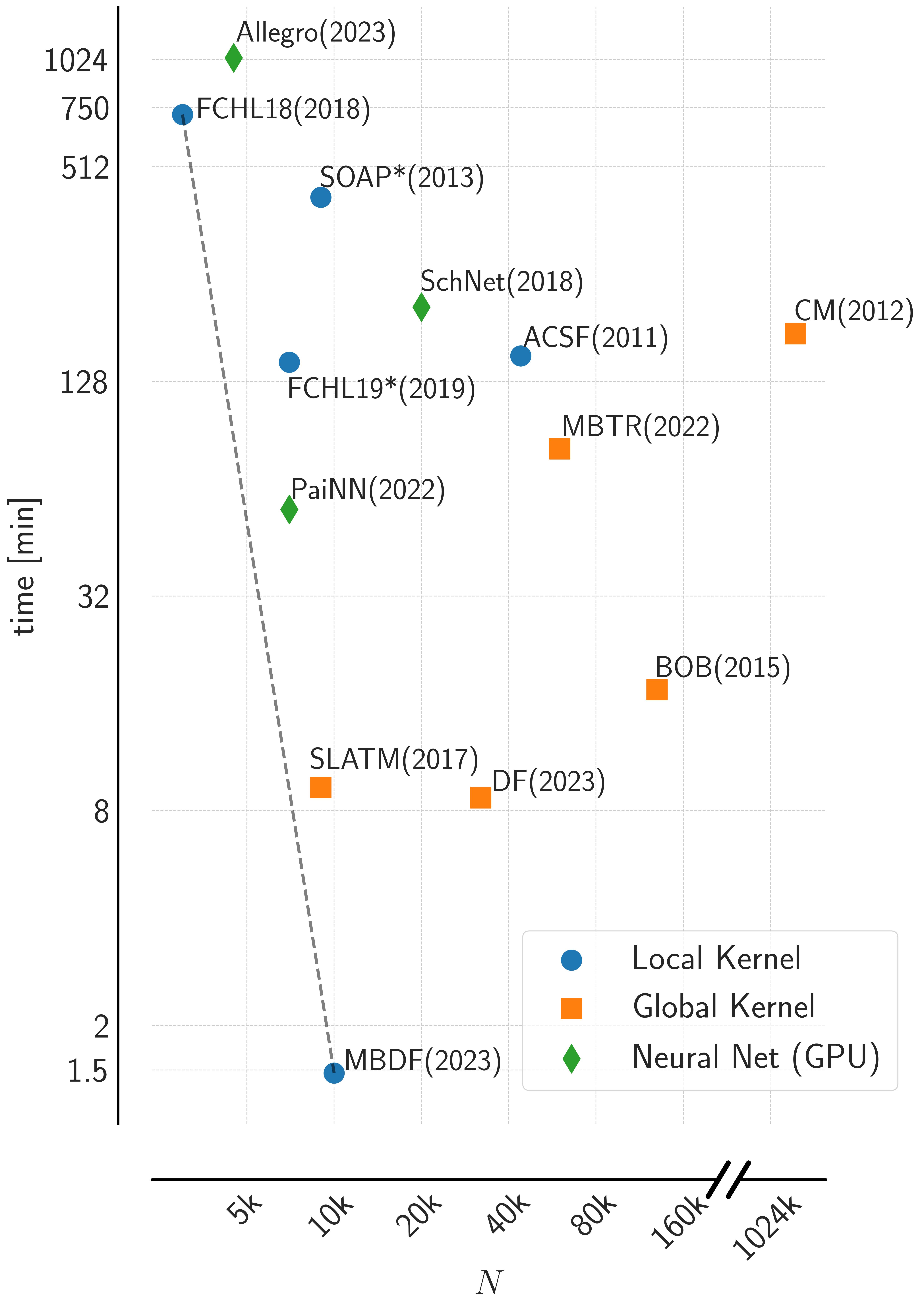}
          \caption{MBDF/DF performance and comparison on atomization energies from QM9 data set (drawn at random from $\sim$134k organic molecules)~\cite{qm9}. 
          a) (Left) Mean absolute error (MAE) of prediction as a function of training set size for representations CM~\cite{CM}, BOB~\cite{bob}, MBTR~\cite{mbtr}, SLATM~\cite{amons_slatm}, PI~\cite{PI_townsend}, ACSF~\cite{acsf,mbsf}, FCHL19~\cite{fchl19}, SOAP~\cite{soap}. Numbers in legend denote representation size (feature vector dimensions), G and L denote Global and Local kernels, respectively.
          b) (Right) Timing for training and testing as a function of training set size  
          required for making chemically accurate (MAE = 1 kcal/mol) predictions on 100k molecules. 
          Blue, red, and green points indicate local kernels, global kernels, and neural network, respectively. Dashed gray line corresponds to the optimal Pareto front.
          For SchNet~\cite{schnet}, PaiNN\cite{painn}, Allegro\cite{allegro} an Nvidia RTX 3070 GPU has been used. 
          All other timings 
          were evaluated on a compute node equipped
          with a 24-core AMD EPYC 7402P @ 3.3 GHz CPU and 512 GB RAM. 
          Timings for FCHL18~\cite{fchl18}, BOB, CM are estimated using smaller kernels 
          (not taking into account kernel inversion). 
          Asterisk denotes representations with reduced hyperparameters used in this work. 
          $N$ values for ACSF, MBTR, BOB, CM estimated via extrapolation. Numbers in brackets denote year of publication. 
          }
          %ssand 16-core AMD Ryzen 9 7950x @5.7 GHz CPU.}
     \label{fig:qm9}
 \end{figure*}

\begin{figure}[htb]
          \centering           
          \includegraphics[width=\columnwidth]{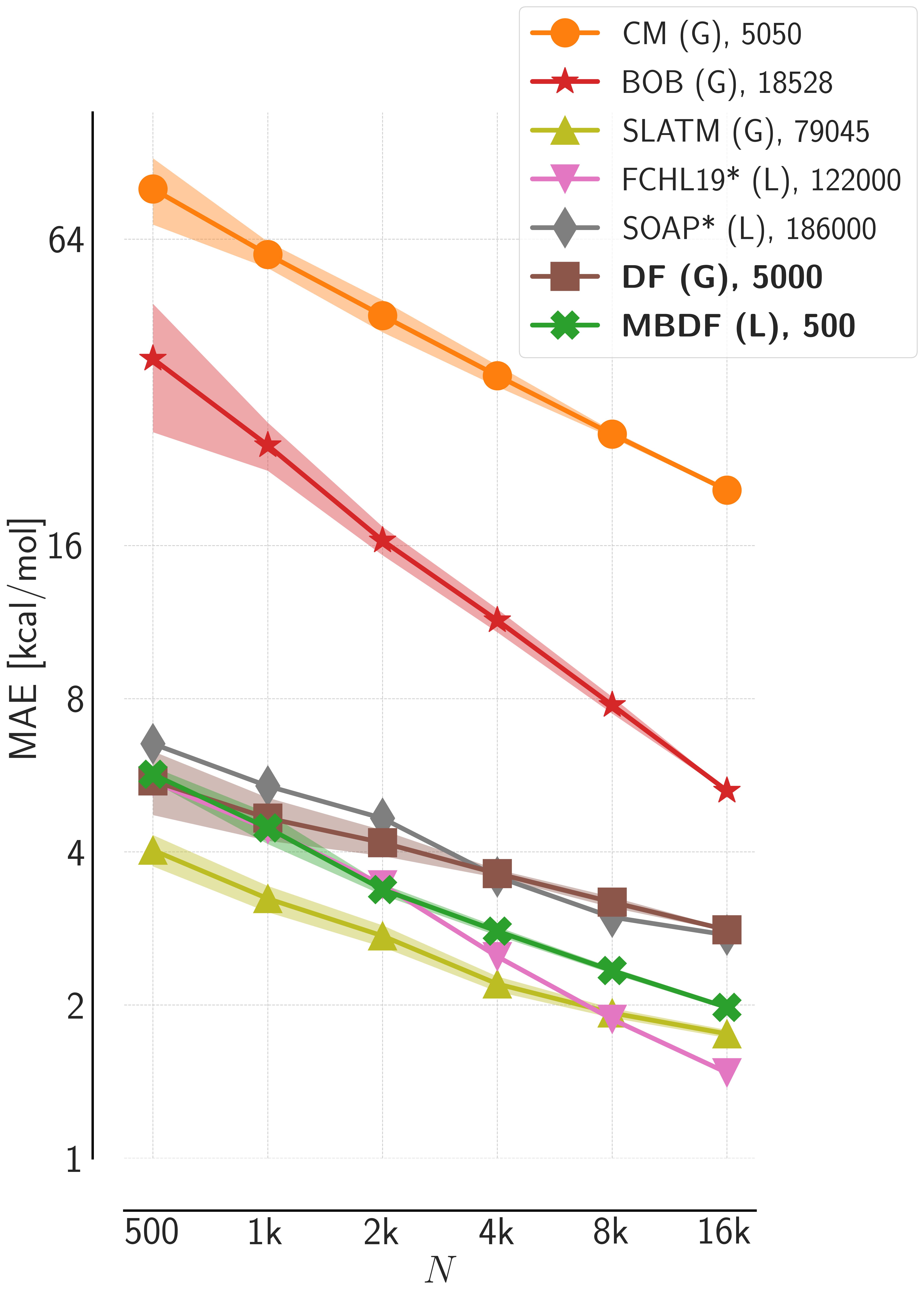}
          \caption{MBDF/DF performance and comparison to CM~\cite{CM}, BOB~\cite{bob}, SLATM~\cite{amons_slatm}, FCHL19~\cite{fchl19}, and SOAP~\cite{soap} representation based atomization energy estimates using the QMugs ($\sim$667k organic molecules with up to one hundred heavy atoms) data set~\cite{QMugs}. Training and testing data drawn at random. 
          Prediction mean absolute errors (MAE) on holdout test set of 4k molecules shown as a function of training set size. Numbers in legend denote representation size (feature vector dimensions), G and L denote Global and Local kernels respectively. Shaded region denotes the standard deviation across 4 different learning curves (except for FCHL19 and SOAP for which only one learning curve was tractable). 
          Asterisk denotes representations with reduced hyperparameters used in this work.}
     \label{fig:qmugs}
 \end{figure}

Figure \ref{fig:qm9} also includes results for QML models based on the 2- (k = 2) and 3-body (k = 3) Many-Body Tensor Representations (MBTR)\cite{mbtr} and a variant\cite{mbsf} of the Atom Centered Symmetry Functions\cite{acsf} (ACSF) as implemented in the QMLcode library\cite{qml}. 
The MBTR representations were generated with the same hyperparameters as those used in Ref.\citenum{reps_review_Rupp} for the 10k training point on the QM9 dataset.

Throughout our work, the FCHL19, SLATM and ACSF representations were generated using the QMLcode library\cite{qml},
SOAP was generated using the Dscribe library\cite{dscribe} with the default gaussian type radial basis functions, 
MBTR was generated using the qmmlpack library\cite{qmmlpack} and SchNet\cite{schnet}, PaiNN\cite{painn} were built and trained using the SchNetPack\cite{schnetpack} libray.
Allegro\cite{allegro} was built and trained using the nequip\cite{nequip} code\cite{nequip_zenodo} based on the E(3)-NN framework\cite{e3nn, e3nn_paper}.
\\
For FCHL19, SOAP and ACSF we employed the local Gaussian (eq. \eqref{eq:local_gaussian}),
for SLATM, MBTR and DF we use the global Gaussian (eq. \eqref{eq:kernel_gaussian}) and for CM, BOB we use the global Laplacian (eq. \eqref{eq:kernel_laplacian}) kernels respectively.
For MBDF we used the local Gaussian kernel on the MD17 dataset and the local exponential (eq. \eqref{eq:local_exponential}) kernel for all other models.
These choices were made based on the best performances for each representation.
All kernel evaluations were performed using the QMLcode library.

\section{Results}
\subsection{Atomization Energies}
\subsubsection{QM9} 
Figure \ref{fig:qm9} a) shows learning curves for QM9 and the size (dimension of feature vector) of the representation arrays in the legend. 
For the task of predicting atomization energies, local representations have previously been shown to be more efficient\cite{fchl19,fchl18}. 
Results for the local representations FCHL19 and SOAP are closely reproduced, and reach chemical accuracy after training on less than 10k molecules~\cite{fchl19}.
Among the global representations, SLATM has previously also been shown\cite{fchl19,amons_slatm} to perform quite well reaching chemical accuracy after training on $\sim$9k molecules although it shows a smaller slope at larger training sizes. 
This is closely followed by MBDF which reaches chemical accuracy after training on $\sim$10k molecules (less than 10\% of the dataset). 
The global DF representation also performs decently well reaching chemical accuracy at $\sim$32k training data. 
The local ACSF representation shows a larger offset but a better slope and it reaches chemical accuracy at $\sim$50k training set size.
We note here that for consistency with the other representations used in this work, we did not optimize the hyperparameters of the MBTR representation on every training point, 
%as was done in Ref.\cite{reps_review_Rupp}, 
but rather kept them fixed throughout.
Only the KRR hyperparameters were optimized at each training set size as with all other representations used here.
The 3-body MBTR reaches chemical accuracy at $\sim$60k training set size while the 2-body MBTR performs better than the other 2-body representations, BOB and CM.
We have also included the recently introduced constant size Persistence Images\cite{PI_townsend}(PI) representation for comparison. 
%unsure whether the default hyperparameters of this representation are suited to learning ground state physical properties of molecules.

Note that MBDF has the smallest size, requiring only 5 numbers per atom (145 dimensions for QM9 molecules).
By contrast, other local representations such as FCHL19, SOAP require $\sim$400 numbers per atom, while ACSF uses a 150 dimensional feature vector per atom. 
Encouragingly, and despite its compact size, MBDF outperforms most of the other larger representations with the exception of SOAP and FCHL.
We note here that while the size of the global DF representation is larger than MBDF, it utilises a global kernel implying training and prediction cost invariance with respect to system size.
%The MBDF representation uses a local kernel which requires a larger number of distance evaluations.

This compactness of the representation translates into faster ML model evaluation timings.
This is shown in Figure~\ref{fig:qm9} b) which plots the trade-off between training and prediction timings vs. training data needs for reaching mean absolute prediction errors of atomization energies of 1 kcal/mol.
%Figure 3 also shows the trade-off between accuracy and computational cost.
%On the x-axis we plot the number of training samples required to make chemically accurate predictions on a held out test set consisting of 100k molecules.
%The y-axis plots the total time required by each ML model for training and prediction on the test set.
We note that there are only two representations located on the Pareto-front, 
FCHL18~\cite{fchl18} and MBDF [this work]. 
We also point out that currently the best performing model on the QM9 dataset is the recently proposed Wigner Kernels method\cite{wigner_kernels} which is not included in this study.
\\
As noted earlier, local kernels based on representations such as FCHL18/19, SOAP, or ACSF exhibit very good training data efficiency,  but this comes at the expense of a larger computational overhead.
The exception is the local MBDF based kernel which achieves the fastest training  timing of $\sim$0.07 compute node minutes (14k training molecules) due to its compact size.
Predictions on 100k QM9 molecules using the local MBDF kernel are made in $\sim$1.46 compute node minutes which translates to an unprecedented chemically accurate navigation rate of $\sim$1140 molecules/second.
SLATM being clost to the Pareto front, and the DF representation, both represent fast global kernel based KRR models. 
While requiring more training data than SLATM in order to reach chemical accuracy, 
 DF has the advantage that it is largely invariant with respect to system size (see below). 
For the sake of completeness, Fig.~\ref{fig:qm9} b) also includes results for the deep learning based models SchNet\cite{schnet}, PaiNN\cite{painn} and the current state-of-the-art, Allegro\cite{allegro} which were all trained on a GPU. 
The reported timing for SchNet refers to 3000 epochs of training on 20k molecules and predictions of 100k molecules taking about 3 h: 27 min and 7 sec respectively.
For PaiNN the reported timing corresponds to 1000 epochs of training on 7k molecules which took 0 h: 56 mins and a prediction time of 9 sec.
Allegro reached chemical accuracy after training on 4500 molecules for 728 epochs, with early stoppping, which took 17 h: 8 min while the prediction on 100k molecules took $\sim$7 mins (using a maximum possible batch size of 5 on the used GPU).
%As such, the prediction times of MBDF based kernel model on a CPU even outperforms a deep neural network on a GPU. 
\\
MBDF achieves the fastest training time out of all models and the fastest prediction rate among the kernel based models.
Numerical results for porting KRR results based on FCHL19 to GPUs~\cite{browning2022gpu} would suggest that it seems likely for the prediction rate to increase significantly once MBDF is reimplemented in CUDA.

\subsubsection{QMugs}
We have tested the generalizability of our method to larger systems and more diverse chemistries using the QMugs~\cite{QMugs} data set. 
Figure \ref{fig:qmugs} shows the atomization energy learning curves.
Due to the large variety in the dataset, the predictive error is larger for all representations than their QM9 counterparts even when predicting on a much smaller test set.
MBDF reaches $~\sim$2 kcal/mol prediction error after training on 16k molecules. 
This is better than the QML based neural network predictions published in Ref.~\citenum{atz2022delta}, 
and similar to the $\Delta$-QML numbers they also report. 
In terms of speed, generating the local MBDF kernel for training and testing on 20k molecules on this dataset takes $\sim$1.8 compute node mins (see below) which corresponds to a navigation rate of $\sim$185 molecules/second.
By comparison, this is substantially faster than the GPU based prediction rates of approximately 50 and 5 molecules per second for the direct and
$\Delta$-learning (using GFN2-xTB~\cite{gfn2xtb}) based ML models, respectively using the convolutional neural network reported in Ref.~\citenum{atz2022delta}.
Only SLATM and FCHL19 exhibit lower off-set than MBDF, 
while the performance for SOAP and DF is similar, albeit slightly worse than MBDF. 
As mentioned before, however, in order to make FCHL19 and SOAP tractable, we
have dramatically reduced the hyper parameters. 
In particular, we believe that the learning efficiency of SOAP for QMugs is 
being reduced due to the use of small basis sets ($n_{max}$~=~$l_{max}$ = 3).
%With a larger basis set we expect it to perform similarly well as FCHL19.
Note that no representation reaches chemical accuracy within 16k training molecules, 
indicating that QMugs possesses substantially more chemical diversity than QM9. 
%This indicates that a more sophisticated selection of the training data could be beneficial in such scenarios\cite{sml_dom}.

%Similar trends in accuracy as QM9 are seen here with FCHL19 performing the best followed by SLATM and MBDF.
%The local representations, again, perform better with FCHL19 achieving the best performance followed by MBDF.
%The global DF representation also shows competetitive accuracy given that its size changes the least when moving to these larger molecules.
%The large disparity between CM/BOB and the other representations is, again, likely due to their lack of 3-body terms. 
In terms of representation sizes, 
MBDF again remains the smallest representation since it still requires only 5 dimensions per atom regardless of the chemical composition. 
However, being a local representation,
on average its size increases $\sim$ 3.4 times compared to QM9.
FCHL19 and SOAP, on the other hand, now
require more than 1000 dimensions to represent each atom for this larger dataset. 
CM, BOB, FCHL19, and SOAP show larger than 10 fold increase in the representation size compared to QM9, followed by SLATM which shows an increase of $\sim$ 6.6 times. 
This results in a considerable increase in the train/test time ({\em vide infra}) 
which precludes the straightforward application of these representations to the entire QMugs dataset.
The DF representation changes the least in size since it does not formally scale with number of atoms but only with number of different chemical elements. 
Consequently, its size doubles compared to QM9 since  a separate density function is generated for each unique chemical element in the dataset using eq. \eqref{eq:kde2} as mentioned earlier.

\subsubsection{QM7b and MD17}
Figure 2 in the SI shows learning curves for the QM7b\cite{qm7b,970m_qm7b} atomization energies and the
size (dimension of feature vector) of the representation
arrays in the legend.
Similar trends in performance and representation size noted so far are observed on this dataset as well.
\\
Figure 3 in the SI shows learning curves of energies for a few molecules from the revised\cite{anders_gradients_role,rmd17} MD17\cite{gdml,sgdml} molecular dynamics dataset.
Although the comparative performance trend on this dataset is similar to the others, we note that the chosen functionals for MBDF and the current representation hyperparameters are not optimized for potential energy surface learning.
Furthermore, the implementation of gradients for MBDF to enable force based learning should significantly enhance the performance for such tasks as was shown\cite{anders_gradients_role} for the FCHL19 representation.
However, the relatively larger difference in performance between MBDF and FCHL19 on this dataset might indicate that  compact representations are sufficient for regression through equilibrium structures across CCS while more verbose representations are required for potential energy surface learning.

\subsection{Timings}
\begin{figure}[htb]
          \centering           
          \includegraphics[width=\columnwidth]{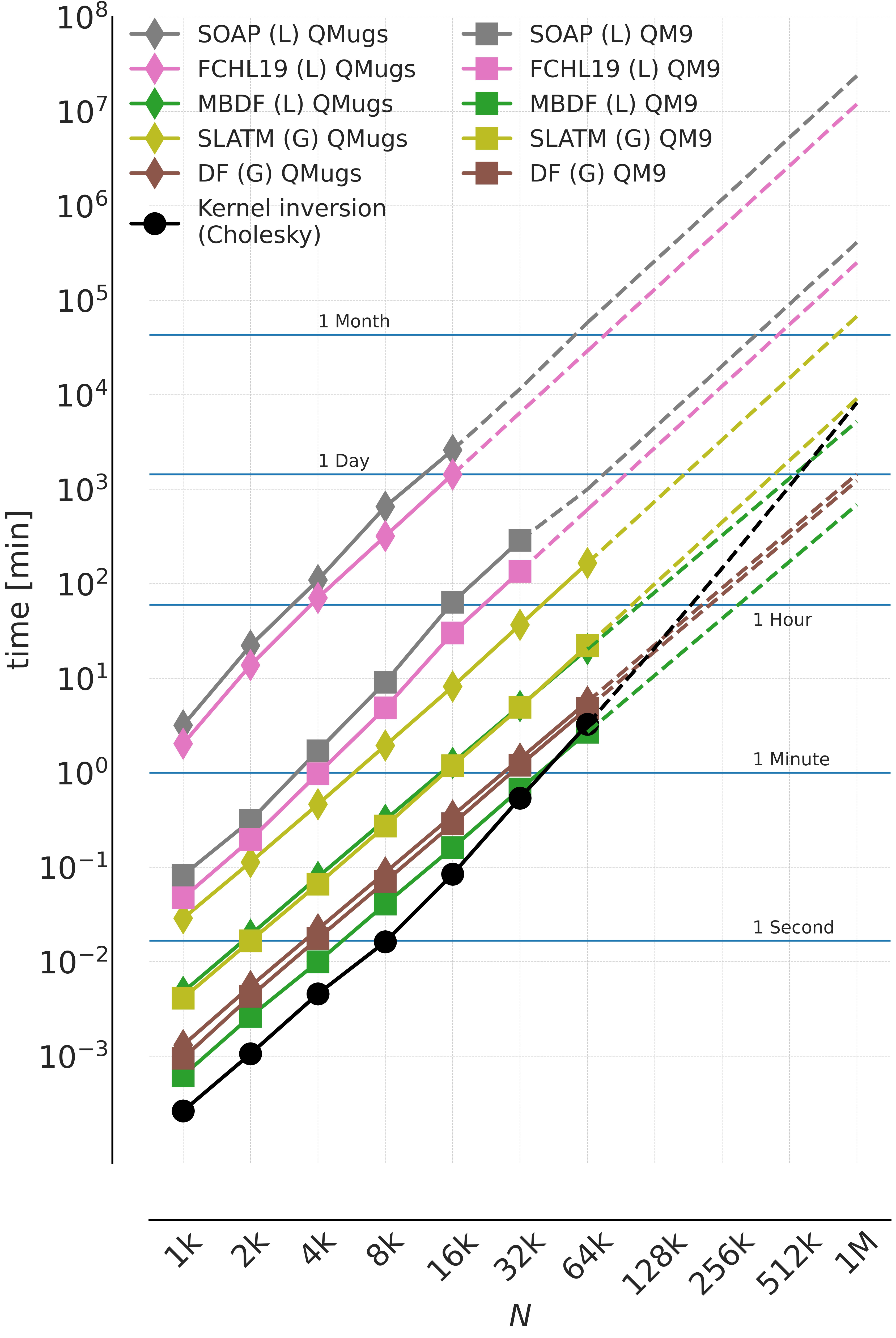}
          \caption{Compute node time required for kernel inversion and evaluation 
           as a function training set sizes $N$ drawn from QM9 (squares) and QMugs (diamonds) datasets. QML results shown for SOAP~\cite{soap}, FCHL19~\cite{fchl19}, SLATM~\cite{amons_slatm}, and MBDF and DF. 
           Dotted lines indicate extrapolation (using quadratic (kernel evaluation) and cubic (kernel inversion) polynomial fits).  G and L denote Global and Local kernels, respectively. Compute node: 24-core AMD EPYC 7402P @ 3.3 GHz CPU with 512 GB RAM.
          }
     \label{fig:fig6}
 \end{figure}

Figure \ref{fig:fig6} shows scaling plots of kernel evaluation timings across both QM9 and QMugs datasets for various representations and as a function of training set size.
As one would expect, the off-set increases systematically with increasing representation and training molecule size.
More specifically, for the larger molecules of QMugs, 
the FCHL19 and SOAP kernel evaluations become rapidly 
intractable very quickly with the 16k kernel 
(Fig.~\ref{fig:qmugs}) 
already taking an entire compute node day.
Encouragingly and in stark contrast, DF, being a size invariant representation, 
shows hardly any change in computational overhead when moving from the small QM9 molecules to the much larger QMugs molecules.

For context, the time required in the kernel inversion step for each training kernel is shown as well.
The bottleneck crossover from kernel generation ($\mathcal{O}(n^2)$) to the inversion step ($\mathcal{O}(n^3)$) occurs rather late. 
When using Cholesky decomposition it occurs for MBDF at $N \sim$ 64k training molecules. For less compact representations (for SLATM and larger) the same cross-over occurs at training set sizes that exceed $\sim$1 M. 
As demonstrated above, chemical accuracy is already achieved at substantially smaller training set sizes. 
Consequently and contrary to popular belief, for any of the more modern and accurate representations, kernel inversion does not constitute a bottleneck. 

Table 2 reports the kernel evaluation and representation generation timings for both the QM9 dataset (130k molecules) and the QMugs subset (20k molecules) used in our work.
It can be seen that MBDF reduces the local kernel evaluation timings from days to a few minutes for both small and large molecules.
For the representation generation step we note that our code is currently written in Python and uses the Numba\cite{numba} library and could be further optimized with a low-level implementation.
However, the current timings as well do not affect the overall QML model cost much given the kernel evaluation bottleneck.
\begin{table*}[htb]
\begin{tabular}{l|
S[table-number-alignment = right]
S[table-number-alignment = right]
r|
S[table-number-alignment = right]
S[table-number-alignment = right]
r}
& \multicolumn{3}{|c|}{\textbf{QM9 130k Molecules}} & \multicolumn{3}{c}{\textbf{QMugs 20k Molecules}} \\
\hline
Representation   & \multicolumn{1}{l}{$t_\mathrm{rep}$ [min]} &  \multicolumn{1}{l}{$t_\mathrm{kernel}$ [min]}   &  \multicolumn{1}{l|}{Dimension} & \multicolumn{1}{l}{$t_\mathrm{rep}$ [min]}  &  \multicolumn{1}{l}{$t_\mathrm{kernel}$ [min]} &  \multicolumn{1}{l}{Dimension}\\ 
\hline
CM$^{a}$ (G)      &  0.186              & 2.862            &      435       & 0.012               & 1.146             &    5050 \\
BoB$^{a}$ (G)     &  0.216              & 7.296            &     1128       & 1.362               & 3.396             &   18528 \\
SLATM$^{a}$ (G)   &  18.60              & 86.32            &    11960       & 15.76               & 14.58             &   79045 \\
FCHL19$^{a}$ (L)  &  0.846              & 1071             &    10440       & 1.764               & 1566              &  122000 \\
SOAP$^{b}$ (L)    &  0.216              & 1873             &    13920       & 0.246               & 2925              &  186000 \\
\rowcolor{Gray}
\textbf{MBDF (L)} &  1.626              & 11.81            & 145            & 4.182               & 1.848             & 500 \\
\rowcolor{Gray}
\textbf{DF (G)}   &  2.262              & 12.16            & 2500           &  2.442              & 0.996             & 5000\\
\hline
\end{tabular}
\caption{Compute node times for generating representations  ($t_\mathrm{rep}$) and kernel matrices ($t_\mathrm{kernel}$) for 130k molecules from the QM9 dataset and 20k molecules from the QMugs dataset. 
Global and Local kernels are again denoted by (G) and (L) respectively. Representations with superscript $a$ were generated with QMLcode\cite{qml} library and $b$ with the Dscribe\cite{dscribe} library. 
Compute node: 24-core AMD EPYC 7402P @ 3.3 GHz CPU and 512 GB RAM.}
\label{table}
\end{table*}

\subsection{Performance for molecular quantum properties}
\begin{figure*}[htb]
          \centering           
          \includegraphics[width=\linewidth]{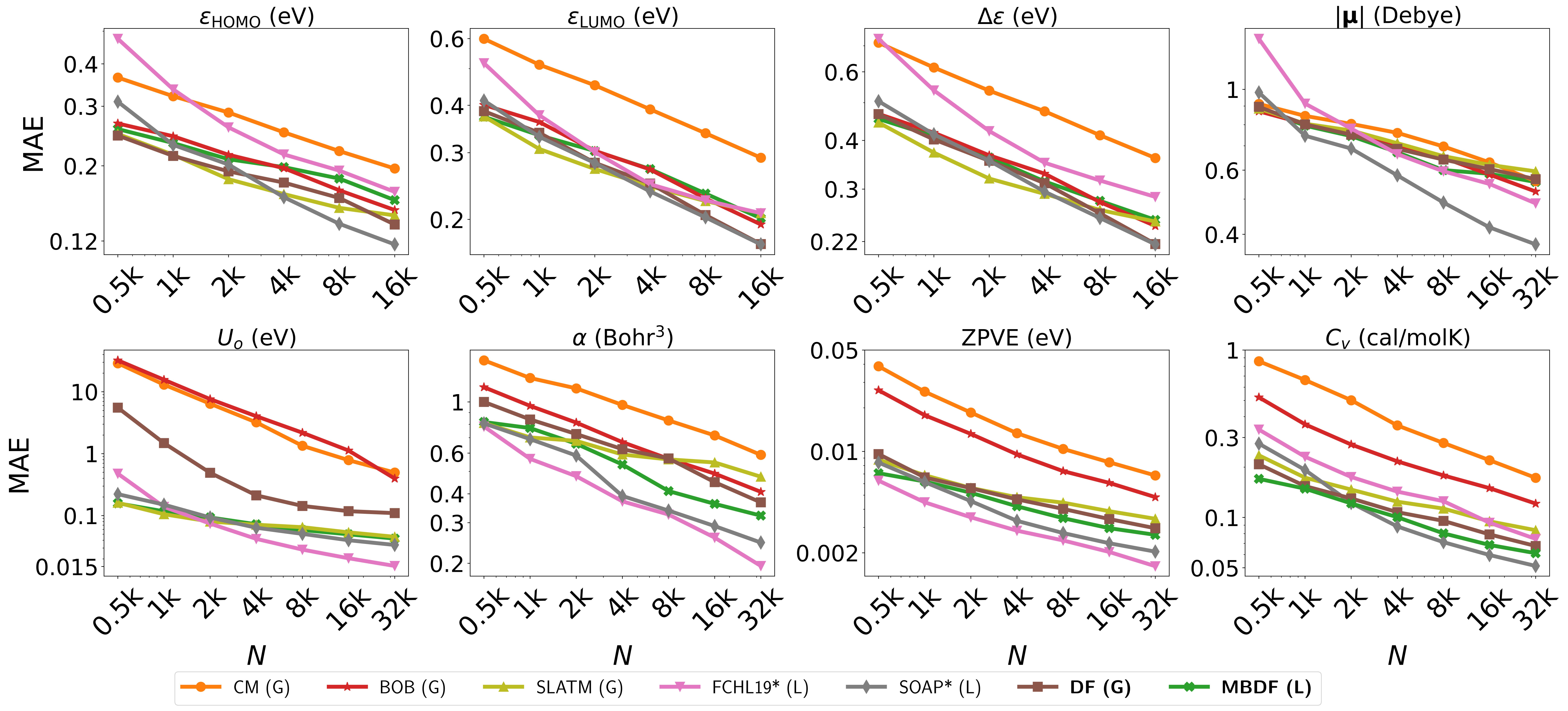}
          \caption{Learning curves for various representations in QML models of highest occupied molecular orbital (HOMO), lowest unoccupied molecular orbital (LUMO) eigenvalues, HOMO-LUMO gap ($\Delta \epsilon$), internal energy at 0 K ($U_{0}$), dipole moment norm ($|\boldsymbol{\mu}|$), static isotropic polarizability ($\alpha$), zero point vibrational energy (ZPVE), and heat capacity ($C_{v}$) from the QM9 dataset. G and L denote Global and Local kernels, respectively.
          Asterisk denotes representations with reduced hyperparameters used in this work.}
     \label{fig:other_props}
 \end{figure*}

We assessed the generalization capacities of MBDF/DF on physical properties other than atomization energies.
Figure \ref{fig:other_props} shows the learning curves for the task of predicting 8 important molecular quantum properties from the QM9 dataset. 
These include the highest occupied molecular
orbital (HOMO), lowest unoccupied molecular orbital
(LUMO) eigenvalues and the HOMO-LUMO gap, internal energy at 0 K ($U_{0}$),
dipole moment norm ($|\boldsymbol{\mu}|$), static isotropic polarizability ($\alpha$), zero point vibrational energy
(ZPVE) and heat capacity ($C_{v}$).
Due to substantial computational costs, the KRR hyper-parameters were not optimized at each training size for the FCHL19 and SOAP representations, and we picked the same parameters as those used for learning atomization energies in figure \ref{fig:qm9}.
We reproduce earlier trends among intensive and extensive properties when using local/global kernels~\cite{op_response_anders,felix_google}. 
MBDF and DF match this trend: They perform better on extensive and on intensive properties, respectively.

Again, note that the performance on these properties of MBDF/DF could be further improved by including different functionals suited to the property, or by augmenting them with response terms as was done for the FCHL19 representation\cite{op_response_anders}.
It would also be interesting to see how the learning capacities across these different physical properties is affected by the inclusion of higher order functional terms. 

\section{Conclusion}
We have introduced ultra-compact atomic local many body distribution functional (MBDF) and global density of functionals (DF) representations for use within kernel ridge regression based Quantum Machine Learning (QML) models for rapid sampling of chemical compound space. 
MBDF/DF can accurately describe any atom/molecule using a minimal number of discrete elements and thereby reduce QML training and prediction times by multiple orders of magnitude for small and large molecules.
MBDF and DF correspond to functionals of analytical weighted many body distributions in interatomic distances and angles, as well as their derivatives. 
Chemical identity is encoded as a prefactor to the atomic functionals. 
DF is a weighted density estimations of MBDF, i.e.~a global molecular fingerprint that is invariant with respect to number of atoms (not number of chemical elements).

We have demonstrated predictive power competitive with the state-of-the-art for a variety of quantum physical properties, as encoded in the QM9 dataset~\cite{qm9}. 
On the QM9 dataset MBDF reaches a MAE of atomization energies of only 0.69 kcal/mol after training on 32k molecules while using only 5 dimensions to describe an atom. 
Regarding different molecular properties, it is beneficial to use DF representation along with a global kernel for intensive properties, and MBDF along with a local kernel for the extensive properties.
MBDF and DF generalize also well to other compound spaces, as evinced for the  chemically more diverse QMugs dataset~\cite{QMugs} consisting of substantially larger molecules: After training on 16k molecules, MBDF reaches a peak MAE for atomization energies of 1.97 kcal/mol while still using only 5 dimensions per atom. 
Corresponding training and prediction times for both data-sets is $\sim$2 compute node time minutes for MBDF based models. 
Furthermore, our results indicate that using MBDF/DF brings the train/test timings of kernel based ML models to their "lower-bound" as imposed by the kernel inversion bottleneck for both small and large molecules.

We have analyzed the comparative performance for the sampling cost vs.~training set size needs to reach chemical accuracy for predicting atomization 
energies in the QM9 data-set. 
MBDF has extended the corresponding optimal Pareto front towards minimal time requirements.
%whereas FCHL19 reaches an accuracy of 1.49 kcal/mol while requiring 1220 elements to describe the atoms, with the cutoff radii still at 6 Å. 

While the numerical evidence collected is indicative of a surprising effectiveness of MBDF, it is clear that the truncations used in this work may lead to lack of uniqueness. However, neither within QM9 nor within the QMugs subset we sample, have we encountered a case where using this small set of functionals maps two different molecular structures to the same feature vector.
Furthermore, the likelihood of uniqueness can easily be increased by inclusion of  higher order derivatives and many-body terms. 
More rigorous numerical analysis would be desirable to quantify this trend.

%Our compact global representation shows peak performances on the two datasets that are better than all the other global representations tested in our work. 

Thanks to its numerical efficiency, we believe that this approach holds great promise for further accelerating the virtual discovery of novel molecules and materials. 
Furthermore, this framework provides a possible solution to the general problem of unfavourable scaling due to i) inclusion of higher order many body and long range physics and ii) applying these ML models to larger molecules with greater chemical diversity. 
Future work will deal with extension of the representations as described to deal with various types of conformers, and an assessment of its sensitivity to changes in molecular structures.

\section{Acknowledgments}
D.K. is thankful for discussions with B.~Huang, S.~Krug, D.~Lemm, M.~Sahre, and J.~Weinreich. 
O.A.v.L. has received funding from the European Research Council (ERC) 
under the European Union’s Horizon 2020 research and innovation programme (grant agreement No. 772834).
O.A.v.L. has received support as the Ed Clark Chair of Advanced Materials and as a Canada CIFAR AI Chair.

\section{Supplementary material}
See supplementary material for QM7b\cite{qm7b}, MD17\cite{gdml, sgdml, rmd17, anders_gradients_role} learning curves and a figure showing the transition of cyclohexane from chair to boat conformation alongside the response by its DF fingerprint. 
It also contains heat maps showing ideal KRR hyperparameters to be used with the MBDF representation and kernel PCAs of some molecules from the QM7b dataset.

\section{Code and Data Availability}
Python script for generating the MBDF and DF representations is available at \url{https://github.com/dkhan42/MBDF}. Train/test split of the QM7 dataset used for optimizing representation hyperparameters, other
data and code for reproducing the results in this study and some tools for performing KRR are openly available at \url{https://github.com/dkhan42/QML}.
\bibliographystyle{apsrev4-1}
\bibliography{literature.bib}

\begin{thebibliography}{93}%
\makeatletter
\providecommand \@ifxundefined [1]{%
 \@ifx{#1\undefined}
}%
\providecommand \@ifnum [1]{%
 \ifnum #1\expandafter \@firstoftwo
 \else \expandafter \@secondoftwo
 \fi
}%
\providecommand \@ifx [1]{%
 \ifx #1\expandafter \@firstoftwo
 \else \expandafter \@secondoftwo
 \fi
}%
\providecommand \natexlab [1]{#1}%
\providecommand \enquote  [1]{``#1''}%
\providecommand \bibnamefont  [1]{#1}%
\providecommand \bibfnamefont [1]{#1}%
\providecommand \citenamefont [1]{#1}%
\providecommand \href@noop [0]{\@secondoftwo}%
\providecommand \href [0]{\begingroup \@sanitize@url \@href}%
\providecommand \@href[1]{\@@startlink{#1}\@@href}%
\providecommand \@@href[1]{\endgroup#1\@@endlink}%
\providecommand \@sanitize@url [0]{\catcode `\\12\catcode `\$12\catcode
  `\&12\catcode `\#12\catcode `\^12\catcode `\_12\catcode `\%12\relax}%
\providecommand \@@startlink[1]{}%
\providecommand \@@endlink[0]{}%
\providecommand \url  [0]{\begingroup\@sanitize@url \@url }%
\providecommand \@url [1]{\endgroup\@href {#1}{\urlprefix }}%
\providecommand \urlprefix  [0]{URL }%
\providecommand \Eprint [0]{\href }%
\providecommand \doibase [0]{http://dx.doi.org/}%
\providecommand \selectlanguage [0]{\@gobble}%
\providecommand \bibinfo  [0]{\@secondoftwo}%
\providecommand \bibfield  [0]{\@secondoftwo}%
\providecommand \translation [1]{[#1]}%
\providecommand \BibitemOpen [0]{}%
\providecommand \bibitemStop [0]{}%
\providecommand \bibitemNoStop [0]{.\EOS\space}%
\providecommand \EOS [0]{\spacefactor3000\relax}%
\providecommand \BibitemShut  [1]{\csname bibitem#1\endcsname}%
\let\auto@bib@innerbib\@empty
%</preamble>
\bibitem [{\citenamefont {Rupp}\ \emph
  {et~al.}(2012{\natexlab{a}})\citenamefont {Rupp}, \citenamefont {Tkatchenko},
  \citenamefont {M\"uller},\ and\ \citenamefont {von Lilienfeld}}]{CM}%
  \BibitemOpen
  \bibfield  {author} {\bibinfo {author} {\bibfnamefont {M.}~\bibnamefont
  {Rupp}}, \bibinfo {author} {\bibfnamefont {A.}~\bibnamefont {Tkatchenko}},
  \bibinfo {author} {\bibfnamefont {K.-R.}\ \bibnamefont {M\"uller}}, \ and\
  \bibinfo {author} {\bibfnamefont {O.~A.}\ \bibnamefont {von Lilienfeld}},\
  }\href {\doibase 10.1103/PhysRevLett.108.058301} {\bibfield  {journal}
  {\bibinfo  {journal} {Phys. Rev. Lett.}\ }\textbf {\bibinfo {volume} {108}},\
  \bibinfo {pages} {058301} (\bibinfo {year} {2012}{\natexlab{a}})}\BibitemShut
  {NoStop}%
\bibitem [{\citenamefont {von Lilienfeld}(2018)}]{QMLessayAnatole}%
  \BibitemOpen
  \bibfield  {author} {\bibinfo {author} {\bibfnamefont {O.~A.}\ \bibnamefont
  {von Lilienfeld}},\ }\href@noop {} {\bibfield  {journal} {\bibinfo  {journal}
  {Angewandte Chemie International Edition}\ }\textbf {\bibinfo {volume}
  {57}},\ \bibinfo {pages} {4164} (\bibinfo {year} {2018})},\ \bibinfo {note}
  {http://dx.doi.org/10.1002/anie.201709686}\BibitemShut {NoStop}%
\bibitem [{\citenamefont {Faber}\ \emph {et~al.}(2017)\citenamefont {Faber},
  \citenamefont {Hutchison}, \citenamefont {Huang}, \citenamefont {Gilmer},
  \citenamefont {Schoenholz}, \citenamefont {Dahl}, \citenamefont {Vinyals},
  \citenamefont {Kearnes}, \citenamefont {Riley},\ and\ \citenamefont {von
  Lilienfeld}}]{felix_google}%
  \BibitemOpen
  \bibfield  {author} {\bibinfo {author} {\bibfnamefont {F.~A.}\ \bibnamefont
  {Faber}}, \bibinfo {author} {\bibfnamefont {L.}~\bibnamefont {Hutchison}},
  \bibinfo {author} {\bibfnamefont {B.}~\bibnamefont {Huang}}, \bibinfo
  {author} {\bibfnamefont {J.}~\bibnamefont {Gilmer}}, \bibinfo {author}
  {\bibfnamefont {S.~S.}\ \bibnamefont {Schoenholz}}, \bibinfo {author}
  {\bibfnamefont {G.~E.}\ \bibnamefont {Dahl}}, \bibinfo {author}
  {\bibfnamefont {O.}~\bibnamefont {Vinyals}}, \bibinfo {author} {\bibfnamefont
  {S.}~\bibnamefont {Kearnes}}, \bibinfo {author} {\bibfnamefont {P.~F.}\
  \bibnamefont {Riley}}, \ and\ \bibinfo {author} {\bibfnamefont {O.~A.}\
  \bibnamefont {von Lilienfeld}},\ }\href {\doibase 10.1021/acs.jctc.7b00577}
  {\bibfield  {journal} {\bibinfo  {journal} {Journal of Chemical Theory and
  Computation}\ }\textbf {\bibinfo {volume} {13}},\ \bibinfo {pages} {5255}
  (\bibinfo {year} {2017})},\ \bibinfo {note} {pMID: 28926232},\ \Eprint
  {http://arxiv.org/abs/https://doi.org/10.1021/acs.jctc.7b00577}
  {https://doi.org/10.1021/acs.jctc.7b00577} \BibitemShut {NoStop}%
\bibitem [{\citenamefont {Ho}\ and\ \citenamefont {Rabitz}(1996)}]{Rabitz1996}%
  \BibitemOpen
  \bibfield  {author} {\bibinfo {author} {\bibfnamefont {T.}~\bibnamefont
  {Ho}}\ and\ \bibinfo {author} {\bibfnamefont {H.}~\bibnamefont {Rabitz}},\
  }\href {\doibase http://dx.doi.org/10.1063/1.470984} {\bibfield  {journal}
  {\bibinfo  {journal} {J. Chem. Phys.}\ }\textbf {\bibinfo {volume} {104}},\
  \bibinfo {pages} {2584} (\bibinfo {year} {1996})}\BibitemShut {NoStop}%
\bibitem [{\citenamefont {Manzhos}\ and\ \citenamefont {{Carrington,
  Jr.}}(2006)}]{NN_Tucker2006}%
  \BibitemOpen
  \bibfield  {author} {\bibinfo {author} {\bibfnamefont {S.}~\bibnamefont
  {Manzhos}}\ and\ \bibinfo {author} {\bibfnamefont {T.}~\bibnamefont
  {{Carrington, Jr.}}},\ }\href@noop {} {\bibfield  {journal} {\bibinfo
  {journal} {J. Chem. Phys.}\ }\textbf {\bibinfo {volume} {125}},\ \bibinfo
  {pages} {084109} (\bibinfo {year} {2006})}\BibitemShut {NoStop}%
\bibitem [{\citenamefont {Behler}\ and\ \citenamefont
  {Parrinello}(2007)}]{Behler-Parrinello_NN}%
  \BibitemOpen
  \bibfield  {author} {\bibinfo {author} {\bibfnamefont {J.}~\bibnamefont
  {Behler}}\ and\ \bibinfo {author} {\bibfnamefont {M.}~\bibnamefont
  {Parrinello}},\ }\href {\doibase 10.1103/PhysRevLett.98.146401} {\bibfield
  {journal} {\bibinfo  {journal} {Phys. Rev. Lett.}\ }\textbf {\bibinfo
  {volume} {98}},\ \bibinfo {pages} {146401} (\bibinfo {year}
  {2007})}\BibitemShut {NoStop}%
\bibitem [{\citenamefont {Bart\'ok}\ \emph {et~al.}(2010)\citenamefont
  {Bart\'ok}, \citenamefont {Payne}, \citenamefont {Kondor},\ and\
  \citenamefont {Cs\'anyi}}]{GAP}%
  \BibitemOpen
  \bibfield  {author} {\bibinfo {author} {\bibfnamefont {A.~P.}\ \bibnamefont
  {Bart\'ok}}, \bibinfo {author} {\bibfnamefont {M.~C.}\ \bibnamefont {Payne}},
  \bibinfo {author} {\bibfnamefont {R.}~\bibnamefont {Kondor}}, \ and\ \bibinfo
  {author} {\bibfnamefont {G.}~\bibnamefont {Cs\'anyi}},\ }\href {\doibase
  10.1103/PhysRevLett.104.136403} {\bibfield  {journal} {\bibinfo  {journal}
  {Phys. Rev. Lett.}\ }\textbf {\bibinfo {volume} {104}},\ \bibinfo {pages}
  {136403} (\bibinfo {year} {2010})}\BibitemShut {NoStop}%
\bibitem [{\citenamefont {Handley}\ and\ \citenamefont
  {Popelier}(2010)}]{PES_NN}%
  \BibitemOpen
  \bibfield  {author} {\bibinfo {author} {\bibfnamefont {C.~M.}\ \bibnamefont
  {Handley}}\ and\ \bibinfo {author} {\bibfnamefont {P.~L.~A.}\ \bibnamefont
  {Popelier}},\ }\href {\doibase 10.1021/jp9105585} {\bibfield  {journal}
  {\bibinfo  {journal} {The Journal of Physical Chemistry A}\ }\textbf
  {\bibinfo {volume} {114}},\ \bibinfo {pages} {3371} (\bibinfo {year}
  {2010})},\ \bibinfo {note} {pMID: 20131763},\ \Eprint
  {http://arxiv.org/abs/https://doi.org/10.1021/jp9105585}
  {https://doi.org/10.1021/jp9105585} \BibitemShut {NoStop}%
\bibitem [{\citenamefont {Smith}\ \emph {et~al.}(2017)\citenamefont {Smith},
  \citenamefont {Isayev},\ and\ \citenamefont {Roitberg}}]{ANI-1}%
  \BibitemOpen
  \bibfield  {author} {\bibinfo {author} {\bibfnamefont {J.~S.}\ \bibnamefont
  {Smith}}, \bibinfo {author} {\bibfnamefont {O.}~\bibnamefont {Isayev}}, \
  and\ \bibinfo {author} {\bibfnamefont {A.~E.}\ \bibnamefont {Roitberg}},\
  }\href {\doibase 10.1039/C6SC05720A} {\bibfield  {journal} {\bibinfo
  {journal} {Chem. Sci.}\ }\textbf {\bibinfo {volume} {8}},\ \bibinfo {pages}
  {3192} (\bibinfo {year} {2017})}\BibitemShut {NoStop}%
\bibitem [{\citenamefont {Snyder}\ \emph {et~al.}(2012)\citenamefont {Snyder},
  \citenamefont {Rupp}, \citenamefont {Hansen}, \citenamefont {M\"uller},\ and\
  \citenamefont {Burke}}]{ML4Kieron2012}%
  \BibitemOpen
  \bibfield  {author} {\bibinfo {author} {\bibfnamefont {J.~C.}\ \bibnamefont
  {Snyder}}, \bibinfo {author} {\bibfnamefont {M.}~\bibnamefont {Rupp}},
  \bibinfo {author} {\bibfnamefont {K.}~\bibnamefont {Hansen}}, \bibinfo
  {author} {\bibfnamefont {K.-R.}\ \bibnamefont {M\"uller}}, \ and\ \bibinfo
  {author} {\bibfnamefont {K.}~\bibnamefont {Burke}},\ }\href@noop {}
  {\bibfield  {journal} {\bibinfo  {journal} {Phys. Rev. Lett.}\ }\textbf
  {\bibinfo {volume} {108}},\ \bibinfo {pages} {253002} (\bibinfo {year}
  {2012})}\BibitemShut {NoStop}%
\bibitem [{\citenamefont {Pozun}\ \emph {et~al.}(2012)\citenamefont {Pozun},
  \citenamefont {Hansen}, \citenamefont {Sheppard}, \citenamefont {Rupp},
  \citenamefont {M\"uller},\ and\ \citenamefont {Henkelman}}]{ML4Graeme2012}%
  \BibitemOpen
  \bibfield  {author} {\bibinfo {author} {\bibfnamefont {Z.~D.}\ \bibnamefont
  {Pozun}}, \bibinfo {author} {\bibfnamefont {K.}~\bibnamefont {Hansen}},
  \bibinfo {author} {\bibfnamefont {D.}~\bibnamefont {Sheppard}}, \bibinfo
  {author} {\bibfnamefont {M.}~\bibnamefont {Rupp}}, \bibinfo {author}
  {\bibfnamefont {K.-R.}\ \bibnamefont {M\"uller}}, \ and\ \bibinfo {author}
  {\bibfnamefont {G.}~\bibnamefont {Henkelman}},\ }\href@noop {} {\bibfield
  {journal} {\bibinfo  {journal} {J. Chem. Phys.}\ }\textbf {\bibinfo {volume}
  {136}},\ \bibinfo {pages} {174101} (\bibinfo {year} {2012})}\BibitemShut
  {NoStop}%
\bibitem [{\citenamefont {Ceriotti}\ \emph
  {et~al.}(2021{\natexlab{a}})\citenamefont {Ceriotti}, \citenamefont
  {Clementi},\ and\ \citenamefont {Anatole~von
  Lilienfeld}}]{ceriotti2021editorial}%
  \BibitemOpen
  \bibfield  {author} {\bibinfo {author} {\bibfnamefont {M.}~\bibnamefont
  {Ceriotti}}, \bibinfo {author} {\bibfnamefont {C.}~\bibnamefont {Clementi}},
  \ and\ \bibinfo {author} {\bibfnamefont {O.}~\bibnamefont {Anatole~von
  Lilienfeld}},\ }\href@noop {} {\bibfield  {journal} {\bibinfo  {journal}
  {Chem. Rev.}\ }\textbf {\bibinfo {volume} {121}},\ \bibinfo {pages} {9719}
  (\bibinfo {year} {2021}{\natexlab{a}})}\BibitemShut {NoStop}%
\bibitem [{\citenamefont {Ceriotti}\ \emph
  {et~al.}(2021{\natexlab{b}})\citenamefont {Ceriotti}, \citenamefont
  {Clementi},\ and\ \citenamefont {Anatole~von
  Lilienfeld}}]{ceriotti_clementi_anatole_jcp2021}%
  \BibitemOpen
  \bibfield  {author} {\bibinfo {author} {\bibfnamefont {M.}~\bibnamefont
  {Ceriotti}}, \bibinfo {author} {\bibfnamefont {C.}~\bibnamefont {Clementi}},
  \ and\ \bibinfo {author} {\bibfnamefont {O.}~\bibnamefont {Anatole~von
  Lilienfeld}},\ }\href {\doibase 10.1063/5.0051418} {\bibfield  {journal}
  {\bibinfo  {journal} {The Journal of Chemical Physics}\ }\textbf {\bibinfo
  {volume} {154}},\ \bibinfo {pages} {160401} (\bibinfo {year}
  {2021}{\natexlab{b}})},\ \Eprint
  {http://arxiv.org/abs/https://doi.org/10.1063/5.0051418}
  {https://doi.org/10.1063/5.0051418} \BibitemShut {NoStop}%
\bibitem [{\citenamefont {von Lilienfeld}\ \emph {et~al.}(2015)\citenamefont
  {von Lilienfeld}, \citenamefont {Ramakrishnan}, \citenamefont {Rupp},\ and\
  \citenamefont {Knoll}}]{FourierDesc.}%
  \BibitemOpen
  \bibfield  {author} {\bibinfo {author} {\bibfnamefont {O.~A.}\ \bibnamefont
  {von Lilienfeld}}, \bibinfo {author} {\bibfnamefont {R.}~\bibnamefont
  {Ramakrishnan}}, \bibinfo {author} {\bibfnamefont {M.}~\bibnamefont {Rupp}},
  \ and\ \bibinfo {author} {\bibfnamefont {A.}~\bibnamefont {Knoll}},\ }\href
  {\doibase https://doi.org/10.1002/qua.24912} {\bibfield  {journal} {\bibinfo
  {journal} {International Journal of Quantum Chemistry}\ }\textbf {\bibinfo
  {volume} {115}},\ \bibinfo {pages} {1084} (\bibinfo {year} {2015})},\ \Eprint
  {http://arxiv.org/abs/https://onlinelibrary.wiley.com/doi/pdf/10.1002/qua.24912}
  {https://onlinelibrary.wiley.com/doi/pdf/10.1002/qua.24912} \BibitemShut
  {NoStop}%
\bibitem [{\citenamefont {Ghiringhelli}\ \emph {et~al.}(2015)\citenamefont
  {Ghiringhelli}, \citenamefont {Vybiral}, \citenamefont {Levchenko},
  \citenamefont {Draxl},\ and\ \citenamefont
  {Scheffler}}]{desc_role_Scheffler}%
  \BibitemOpen
  \bibfield  {author} {\bibinfo {author} {\bibfnamefont {L.~M.}\ \bibnamefont
  {Ghiringhelli}}, \bibinfo {author} {\bibfnamefont {J.}~\bibnamefont
  {Vybiral}}, \bibinfo {author} {\bibfnamefont {S.~V.}\ \bibnamefont
  {Levchenko}}, \bibinfo {author} {\bibfnamefont {C.}~\bibnamefont {Draxl}}, \
  and\ \bibinfo {author} {\bibfnamefont {M.}~\bibnamefont {Scheffler}},\ }\href
  {\doibase 10.1103/PhysRevLett.114.105503} {\bibfield  {journal} {\bibinfo
  {journal} {Phys. Rev. Lett.}\ }\textbf {\bibinfo {volume} {114}},\ \bibinfo
  {pages} {105503} (\bibinfo {year} {2015})}\BibitemShut {NoStop}%
\bibitem [{\citenamefont {Musil}\ \emph {et~al.}(2021)\citenamefont {Musil},
  \citenamefont {Grisafi}, \citenamefont {Bartók}, \citenamefont {Ortner},
  \citenamefont {Csányi},\ and\ \citenamefont
  {Ceriotti}}]{physics-inspired-reps-ceriotti}%
  \BibitemOpen
  \bibfield  {author} {\bibinfo {author} {\bibfnamefont {F.}~\bibnamefont
  {Musil}}, \bibinfo {author} {\bibfnamefont {A.}~\bibnamefont {Grisafi}},
  \bibinfo {author} {\bibfnamefont {A.~P.}\ \bibnamefont {Bartók}}, \bibinfo
  {author} {\bibfnamefont {C.}~\bibnamefont {Ortner}}, \bibinfo {author}
  {\bibfnamefont {G.}~\bibnamefont {Csányi}}, \ and\ \bibinfo {author}
  {\bibfnamefont {M.}~\bibnamefont {Ceriotti}},\ }\href {\doibase
  10.1021/acs.chemrev.1c00021} {\bibfield  {journal} {\bibinfo  {journal}
  {Chemical Reviews}\ }\textbf {\bibinfo {volume} {121}},\ \bibinfo {pages}
  {9759} (\bibinfo {year} {2021})},\ \bibinfo {note} {pMID: 34310133},\ \Eprint
  {http://arxiv.org/abs/https://doi.org/10.1021/acs.chemrev.1c00021}
  {https://doi.org/10.1021/acs.chemrev.1c00021} \BibitemShut {NoStop}%
\bibitem [{\citenamefont {Ramakrishnan}\ and\ \citenamefont {von
  Lilienfeld}(2015)}]{Ramakrishnan_vonLilienfeld_2015}%
  \BibitemOpen
  \bibfield  {author} {\bibinfo {author} {\bibfnamefont {R.}~\bibnamefont
  {Ramakrishnan}}\ and\ \bibinfo {author} {\bibfnamefont {O.~A.}\ \bibnamefont
  {von Lilienfeld}},\ }\href {\doibase 10.2533/chimia.2015.182} {\bibfield
  {journal} {\bibinfo  {journal} {CHIMIA}\ }\textbf {\bibinfo {volume} {69}},\
  \bibinfo {pages} {182} (\bibinfo {year} {2015})}\BibitemShut {NoStop}%
\bibitem [{\citenamefont {Vapnik}(1998)}]{Vapnik1998}%
  \BibitemOpen
  \bibfield  {author} {\bibinfo {author} {\bibfnamefont {V.~N.}\ \bibnamefont
  {Vapnik}},\ }\href@noop {} {\emph {\bibinfo {title} {Statistical Learning
  Theory}}}\ (\bibinfo  {publisher} {Wiley-Interscience},\ \bibinfo {year}
  {1998})\BibitemShut {NoStop}%
\bibitem [{\citenamefont {Collins}\ \emph {et~al.}(2018)\citenamefont
  {Collins}, \citenamefont {Gordon}, \citenamefont {von Lilienfeld},\ and\
  \citenamefont {Yaron}}]{const_size}%
  \BibitemOpen
  \bibfield  {author} {\bibinfo {author} {\bibfnamefont {C.~R.}\ \bibnamefont
  {Collins}}, \bibinfo {author} {\bibfnamefont {G.~J.}\ \bibnamefont {Gordon}},
  \bibinfo {author} {\bibfnamefont {O.~A.}\ \bibnamefont {von Lilienfeld}}, \
  and\ \bibinfo {author} {\bibfnamefont {D.~J.}\ \bibnamefont {Yaron}},\ }\href
  {\doibase 10.1063/1.5020441} {\bibfield  {journal} {\bibinfo  {journal} {The
  Journal of Chemical Physics}\ }\textbf {\bibinfo {volume} {148}},\ \bibinfo
  {pages} {241718} (\bibinfo {year} {2018})},\ \Eprint
  {http://arxiv.org/abs/https://doi.org/10.1063/1.5020441}
  {https://doi.org/10.1063/1.5020441} \BibitemShut {NoStop}%
\bibitem [{\citenamefont {Behler}(2011)}]{acsf}%
  \BibitemOpen
  \bibfield  {author} {\bibinfo {author} {\bibfnamefont {J.}~\bibnamefont
  {Behler}},\ }\href {\doibase 10.1063/1.3553717} {\bibfield  {journal}
  {\bibinfo  {journal} {The Journal of Chemical Physics}\ }\textbf {\bibinfo
  {volume} {134}},\ \bibinfo {pages} {074106} (\bibinfo {year} {2011})},\
  \Eprint {http://arxiv.org/abs/https://doi.org/10.1063/1.3553717}
  {https://doi.org/10.1063/1.3553717} \BibitemShut {NoStop}%
\bibitem [{\citenamefont {Zhu}\ \emph {et~al.}(2016)\citenamefont {Zhu},
  \citenamefont {Amsler}, \citenamefont {Fuhrer}, \citenamefont {Schaefer},
  \citenamefont {Faraji}, \citenamefont {Rostami}, \citenamefont {Ghasemi},
  \citenamefont {Sadeghi}, \citenamefont {Grauzinyte}, \citenamefont
  {Wolverton},\ and\ \citenamefont {Goedecker}}]{OM}%
  \BibitemOpen
  \bibfield  {author} {\bibinfo {author} {\bibfnamefont {L.}~\bibnamefont
  {Zhu}}, \bibinfo {author} {\bibfnamefont {M.}~\bibnamefont {Amsler}},
  \bibinfo {author} {\bibfnamefont {T.}~\bibnamefont {Fuhrer}}, \bibinfo
  {author} {\bibfnamefont {B.}~\bibnamefont {Schaefer}}, \bibinfo {author}
  {\bibfnamefont {S.}~\bibnamefont {Faraji}}, \bibinfo {author} {\bibfnamefont
  {S.}~\bibnamefont {Rostami}}, \bibinfo {author} {\bibfnamefont {S.~A.}\
  \bibnamefont {Ghasemi}}, \bibinfo {author} {\bibfnamefont {A.}~\bibnamefont
  {Sadeghi}}, \bibinfo {author} {\bibfnamefont {M.}~\bibnamefont {Grauzinyte}},
  \bibinfo {author} {\bibfnamefont {C.}~\bibnamefont {Wolverton}}, \ and\
  \bibinfo {author} {\bibfnamefont {S.}~\bibnamefont {Goedecker}},\ }\href
  {\doibase 10.1063/1.4940026} {\bibfield  {journal} {\bibinfo  {journal} {The
  Journal of Chemical Physics}\ }\textbf {\bibinfo {volume} {144}},\ \bibinfo
  {pages} {034203} (\bibinfo {year} {2016})},\ \Eprint
  {http://arxiv.org/abs/https://doi.org/10.1063/1.4940026}
  {https://doi.org/10.1063/1.4940026} \BibitemShut {NoStop}%
\bibitem [{\citenamefont {Huo}\ and\ \citenamefont {Rupp}(2022)}]{mbtr}%
  \BibitemOpen
  \bibfield  {author} {\bibinfo {author} {\bibfnamefont {H.}~\bibnamefont
  {Huo}}\ and\ \bibinfo {author} {\bibfnamefont {M.}~\bibnamefont {Rupp}},\
  }\href {\doibase 10.1088/2632-2153/aca005} {\bibfield  {journal} {\bibinfo
  {journal} {Machine Learning: Science and Technology}\ }\textbf {\bibinfo
  {volume} {3}},\ \bibinfo {pages} {045017} (\bibinfo {year}
  {2022})}\BibitemShut {NoStop}%
\bibitem [{\citenamefont {Gastegger}\ \emph {et~al.}(2018)\citenamefont
  {Gastegger}, \citenamefont {Schwiedrzik}, \citenamefont {Bittermann},
  \citenamefont {Berzsenyi},\ and\ \citenamefont {Marquetand}}]{wacsf}%
  \BibitemOpen
  \bibfield  {author} {\bibinfo {author} {\bibfnamefont {M.}~\bibnamefont
  {Gastegger}}, \bibinfo {author} {\bibfnamefont {L.}~\bibnamefont
  {Schwiedrzik}}, \bibinfo {author} {\bibfnamefont {M.}~\bibnamefont
  {Bittermann}}, \bibinfo {author} {\bibfnamefont {F.}~\bibnamefont
  {Berzsenyi}}, \ and\ \bibinfo {author} {\bibfnamefont {P.}~\bibnamefont
  {Marquetand}},\ }\href {\doibase 10.1063/1.5019667} {\bibfield  {journal}
  {\bibinfo  {journal} {The Journal of Chemical Physics}\ }\textbf {\bibinfo
  {volume} {148}},\ \bibinfo {pages} {241709} (\bibinfo {year} {2018})},\
  \Eprint {http://arxiv.org/abs/https://doi.org/10.1063/1.5019667}
  {https://doi.org/10.1063/1.5019667} \BibitemShut {NoStop}%
\bibitem [{\citenamefont {Drautz}(2019)}]{ace}%
  \BibitemOpen
  \bibfield  {author} {\bibinfo {author} {\bibfnamefont {R.}~\bibnamefont
  {Drautz}},\ }\href {\doibase 10.1103/PhysRevB.99.014104} {\bibfield
  {journal} {\bibinfo  {journal} {Phys. Rev. B}\ }\textbf {\bibinfo {volume}
  {99}},\ \bibinfo {pages} {014104} (\bibinfo {year} {2019})}\BibitemShut
  {NoStop}%
\bibitem [{\citenamefont {Braams}\ and\ \citenamefont {Bowman}(2009)}]{pip}%
  \BibitemOpen
  \bibfield  {author} {\bibinfo {author} {\bibfnamefont {B.~J.}\ \bibnamefont
  {Braams}}\ and\ \bibinfo {author} {\bibfnamefont {J.~M.}\ \bibnamefont
  {Bowman}},\ }\href {\doibase 10.1080/01442350903234923} {\bibfield  {journal}
  {\bibinfo  {journal} {International Reviews in Physical Chemistry}\ }\textbf
  {\bibinfo {volume} {28}},\ \bibinfo {pages} {577} (\bibinfo {year} {2009})},\
  \Eprint {http://arxiv.org/abs/https://doi.org/10.1080/01442350903234923}
  {https://doi.org/10.1080/01442350903234923} \BibitemShut {NoStop}%
\bibitem [{\citenamefont {Hirn}\ \emph {et~al.}(2015)\citenamefont {Hirn},
  \citenamefont {Poilvert},\ and\ \citenamefont {Mallat}}]{wst}%
  \BibitemOpen
  \bibfield  {author} {\bibinfo {author} {\bibfnamefont {M.~J.}\ \bibnamefont
  {Hirn}}, \bibinfo {author} {\bibfnamefont {N.}~\bibnamefont {Poilvert}}, \
  and\ \bibinfo {author} {\bibfnamefont {S.}~\bibnamefont {Mallat}},\ }\href
  {http://arxiv.org/abs/1502.02077} {\bibfield  {journal} {\bibinfo  {journal}
  {CoRR}\ }\textbf {\bibinfo {volume} {abs/1502.02077}} (\bibinfo {year}
  {2015})},\ \Eprint {http://arxiv.org/abs/1502.02077} {1502.02077}
  \BibitemShut {NoStop}%
\bibitem [{\citenamefont {Shapeev}(2016)}]{mtp}%
  \BibitemOpen
  \bibfield  {author} {\bibinfo {author} {\bibfnamefont {A.~V.}\ \bibnamefont
  {Shapeev}},\ }\href {\doibase 10.1137/15M1054183} {\bibfield  {journal}
  {\bibinfo  {journal} {Multiscale Modeling \& Simulation}\ }\textbf {\bibinfo
  {volume} {14}},\ \bibinfo {pages} {1153} (\bibinfo {year} {2016})},\ \Eprint
  {http://arxiv.org/abs/https://doi.org/10.1137/15M1054183}
  {https://doi.org/10.1137/15M1054183} \BibitemShut {NoStop}%
\bibitem [{\citenamefont {Nigam}\ \emph {et~al.}(2020)\citenamefont {Nigam},
  \citenamefont {Pozdnyakov},\ and\ \citenamefont {Ceriotti}}]{nice}%
  \BibitemOpen
  \bibfield  {author} {\bibinfo {author} {\bibfnamefont {J.}~\bibnamefont
  {Nigam}}, \bibinfo {author} {\bibfnamefont {S.}~\bibnamefont {Pozdnyakov}}, \
  and\ \bibinfo {author} {\bibfnamefont {M.}~\bibnamefont {Ceriotti}},\ }\href
  {\doibase 10.1063/5.0021116} {\bibfield  {journal} {\bibinfo  {journal} {The
  Journal of Chemical Physics}\ }\textbf {\bibinfo {volume} {153}},\ \bibinfo
  {pages} {121101} (\bibinfo {year} {2020})},\ \Eprint
  {http://arxiv.org/abs/https://doi.org/10.1063/5.0021116}
  {https://doi.org/10.1063/5.0021116} \BibitemShut {NoStop}%
\bibitem [{\citenamefont {Zaverkin}\ and\ \citenamefont
  {Kästner}(2020)}]{gaussian_moments}%
  \BibitemOpen
  \bibfield  {author} {\bibinfo {author} {\bibfnamefont {V.}~\bibnamefont
  {Zaverkin}}\ and\ \bibinfo {author} {\bibfnamefont {J.}~\bibnamefont
  {Kästner}},\ }\href {\doibase 10.1021/acs.jctc.0c00347} {\bibfield
  {journal} {\bibinfo  {journal} {Journal of Chemical Theory and Computation}\
  }\textbf {\bibinfo {volume} {16}},\ \bibinfo {pages} {5410} (\bibinfo {year}
  {2020})},\ \bibinfo {note} {pMID: 32672968},\ \Eprint
  {http://arxiv.org/abs/https://doi.org/10.1021/acs.jctc.0c00347}
  {https://doi.org/10.1021/acs.jctc.0c00347} \BibitemShut {NoStop}%
\bibitem [{\citenamefont {Jinnouchi}\ \emph {et~al.}(2020)\citenamefont
  {Jinnouchi}, \citenamefont {Karsai}, \citenamefont {Verdi}, \citenamefont
  {Asahi},\ and\ \citenamefont {Kresse}}]{kresse2020}%
  \BibitemOpen
  \bibfield  {author} {\bibinfo {author} {\bibfnamefont {R.}~\bibnamefont
  {Jinnouchi}}, \bibinfo {author} {\bibfnamefont {F.}~\bibnamefont {Karsai}},
  \bibinfo {author} {\bibfnamefont {C.}~\bibnamefont {Verdi}}, \bibinfo
  {author} {\bibfnamefont {R.}~\bibnamefont {Asahi}}, \ and\ \bibinfo {author}
  {\bibfnamefont {G.}~\bibnamefont {Kresse}},\ }\href {\doibase
  10.1063/5.0009491} {\bibfield  {journal} {\bibinfo  {journal} {The Journal of
  Chemical Physics}\ }\textbf {\bibinfo {volume} {152}},\ \bibinfo {pages}
  {234102} (\bibinfo {year} {2020})},\ \Eprint
  {http://arxiv.org/abs/https://doi.org/10.1063/5.0009491}
  {https://doi.org/10.1063/5.0009491} \BibitemShut {NoStop}%
\bibitem [{\citenamefont {Thompson}\ \emph {et~al.}(2015)\citenamefont
  {Thompson}, \citenamefont {Swiler}, \citenamefont {Trott}, \citenamefont
  {Foiles},\ and\ \citenamefont {Tucker}}]{snap}%
  \BibitemOpen
  \bibfield  {author} {\bibinfo {author} {\bibfnamefont {A.}~\bibnamefont
  {Thompson}}, \bibinfo {author} {\bibfnamefont {L.}~\bibnamefont {Swiler}},
  \bibinfo {author} {\bibfnamefont {C.}~\bibnamefont {Trott}}, \bibinfo
  {author} {\bibfnamefont {S.}~\bibnamefont {Foiles}}, \ and\ \bibinfo {author}
  {\bibfnamefont {G.}~\bibnamefont {Tucker}},\ }\href {\doibase
  https://doi.org/10.1016/j.jcp.2014.12.018} {\bibfield  {journal} {\bibinfo
  {journal} {Journal of Computational Physics}\ }\textbf {\bibinfo {volume}
  {285}},\ \bibinfo {pages} {316} (\bibinfo {year} {2015})}\BibitemShut
  {NoStop}%
\bibitem [{\citenamefont {Willatt}\ \emph {et~al.}(2018)\citenamefont
  {Willatt}, \citenamefont {Musil},\ and\ \citenamefont {Ceriotti}}]{soap2018}%
  \BibitemOpen
  \bibfield  {author} {\bibinfo {author} {\bibfnamefont {M.~J.}\ \bibnamefont
  {Willatt}}, \bibinfo {author} {\bibfnamefont {F.}~\bibnamefont {Musil}}, \
  and\ \bibinfo {author} {\bibfnamefont {M.}~\bibnamefont {Ceriotti}},\ }\href
  {\doibase 10.1039/C8CP05921G} {\bibfield  {journal} {\bibinfo  {journal}
  {Phys. Chem. Chem. Phys.}\ }\textbf {\bibinfo {volume} {20}},\ \bibinfo
  {pages} {29661} (\bibinfo {year} {2018})}\BibitemShut {NoStop}%
\bibitem [{\citenamefont {Langer}\ \emph {et~al.}(2022)\citenamefont {Langer},
  \citenamefont {Goe{\ss}mann},\ and\ \citenamefont {Rupp}}]{reps_review_Rupp}%
  \BibitemOpen
  \bibfield  {author} {\bibinfo {author} {\bibfnamefont {M.~F.}\ \bibnamefont
  {Langer}}, \bibinfo {author} {\bibfnamefont {A.}~\bibnamefont
  {Goe{\ss}mann}}, \ and\ \bibinfo {author} {\bibfnamefont {M.}~\bibnamefont
  {Rupp}},\ }\href {\doibase 10.1038/s41524-022-00721-x} {\bibfield  {journal}
  {\bibinfo  {journal} {npj Computational Materials}\ }\textbf {\bibinfo
  {volume} {8}},\ \bibinfo {pages} {41} (\bibinfo {year} {2022})}\BibitemShut
  {NoStop}%
\bibitem [{\citenamefont {Christensen}\ \emph {et~al.}(2020)\citenamefont
  {Christensen}, \citenamefont {Bratholm}, \citenamefont {Faber},\ and\
  \citenamefont {Anatole~von Lilienfeld}}]{fchl19}%
  \BibitemOpen
  \bibfield  {author} {\bibinfo {author} {\bibfnamefont {A.~S.}\ \bibnamefont
  {Christensen}}, \bibinfo {author} {\bibfnamefont {L.~A.}\ \bibnamefont
  {Bratholm}}, \bibinfo {author} {\bibfnamefont {F.~A.}\ \bibnamefont {Faber}},
  \ and\ \bibinfo {author} {\bibfnamefont {O.}~\bibnamefont {Anatole~von
  Lilienfeld}},\ }\href {\doibase 10.1063/1.5126701} {\bibfield  {journal}
  {\bibinfo  {journal} {The Journal of Chemical Physics}\ }\textbf {\bibinfo
  {volume} {152}},\ \bibinfo {pages} {044107} (\bibinfo {year} {2020})},\
  \Eprint {http://arxiv.org/abs/https://doi.org/10.1063/1.5126701}
  {https://doi.org/10.1063/1.5126701} \BibitemShut {NoStop}%
\bibitem [{\citenamefont {Faber}\ \emph {et~al.}(2018)\citenamefont {Faber},
  \citenamefont {Christensen}, \citenamefont {Huang},\ and\ \citenamefont {von
  Lilienfeld}}]{fchl18}%
  \BibitemOpen
  \bibfield  {author} {\bibinfo {author} {\bibfnamefont {F.~A.}\ \bibnamefont
  {Faber}}, \bibinfo {author} {\bibfnamefont {A.~S.}\ \bibnamefont
  {Christensen}}, \bibinfo {author} {\bibfnamefont {B.}~\bibnamefont {Huang}},
  \ and\ \bibinfo {author} {\bibfnamefont {O.~A.}\ \bibnamefont {von
  Lilienfeld}},\ }\href {\doibase 10.1063/1.5020710} {\bibfield  {journal}
  {\bibinfo  {journal} {The Journal of Chemical Physics}\ }\textbf {\bibinfo
  {volume} {148}},\ \bibinfo {pages} {241717} (\bibinfo {year} {2018})},\
  \Eprint {http://arxiv.org/abs/https://doi.org/10.1063/1.5020710}
  {https://doi.org/10.1063/1.5020710} \BibitemShut {NoStop}%
\bibitem [{\citenamefont {Bart\'ok}\ \emph {et~al.}(2013)\citenamefont
  {Bart\'ok}, \citenamefont {Kondor},\ and\ \citenamefont {Cs\'anyi}}]{soap}%
  \BibitemOpen
  \bibfield  {author} {\bibinfo {author} {\bibfnamefont {A.~P.}\ \bibnamefont
  {Bart\'ok}}, \bibinfo {author} {\bibfnamefont {R.}~\bibnamefont {Kondor}}, \
  and\ \bibinfo {author} {\bibfnamefont {G.}~\bibnamefont {Cs\'anyi}},\ }\href
  {\doibase 10.1103/PhysRevB.87.184115} {\bibfield  {journal} {\bibinfo
  {journal} {Phys. Rev. B}\ }\textbf {\bibinfo {volume} {87}},\ \bibinfo
  {pages} {184115} (\bibinfo {year} {2013})}\BibitemShut {NoStop}%
\bibitem [{\citenamefont {Huang}\ and\ \citenamefont {von
  Lilienfeld}(2017)}]{amons_slatm}%
  \BibitemOpen
  \bibfield  {author} {\bibinfo {author} {\bibfnamefont {B.}~\bibnamefont
  {Huang}}\ and\ \bibinfo {author} {\bibfnamefont {O.~A.}\ \bibnamefont {von
  Lilienfeld}},\ }\href {\doibase 10.48550/ARXIV.1707.04146} {\enquote
  {\bibinfo {title} {Quantum machine learning using atom-in-molecule-based
  fragments selected on-the-fly},}\ } (\bibinfo {year} {2017})\BibitemShut
  {NoStop}%
\bibitem [{\citenamefont {Hansen}\ \emph {et~al.}(2015)\citenamefont {Hansen},
  \citenamefont {Biegler}, \citenamefont {Ramakrishnan}, \citenamefont
  {Pronobis}, \citenamefont {von Lilienfeld}, \citenamefont {Müller},\ and\
  \citenamefont {Tkatchenko}}]{bob}%
  \BibitemOpen
  \bibfield  {author} {\bibinfo {author} {\bibfnamefont {K.}~\bibnamefont
  {Hansen}}, \bibinfo {author} {\bibfnamefont {F.}~\bibnamefont {Biegler}},
  \bibinfo {author} {\bibfnamefont {R.}~\bibnamefont {Ramakrishnan}}, \bibinfo
  {author} {\bibfnamefont {W.}~\bibnamefont {Pronobis}}, \bibinfo {author}
  {\bibfnamefont {O.~A.}\ \bibnamefont {von Lilienfeld}}, \bibinfo {author}
  {\bibfnamefont {K.-R.}\ \bibnamefont {Müller}}, \ and\ \bibinfo {author}
  {\bibfnamefont {A.}~\bibnamefont {Tkatchenko}},\ }\href {\doibase
  10.1021/acs.jpclett.5b00831} {\bibfield  {journal} {\bibinfo  {journal} {The
  Journal of Physical Chemistry Letters}\ }\textbf {\bibinfo {volume} {6}},\
  \bibinfo {pages} {2326} (\bibinfo {year} {2015})},\ \bibinfo {note} {pMID:
  26113956},\ \Eprint
  {http://arxiv.org/abs/https://doi.org/10.1021/acs.jpclett.5b00831}
  {https://doi.org/10.1021/acs.jpclett.5b00831} \BibitemShut {NoStop}%
\bibitem [{\citenamefont {Rasmussen}\ and\ \citenamefont
  {Williams}(2006)}]{gpr_rasmussen}%
  \BibitemOpen
  \bibfield  {author} {\bibinfo {author} {\bibfnamefont {C.~E.}\ \bibnamefont
  {Rasmussen}}\ and\ \bibinfo {author} {\bibfnamefont {C.~K.~I.}\ \bibnamefont
  {Williams}},\ }\href@noop {} {\emph {\bibinfo {title} {Gaussian processes for
  machine learning.}}},\ Adaptive computation and machine learning\ (\bibinfo
  {publisher} {MIT Press},\ \bibinfo {year} {2006})\ pp.\ \bibinfo {pages}
  {I--XVIII, 1--248}\BibitemShut {NoStop}%
\bibitem [{\citenamefont {De}\ \emph {et~al.}(2016)\citenamefont {De},
  \citenamefont {Bartók}, \citenamefont {Csányi},\ and\ \citenamefont
  {Ceriotti}}]{local_kernels_ceriotti}%
  \BibitemOpen
  \bibfield  {author} {\bibinfo {author} {\bibfnamefont {S.}~\bibnamefont
  {De}}, \bibinfo {author} {\bibfnamefont {A.~P.}\ \bibnamefont {Bartók}},
  \bibinfo {author} {\bibfnamefont {G.}~\bibnamefont {Csányi}}, \ and\
  \bibinfo {author} {\bibfnamefont {M.}~\bibnamefont {Ceriotti}},\ }\href
  {\doibase 10.1039/C6CP00415F} {\bibfield  {journal} {\bibinfo  {journal}
  {Phys. Chem. Chem. Phys.}\ }\textbf {\bibinfo {volume} {18}},\ \bibinfo
  {pages} {13754} (\bibinfo {year} {2016})}\BibitemShut {NoStop}%
\bibitem [{\citenamefont {Bart{\'{o}}k}\ and\ \citenamefont
  {Cs{\'{a}}nyi}(2015)}]{sparse_GP_bartok}%
  \BibitemOpen
  \bibfield  {author} {\bibinfo {author} {\bibfnamefont {A.~P.}\ \bibnamefont
  {Bart{\'{o}}k}}\ and\ \bibinfo {author} {\bibfnamefont {G.}~\bibnamefont
  {Cs{\'{a}}nyi}},\ }\href {\doibase 10.1002/qua.24927} {\bibfield  {journal}
  {\bibinfo  {journal} {International Journal of Quantum Chemistry}\ }\textbf
  {\bibinfo {volume} {115}},\ \bibinfo {pages} {1051} (\bibinfo {year}
  {2015})}\BibitemShut {NoStop}%
\bibitem [{\citenamefont {Chmiela}\ \emph {et~al.}(2023)\citenamefont
  {Chmiela}, \citenamefont {Vassilev-Galindo}, \citenamefont {Unke},
  \citenamefont {Kabylda}, \citenamefont {Sauceda}, \citenamefont
  {Tkatchenko},\ and\ \citenamefont {Müller}}]{conjugate_gradient_kernels}%
  \BibitemOpen
  \bibfield  {author} {\bibinfo {author} {\bibfnamefont {S.}~\bibnamefont
  {Chmiela}}, \bibinfo {author} {\bibfnamefont {V.}~\bibnamefont
  {Vassilev-Galindo}}, \bibinfo {author} {\bibfnamefont {O.~T.}\ \bibnamefont
  {Unke}}, \bibinfo {author} {\bibfnamefont {A.}~\bibnamefont {Kabylda}},
  \bibinfo {author} {\bibfnamefont {H.~E.}\ \bibnamefont {Sauceda}}, \bibinfo
  {author} {\bibfnamefont {A.}~\bibnamefont {Tkatchenko}}, \ and\ \bibinfo
  {author} {\bibfnamefont {K.-R.}\ \bibnamefont {Müller}},\ }\href {\doibase
  10.1126/sciadv.adf0873} {\bibfield  {journal} {\bibinfo  {journal} {Science
  Advances}\ }\textbf {\bibinfo {volume} {9}},\ \bibinfo {pages} {eadf0873}
  (\bibinfo {year} {2023})},\ \Eprint
  {http://arxiv.org/abs/https://www.science.org/doi/pdf/10.1126/sciadv.adf0873}
  {https://www.science.org/doi/pdf/10.1126/sciadv.adf0873} \BibitemShut
  {NoStop}%
\bibitem [{\citenamefont {Faber}\ \emph {et~al.}(2016)\citenamefont {Faber},
  \citenamefont {Lindmaa}, \citenamefont {von Lilienfeld},\ and\ \citenamefont
  {Armiento}}]{Elpasolite_2016}%
  \BibitemOpen
  \bibfield  {author} {\bibinfo {author} {\bibfnamefont {F.~A.}\ \bibnamefont
  {Faber}}, \bibinfo {author} {\bibfnamefont {A.}~\bibnamefont {Lindmaa}},
  \bibinfo {author} {\bibfnamefont {O.~A.}\ \bibnamefont {von Lilienfeld}}, \
  and\ \bibinfo {author} {\bibfnamefont {R.}~\bibnamefont {Armiento}},\ }\href
  {\doibase 10.1103/PhysRevLett.117.135502} {\bibfield  {journal} {\bibinfo
  {journal} {Phys. Rev. Lett.}\ }\textbf {\bibinfo {volume} {117}},\ \bibinfo
  {pages} {135502} (\bibinfo {year} {2016})}\BibitemShut {NoStop}%
\bibitem [{\citenamefont {Morse}(1929)}]{morse}%
  \BibitemOpen
  \bibfield  {author} {\bibinfo {author} {\bibfnamefont {P.~M.}\ \bibnamefont
  {Morse}},\ }\href {\doibase 10.1103/PhysRev.34.57} {\bibfield  {journal}
  {\bibinfo  {journal} {Phys. Rev.}\ }\textbf {\bibinfo {volume} {34}},\
  \bibinfo {pages} {57} (\bibinfo {year} {1929})}\BibitemShut {NoStop}%
\bibitem [{\citenamefont {Jones}(1924)}]{lennard-jones2}%
  \BibitemOpen
  \bibfield  {author} {\bibinfo {author} {\bibfnamefont {J.~E.}\ \bibnamefont
  {Jones}},\ }\href
  {https://royalsocietypublishing.org/doi/10.1098/rspa.1924.0081} {\bibfield
  {journal} {\bibinfo  {journal} {Proceedings of the Royal Society of London.
  Series A}\ } (\bibinfo {year} {1924})}\BibitemShut {NoStop}%
\bibitem [{\citenamefont {Axilrod}\ and\ \citenamefont
  {Teller}(1943)}]{axilrod_teller}%
  \BibitemOpen
  \bibfield  {author} {\bibinfo {author} {\bibfnamefont {B.~M.}\ \bibnamefont
  {Axilrod}}\ and\ \bibinfo {author} {\bibfnamefont {E.}~\bibnamefont
  {Teller}},\ }\href {\doibase 10.1063/1.1723844} {\bibfield  {journal}
  {\bibinfo  {journal} {The Journal of Chemical Physics}\ }\textbf {\bibinfo
  {volume} {11}},\ \bibinfo {pages} {299} (\bibinfo {year} {1943})},\ \Eprint
  {http://arxiv.org/abs/https://doi.org/10.1063/1.1723844}
  {https://doi.org/10.1063/1.1723844} \BibitemShut {NoStop}%
\bibitem [{\citenamefont {Stillinger}\ and\ \citenamefont
  {Weber}(1985)}]{stillinger-weber}%
  \BibitemOpen
  \bibfield  {author} {\bibinfo {author} {\bibfnamefont {F.~H.}\ \bibnamefont
  {Stillinger}}\ and\ \bibinfo {author} {\bibfnamefont {T.~A.}\ \bibnamefont
  {Weber}},\ }\href {\doibase 10.1103/PhysRevB.31.5262} {\bibfield  {journal}
  {\bibinfo  {journal} {Phys. Rev. B}\ }\textbf {\bibinfo {volume} {31}},\
  \bibinfo {pages} {5262} (\bibinfo {year} {1985})}\BibitemShut {NoStop}%
\bibitem [{\citenamefont {Ramakrishnan}\ \emph {et~al.}(2014)\citenamefont
  {Ramakrishnan}, \citenamefont {Dral}, \citenamefont {Rupp},\ and\
  \citenamefont {von Lilienfeld}}]{qm9}%
  \BibitemOpen
  \bibfield  {author} {\bibinfo {author} {\bibfnamefont {R.}~\bibnamefont
  {Ramakrishnan}}, \bibinfo {author} {\bibfnamefont {P.~O.}\ \bibnamefont
  {Dral}}, \bibinfo {author} {\bibfnamefont {M.}~\bibnamefont {Rupp}}, \ and\
  \bibinfo {author} {\bibfnamefont {O.~A.}\ \bibnamefont {von Lilienfeld}},\
  }\href@noop {} {\bibfield  {journal} {\bibinfo  {journal} {Scientific Data}\
  }\textbf {\bibinfo {volume} {1}} (\bibinfo {year} {2014})}\BibitemShut
  {NoStop}%
\bibitem [{\citenamefont {Rupp}\ \emph
  {et~al.}(2012{\natexlab{b}})\citenamefont {Rupp}, \citenamefont {Tkatchenko},
  \citenamefont {M\"uller},\ and\ \citenamefont {von Lilienfeld}}]{qm7}%
  \BibitemOpen
  \bibfield  {author} {\bibinfo {author} {\bibfnamefont {M.}~\bibnamefont
  {Rupp}}, \bibinfo {author} {\bibfnamefont {A.}~\bibnamefont {Tkatchenko}},
  \bibinfo {author} {\bibfnamefont {K.-R.}\ \bibnamefont {M\"uller}}, \ and\
  \bibinfo {author} {\bibfnamefont {O.~A.}\ \bibnamefont {von Lilienfeld}},\
  }\href {\doibase 10.1103/PhysRevLett.108.058301} {\bibfield  {journal}
  {\bibinfo  {journal} {Phys. Rev. Lett.}\ }\textbf {\bibinfo {volume} {108}},\
  \bibinfo {pages} {058301} (\bibinfo {year} {2012}{\natexlab{b}})}\BibitemShut
  {NoStop}%
\bibitem [{\citenamefont {Isert}\ \emph {et~al.}(2022)\citenamefont {Isert},
  \citenamefont {Atz}, \citenamefont {Jim{\'e}nez-Luna},\ and\ \citenamefont
  {Schneider}}]{QMugs}%
  \BibitemOpen
  \bibfield  {author} {\bibinfo {author} {\bibfnamefont {C.}~\bibnamefont
  {Isert}}, \bibinfo {author} {\bibfnamefont {K.}~\bibnamefont {Atz}}, \bibinfo
  {author} {\bibfnamefont {J.}~\bibnamefont {Jim{\'e}nez-Luna}}, \ and\
  \bibinfo {author} {\bibfnamefont {G.}~\bibnamefont {Schneider}},\ }\href
  {\doibase 10.1038/s41597-022-01390-7} {\bibfield  {journal} {\bibinfo
  {journal} {Scientific Data}\ }\textbf {\bibinfo {volume} {9}},\ \bibinfo
  {pages} {273} (\bibinfo {year} {2022})}\BibitemShut {NoStop}%
\bibitem [{\citenamefont {Pozdnyakov}\ \emph {et~al.}(2020)\citenamefont
  {Pozdnyakov}, \citenamefont {Willatt}, \citenamefont {Bart\'ok},
  \citenamefont {Ortner}, \citenamefont {Cs\'anyi},\ and\ \citenamefont
  {Ceriotti}}]{incompleteness_ceriotti}%
  \BibitemOpen
  \bibfield  {author} {\bibinfo {author} {\bibfnamefont {S.~N.}\ \bibnamefont
  {Pozdnyakov}}, \bibinfo {author} {\bibfnamefont {M.~J.}\ \bibnamefont
  {Willatt}}, \bibinfo {author} {\bibfnamefont {A.~P.}\ \bibnamefont
  {Bart\'ok}}, \bibinfo {author} {\bibfnamefont {C.}~\bibnamefont {Ortner}},
  \bibinfo {author} {\bibfnamefont {G.}~\bibnamefont {Cs\'anyi}}, \ and\
  \bibinfo {author} {\bibfnamefont {M.}~\bibnamefont {Ceriotti}},\ }\href
  {\doibase 10.1103/PhysRevLett.125.166001} {\bibfield  {journal} {\bibinfo
  {journal} {Phys. Rev. Lett.}\ }\textbf {\bibinfo {volume} {125}},\ \bibinfo
  {pages} {166001} (\bibinfo {year} {2020})}\BibitemShut {NoStop}%
\bibitem [{\citenamefont {Huang}\ and\ \citenamefont {von
  Lilienfeld}(2016)}]{communication_bing}%
  \BibitemOpen
  \bibfield  {author} {\bibinfo {author} {\bibfnamefont {B.}~\bibnamefont
  {Huang}}\ and\ \bibinfo {author} {\bibfnamefont {O.~A.}\ \bibnamefont {von
  Lilienfeld}},\ }\href {\doibase 10.1063/1.4964627} {\bibfield  {journal}
  {\bibinfo  {journal} {The Journal of Chemical Physics}\ }\textbf {\bibinfo
  {volume} {145}},\ \bibinfo {pages} {161102} (\bibinfo {year} {2016})},\
  \Eprint {http://arxiv.org/abs/https://doi.org/10.1063/1.4964627}
  {https://doi.org/10.1063/1.4964627} \BibitemShut {NoStop}%
\bibitem [{\citenamefont {Parzen}(1962)}]{parzen_density}%
  \BibitemOpen
  \bibfield  {author} {\bibinfo {author} {\bibfnamefont {E.}~\bibnamefont
  {Parzen}},\ }\href {\doibase 10.1214/aoms/1177704472} {\bibfield  {journal}
  {\bibinfo  {journal} {The Annals of Mathematical Statistics}\ }\textbf
  {\bibinfo {volume} {33}},\ \bibinfo {pages} {1065 } (\bibinfo {year}
  {1962})}\BibitemShut {NoStop}%
\bibitem [{\citenamefont {Amari}\ and\ \citenamefont
  {Nagaoka}(2000)}]{amari2000methods}%
  \BibitemOpen
  \bibfield  {author} {\bibinfo {author} {\bibfnamefont {S.-i.}\ \bibnamefont
  {Amari}}\ and\ \bibinfo {author} {\bibfnamefont {H.}~\bibnamefont
  {Nagaoka}},\ }\href@noop {} {\emph {\bibinfo {title} {Methods of information
  geometry}}},\ Vol.\ \bibinfo {volume} {191}\ (\bibinfo  {publisher} {American
  Mathematical Soc.},\ \bibinfo {year} {2000})\BibitemShut {NoStop}%
\bibitem [{\citenamefont {Deisenroth}\ \emph {et~al.}(2020)\citenamefont
  {Deisenroth}, \citenamefont {Faisal},\ and\ \citenamefont
  {Ong}}]{Deisenroth2020}%
  \BibitemOpen
  \bibfield  {author} {\bibinfo {author} {\bibfnamefont {M.~P.}\ \bibnamefont
  {Deisenroth}}, \bibinfo {author} {\bibfnamefont {A.~A.}\ \bibnamefont
  {Faisal}}, \ and\ \bibinfo {author} {\bibfnamefont {C.~S.}\ \bibnamefont
  {Ong}},\ }\href@noop {} {\emph {\bibinfo {title} {Mathematics for Machine
  Learning}}}\ (\bibinfo  {publisher} {Cambridge University Press},\ \bibinfo
  {year} {2020})\BibitemShut {NoStop}%
\bibitem [{\citenamefont {Schütt}\ \emph {et~al.}(2020)\citenamefont
  {Schütt}, \citenamefont {Chmiela}, \citenamefont {von Lilienfeld},
  \citenamefont {Tkatchenko}, \citenamefont {Tsuda},\ and\ \citenamefont
  {Müller}}]{anatolebook}%
  \BibitemOpen
  \bibfield  {author} {\bibinfo {author} {\bibfnamefont {K.~T.}\ \bibnamefont
  {Schütt}}, \bibinfo {author} {\bibfnamefont {S.}~\bibnamefont {Chmiela}},
  \bibinfo {author} {\bibfnamefont {O.~A.}\ \bibnamefont {von Lilienfeld}},
  \bibinfo {author} {\bibfnamefont {A.}~\bibnamefont {Tkatchenko}}, \bibinfo
  {author} {\bibfnamefont {K.}~\bibnamefont {Tsuda}}, \ and\ \bibinfo {author}
  {\bibfnamefont {K.-R.}\ \bibnamefont {Müller}},\ }\href@noop {} {\emph
  {\bibinfo {title} {Machine Learning Meets Quantum Physics}}}\ (\bibinfo
  {publisher} {Springer},\ \bibinfo {year} {2020})\BibitemShut {NoStop}%
\bibitem [{\citenamefont {Çaylak}\ \emph {et~al.}(2020)\citenamefont
  {Çaylak}, \citenamefont {von Lilienfeld},\ and\ \citenamefont
  {Baumeier}}]{wasserstein}%
  \BibitemOpen
  \bibfield  {author} {\bibinfo {author} {\bibfnamefont {O.}~\bibnamefont
  {Çaylak}}, \bibinfo {author} {\bibfnamefont {O.~A.}\ \bibnamefont {von
  Lilienfeld}}, \ and\ \bibinfo {author} {\bibfnamefont {B.}~\bibnamefont
  {Baumeier}},\ }\href {\doibase 10.1088/2632-2153/aba048} {\bibfield
  {journal} {\bibinfo  {journal} {Machine Learning: Science and Technology}\
  }\textbf {\bibinfo {volume} {1}},\ \bibinfo {pages} {03LT01} (\bibinfo {year}
  {2020})}\BibitemShut {NoStop}%
\bibitem [{\citenamefont {Fabregat}\ \emph {et~al.}(2022)\citenamefont
  {Fabregat}, \citenamefont {van Gerwen}, \citenamefont {Haeberle},
  \citenamefont {Eisenbrand},\ and\ \citenamefont
  {Corminboeuf}}]{metric_learning}%
  \BibitemOpen
  \bibfield  {author} {\bibinfo {author} {\bibfnamefont {R.}~\bibnamefont
  {Fabregat}}, \bibinfo {author} {\bibfnamefont {P.}~\bibnamefont {van
  Gerwen}}, \bibinfo {author} {\bibfnamefont {M.}~\bibnamefont {Haeberle}},
  \bibinfo {author} {\bibfnamefont {F.}~\bibnamefont {Eisenbrand}}, \ and\
  \bibinfo {author} {\bibfnamefont {C.}~\bibnamefont {Corminboeuf}},\ }\href
  {\doibase 10.1088/2632-2153/ac8e4f} {\bibfield  {journal} {\bibinfo
  {journal} {Machine Learning: Science and Technology}\ }\textbf {\bibinfo
  {volume} {3}},\ \bibinfo {pages} {035015} (\bibinfo {year}
  {2022})}\BibitemShut {NoStop}%
\bibitem [{\citenamefont {Cortes}\ \emph {et~al.}(1994)\citenamefont {Cortes},
  \citenamefont {Jackel}, \citenamefont {Solla}, \citenamefont {Vapnik},\ and\
  \citenamefont {Denker}}]{vapnik1994learningcurves}%
  \BibitemOpen
  \bibfield  {author} {\bibinfo {author} {\bibfnamefont {C.}~\bibnamefont
  {Cortes}}, \bibinfo {author} {\bibfnamefont {L.~D.}\ \bibnamefont {Jackel}},
  \bibinfo {author} {\bibfnamefont {S.~A.}\ \bibnamefont {Solla}}, \bibinfo
  {author} {\bibfnamefont {V.}~\bibnamefont {Vapnik}}, \ and\ \bibinfo {author}
  {\bibfnamefont {J.~S.}\ \bibnamefont {Denker}},\ }in\ \href@noop {} {\emph
  {\bibinfo {booktitle} {Advances in Neural Information Processing Systems}}}\
  (\bibinfo {year} {1994})\ pp.\ \bibinfo {pages} {327--334}\BibitemShut
  {NoStop}%
\bibitem [{\citenamefont {{M\"uller}}\ \emph {et~al.}(1996)\citenamefont
  {{M\"uller}}, \citenamefont {Finke}, \citenamefont {Murata}, \citenamefont
  {Schulten},\ and\ \citenamefont {Amari}}]{StatError_Muller1996}%
  \BibitemOpen
  \bibfield  {author} {\bibinfo {author} {\bibfnamefont {K.~R.}\ \bibnamefont
  {{M\"uller}}}, \bibinfo {author} {\bibfnamefont {M.}~\bibnamefont {Finke}},
  \bibinfo {author} {\bibfnamefont {N.}~\bibnamefont {Murata}}, \bibinfo
  {author} {\bibfnamefont {K.}~\bibnamefont {Schulten}}, \ and\ \bibinfo
  {author} {\bibfnamefont {S.}~\bibnamefont {Amari}},\ }\href@noop {}
  {\bibfield  {journal} {\bibinfo  {journal} {Neural Comp.}\ }\textbf {\bibinfo
  {volume} {8}},\ \bibinfo {pages} {1085} (\bibinfo {year} {1996})}\BibitemShut
  {NoStop}%
\bibitem [{\citenamefont {Head}\ \emph {et~al.}(2021)\citenamefont {Head},
  \citenamefont {Kumar}, \citenamefont {Nahrstaedt}, \citenamefont {Louppe},\
  and\ \citenamefont {Shcherbatyi}}]{scikit-optimize}%
  \BibitemOpen
  \bibfield  {author} {\bibinfo {author} {\bibfnamefont {T.}~\bibnamefont
  {Head}}, \bibinfo {author} {\bibfnamefont {M.}~\bibnamefont {Kumar}},
  \bibinfo {author} {\bibfnamefont {H.}~\bibnamefont {Nahrstaedt}}, \bibinfo
  {author} {\bibfnamefont {G.}~\bibnamefont {Louppe}}, \ and\ \bibinfo {author}
  {\bibfnamefont {I.}~\bibnamefont {Shcherbatyi}},\ }\href {\doibase
  10.5281/zenodo.5565057} {\enquote {\bibinfo {title}
  {scikit-optimize/scikit-optimize},}\ } (\bibinfo {year} {2021})\BibitemShut
  {NoStop}%
\bibitem [{\citenamefont {Liu}\ and\ \citenamefont {Nocedal}(1989)}]{lbfgs}%
  \BibitemOpen
  \bibfield  {author} {\bibinfo {author} {\bibfnamefont {D.~C.}\ \bibnamefont
  {Liu}}\ and\ \bibinfo {author} {\bibfnamefont {J.}~\bibnamefont {Nocedal}},\
  }\href {\doibase 10.1007/bf01589116} {\bibfield  {journal} {\bibinfo
  {journal} {Mathematical Programming}\ }\textbf {\bibinfo {volume} {45}},\
  \bibinfo {pages} {503} (\bibinfo {year} {1989})}\BibitemShut {NoStop}%
\bibitem [{\citenamefont {Harris}\ \emph {et~al.}(2020)\citenamefont {Harris},
  \citenamefont {Millman}, \citenamefont {van~der Walt}, \citenamefont
  {Gommers}, \citenamefont {Virtanen}, \citenamefont {Cournapeau},
  \citenamefont {Wieser}, \citenamefont {Taylor}, \citenamefont {Berg},
  \citenamefont {Smith}, \citenamefont {Kern}, \citenamefont {Picus},
  \citenamefont {Hoyer}, \citenamefont {van Kerkwijk}, \citenamefont {Brett},
  \citenamefont {Haldane}, \citenamefont {del R{\'{i}}o}, \citenamefont
  {Wiebe}, \citenamefont {Peterson}, \citenamefont {G{\'{e}}rard-Marchant},
  \citenamefont {Sheppard}, \citenamefont {Reddy}, \citenamefont {Weckesser},
  \citenamefont {Abbasi}, \citenamefont {Gohlke},\ and\ \citenamefont
  {Oliphant}}]{numpy}%
  \BibitemOpen
  \bibfield  {author} {\bibinfo {author} {\bibfnamefont {C.~R.}\ \bibnamefont
  {Harris}}, \bibinfo {author} {\bibfnamefont {K.~J.}\ \bibnamefont {Millman}},
  \bibinfo {author} {\bibfnamefont {S.~J.}\ \bibnamefont {van~der Walt}},
  \bibinfo {author} {\bibfnamefont {R.}~\bibnamefont {Gommers}}, \bibinfo
  {author} {\bibfnamefont {P.}~\bibnamefont {Virtanen}}, \bibinfo {author}
  {\bibfnamefont {D.}~\bibnamefont {Cournapeau}}, \bibinfo {author}
  {\bibfnamefont {E.}~\bibnamefont {Wieser}}, \bibinfo {author} {\bibfnamefont
  {J.}~\bibnamefont {Taylor}}, \bibinfo {author} {\bibfnamefont
  {S.}~\bibnamefont {Berg}}, \bibinfo {author} {\bibfnamefont {N.~J.}\
  \bibnamefont {Smith}}, \bibinfo {author} {\bibfnamefont {R.}~\bibnamefont
  {Kern}}, \bibinfo {author} {\bibfnamefont {M.}~\bibnamefont {Picus}},
  \bibinfo {author} {\bibfnamefont {S.}~\bibnamefont {Hoyer}}, \bibinfo
  {author} {\bibfnamefont {M.~H.}\ \bibnamefont {van Kerkwijk}}, \bibinfo
  {author} {\bibfnamefont {M.}~\bibnamefont {Brett}}, \bibinfo {author}
  {\bibfnamefont {A.}~\bibnamefont {Haldane}}, \bibinfo {author} {\bibfnamefont
  {J.~F.}\ \bibnamefont {del R{\'{i}}o}}, \bibinfo {author} {\bibfnamefont
  {M.}~\bibnamefont {Wiebe}}, \bibinfo {author} {\bibfnamefont
  {P.}~\bibnamefont {Peterson}}, \bibinfo {author} {\bibfnamefont
  {P.}~\bibnamefont {G{\'{e}}rard-Marchant}}, \bibinfo {author} {\bibfnamefont
  {K.}~\bibnamefont {Sheppard}}, \bibinfo {author} {\bibfnamefont
  {T.}~\bibnamefont {Reddy}}, \bibinfo {author} {\bibfnamefont
  {W.}~\bibnamefont {Weckesser}}, \bibinfo {author} {\bibfnamefont
  {H.}~\bibnamefont {Abbasi}}, \bibinfo {author} {\bibfnamefont
  {C.}~\bibnamefont {Gohlke}}, \ and\ \bibinfo {author} {\bibfnamefont {T.~E.}\
  \bibnamefont {Oliphant}},\ }\href {\doibase 10.1038/s41586-020-2649-2}
  {\bibfield  {journal} {\bibinfo  {journal} {Nature}\ }\textbf {\bibinfo
  {volume} {585}},\ \bibinfo {pages} {357} (\bibinfo {year}
  {2020})}\BibitemShut {NoStop}%
\bibitem [{\citenamefont {Lam}\ \emph {et~al.}(2015)\citenamefont {Lam},
  \citenamefont {Pitrou},\ and\ \citenamefont {Seibert}}]{numba}%
  \BibitemOpen
  \bibfield  {author} {\bibinfo {author} {\bibfnamefont {S.~K.}\ \bibnamefont
  {Lam}}, \bibinfo {author} {\bibfnamefont {A.}~\bibnamefont {Pitrou}}, \ and\
  \bibinfo {author} {\bibfnamefont {S.}~\bibnamefont {Seibert}},\ }in\ \href
  {\doibase 10.1145/2833157.2833162} {\emph {\bibinfo {booktitle} {Proceedings
  of the Second Workshop on the LLVM Compiler Infrastructure in HPC}}},\
  \bibinfo {series and number} {LLVM '15}\ (\bibinfo  {publisher} {Association
  for Computing Machinery},\ \bibinfo {address} {New York, NY, USA},\ \bibinfo
  {year} {2015})\BibitemShut {NoStop}%
\bibitem [{\citenamefont {Becke}(1993)}]{b3}%
  \BibitemOpen
  \bibfield  {author} {\bibinfo {author} {\bibfnamefont {A.~D.}\ \bibnamefont
  {Becke}},\ }\href {\doibase 10.1063/1.464913} {\bibfield  {journal} {\bibinfo
   {journal} {The Journal of Chemical Physics}\ }\textbf {\bibinfo {volume}
  {98}},\ \bibinfo {pages} {5648} (\bibinfo {year} {1993})},\ \Eprint
  {http://arxiv.org/abs/https://doi.org/10.1063/1.464913}
  {https://doi.org/10.1063/1.464913} \BibitemShut {NoStop}%
\bibitem [{\citenamefont {Lee}\ \emph {et~al.}(1988)\citenamefont {Lee},
  \citenamefont {Yang},\ and\ \citenamefont {Parr}}]{lyp}%
  \BibitemOpen
  \bibfield  {author} {\bibinfo {author} {\bibfnamefont {C.}~\bibnamefont
  {Lee}}, \bibinfo {author} {\bibfnamefont {W.}~\bibnamefont {Yang}}, \ and\
  \bibinfo {author} {\bibfnamefont {R.~G.}\ \bibnamefont {Parr}},\ }\href
  {\doibase 10.1103/PhysRevB.37.785} {\bibfield  {journal} {\bibinfo  {journal}
  {Phys. Rev. B}\ }\textbf {\bibinfo {volume} {37}},\ \bibinfo {pages} {785}
  (\bibinfo {year} {1988})}\BibitemShut {NoStop}%
\bibitem [{\citenamefont {Petersson}\ \emph {et~al.}(1988)\citenamefont
  {Petersson}, \citenamefont {Bennett}, \citenamefont {Tensfeldt},
  \citenamefont {Al‐Laham}, \citenamefont {Shirley},\ and\ \citenamefont
  {Mantzaris}}]{631g2dfp}%
  \BibitemOpen
  \bibfield  {author} {\bibinfo {author} {\bibfnamefont {G.~A.}\ \bibnamefont
  {Petersson}}, \bibinfo {author} {\bibfnamefont {A.}~\bibnamefont {Bennett}},
  \bibinfo {author} {\bibfnamefont {T.~G.}\ \bibnamefont {Tensfeldt}}, \bibinfo
  {author} {\bibfnamefont {M.~A.}\ \bibnamefont {Al‐Laham}}, \bibinfo
  {author} {\bibfnamefont {W.~A.}\ \bibnamefont {Shirley}}, \ and\ \bibinfo
  {author} {\bibfnamefont {J.}~\bibnamefont {Mantzaris}},\ }\href {\doibase
  10.1063/1.455064} {\bibfield  {journal} {\bibinfo  {journal} {The Journal of
  Chemical Physics}\ }\textbf {\bibinfo {volume} {89}},\ \bibinfo {pages}
  {2193} (\bibinfo {year} {1988})},\ \Eprint
  {http://arxiv.org/abs/https://doi.org/10.1063/1.455064}
  {https://doi.org/10.1063/1.455064} \BibitemShut {NoStop}%
\bibitem [{\citenamefont {Chai}\ and\ \citenamefont
  {Head-Gordon}(2008)}]{wb97xd}%
  \BibitemOpen
  \bibfield  {author} {\bibinfo {author} {\bibfnamefont {J.-D.}\ \bibnamefont
  {Chai}}\ and\ \bibinfo {author} {\bibfnamefont {M.}~\bibnamefont
  {Head-Gordon}},\ }\href {\doibase 10.1039/B810189B} {\bibfield  {journal}
  {\bibinfo  {journal} {Phys. Chem. Chem. Phys.}\ }\textbf {\bibinfo {volume}
  {10}},\ \bibinfo {pages} {6615} (\bibinfo {year} {2008})}\BibitemShut
  {NoStop}%
\bibitem [{\citenamefont {Weigend}\ and\ \citenamefont
  {Ahlrichs}(2005)}]{def2}%
  \BibitemOpen
  \bibfield  {author} {\bibinfo {author} {\bibfnamefont {F.}~\bibnamefont
  {Weigend}}\ and\ \bibinfo {author} {\bibfnamefont {R.}~\bibnamefont
  {Ahlrichs}},\ }\href {\doibase 10.1039/B508541A} {\bibfield  {journal}
  {\bibinfo  {journal} {Phys. Chem. Chem. Phys.}\ }\textbf {\bibinfo {volume}
  {7}},\ \bibinfo {pages} {3297} (\bibinfo {year} {2005})}\BibitemShut
  {NoStop}%
\bibitem [{\citenamefont {Yao}\ \emph {et~al.}(2018)\citenamefont {Yao},
  \citenamefont {Herr}, \citenamefont {Toth}, \citenamefont {Mckintyre},\ and\
  \citenamefont {Parkhill}}]{mbsf}%
  \BibitemOpen
  \bibfield  {author} {\bibinfo {author} {\bibfnamefont {K.}~\bibnamefont
  {Yao}}, \bibinfo {author} {\bibfnamefont {J.~E.}\ \bibnamefont {Herr}},
  \bibinfo {author} {\bibfnamefont {D.~W.}\ \bibnamefont {Toth}}, \bibinfo
  {author} {\bibfnamefont {R.}~\bibnamefont {Mckintyre}}, \ and\ \bibinfo
  {author} {\bibfnamefont {J.}~\bibnamefont {Parkhill}},\ }\href {\doibase
  10.1039/c7sc04934j} {\bibfield  {journal} {\bibinfo  {journal} {Chemical
  Science}\ }\textbf {\bibinfo {volume} {9}},\ \bibinfo {pages} {2261}
  (\bibinfo {year} {2018})}\BibitemShut {NoStop}%
\bibitem [{\citenamefont {Christensen}\ \emph {et~al.}(2017)\citenamefont
  {Christensen}, \citenamefont {Bratholm}, \citenamefont {Faber}, \citenamefont
  {Huang}, \citenamefont {Tkatchenko}, \citenamefont {M\"uller},\ and\
  \citenamefont {von Lilienfeld}}]{qml}%
  \BibitemOpen
  \bibfield  {author} {\bibinfo {author} {\bibfnamefont {A.~S.}\ \bibnamefont
  {Christensen}}, \bibinfo {author} {\bibfnamefont {L.~A.}\ \bibnamefont
  {Bratholm}}, \bibinfo {author} {\bibfnamefont {F.~A.}\ \bibnamefont {Faber}},
  \bibinfo {author} {\bibfnamefont {B.}~\bibnamefont {Huang}}, \bibinfo
  {author} {\bibfnamefont {A.}~\bibnamefont {Tkatchenko}}, \bibinfo {author}
  {\bibfnamefont {K.~R.}\ \bibnamefont {M\"uller}}, \ and\ \bibinfo {author}
  {\bibfnamefont {O.~A.}\ \bibnamefont {von Lilienfeld}},\ }\href {\doibase
  10.5281/zenodo.817332} {\enquote {\bibinfo {title} {Qml: A python toolkit for
  quantum machine learning},}\ } (\bibinfo {year} {2017})\BibitemShut {NoStop}%
\bibitem [{\citenamefont {Himanen}\ \emph {et~al.}(2020)\citenamefont
  {Himanen}, \citenamefont {J{\"a}ger}, \citenamefont {Morooka}, \citenamefont
  {Federici~Canova}, \citenamefont {Ranawat}, \citenamefont {Gao},
  \citenamefont {Rinke},\ and\ \citenamefont {Foster}}]{dscribe}%
  \BibitemOpen
  \bibfield  {author} {\bibinfo {author} {\bibfnamefont {L.}~\bibnamefont
  {Himanen}}, \bibinfo {author} {\bibfnamefont {M.~O.~J.}\ \bibnamefont
  {J{\"a}ger}}, \bibinfo {author} {\bibfnamefont {E.~V.}\ \bibnamefont
  {Morooka}}, \bibinfo {author} {\bibfnamefont {F.}~\bibnamefont
  {Federici~Canova}}, \bibinfo {author} {\bibfnamefont {Y.~S.}\ \bibnamefont
  {Ranawat}}, \bibinfo {author} {\bibfnamefont {D.~Z.}\ \bibnamefont {Gao}},
  \bibinfo {author} {\bibfnamefont {P.}~\bibnamefont {Rinke}}, \ and\ \bibinfo
  {author} {\bibfnamefont {A.~S.}\ \bibnamefont {Foster}},\ }\href {\doibase
  10.1016/j.cpc.2019.106949} {\bibfield  {journal} {\bibinfo  {journal}
  {Computer Physics Communications}\ }\textbf {\bibinfo {volume} {247}},\
  \bibinfo {pages} {106949} (\bibinfo {year} {2020})}\BibitemShut {NoStop}%
\bibitem [{\citenamefont {Rupp}(2015)}]{qmmlpack}%
  \BibitemOpen
  \bibfield  {author} {\bibinfo {author} {\bibfnamefont {M.}~\bibnamefont
  {Rupp}},\ }\href {\doibase 10.1002/qua.24954} {\bibfield  {journal} {\bibinfo
   {journal} {International Journal of Quantum Chemistry}\ }\textbf {\bibinfo
  {volume} {115}},\ \bibinfo {pages} {1058} (\bibinfo {year}
  {2015})}\BibitemShut {NoStop}%
\bibitem [{\citenamefont {Sch\"{u}tt}\ \emph {et~al.}(2018)\citenamefont
  {Sch\"{u}tt}, \citenamefont {Kessel}, \citenamefont {Gastegger},
  \citenamefont {Nicoli}, \citenamefont {Tkatchenko},\ and\ \citenamefont
  {M\"{u}ller}}]{schnet}%
  \BibitemOpen
  \bibfield  {author} {\bibinfo {author} {\bibfnamefont {K.~T.}\ \bibnamefont
  {Sch\"{u}tt}}, \bibinfo {author} {\bibfnamefont {P.}~\bibnamefont {Kessel}},
  \bibinfo {author} {\bibfnamefont {M.}~\bibnamefont {Gastegger}}, \bibinfo
  {author} {\bibfnamefont {K.~A.}\ \bibnamefont {Nicoli}}, \bibinfo {author}
  {\bibfnamefont {A.}~\bibnamefont {Tkatchenko}}, \ and\ \bibinfo {author}
  {\bibfnamefont {K.-R.}\ \bibnamefont {M\"{u}ller}},\ }\href {\doibase
  10.1021/acs.jctc.8b00908} {\bibfield  {journal} {\bibinfo  {journal} {Journal
  of Chemical Theory and Computation}\ }\textbf {\bibinfo {volume} {15}},\
  \bibinfo {pages} {448} (\bibinfo {year} {2018})}\BibitemShut {NoStop}%
\bibitem [{\citenamefont {Sch{\"{u}}tt}\ \emph {et~al.}(2021)\citenamefont
  {Sch{\"{u}}tt}, \citenamefont {Unke},\ and\ \citenamefont
  {Gastegger}}]{painn}%
  \BibitemOpen
  \bibfield  {author} {\bibinfo {author} {\bibfnamefont {K.~T.}\ \bibnamefont
  {Sch{\"{u}}tt}}, \bibinfo {author} {\bibfnamefont {O.~T.}\ \bibnamefont
  {Unke}}, \ and\ \bibinfo {author} {\bibfnamefont {M.}~\bibnamefont
  {Gastegger}},\ }\href {https://arxiv.org/abs/2102.03150} {\bibfield
  {journal} {\bibinfo  {journal} {CoRR}\ }\textbf {\bibinfo {volume}
  {abs/2102.03150}} (\bibinfo {year} {2021})},\ \Eprint
  {http://arxiv.org/abs/2102.03150} {2102.03150} \BibitemShut {NoStop}%
\bibitem [{\citenamefont {Sch\"{u}tt}\ \emph {et~al.}(2017)\citenamefont
  {Sch\"{u}tt}, \citenamefont {Kindermans}, \citenamefont {Sauceda~Felix},
  \citenamefont {Chmiela}, \citenamefont {Tkatchenko},\ and\ \citenamefont
  {M\"{u}ller}}]{schnetpack}%
  \BibitemOpen
  \bibfield  {author} {\bibinfo {author} {\bibfnamefont {K.}~\bibnamefont
  {Sch\"{u}tt}}, \bibinfo {author} {\bibfnamefont {P.-J.}\ \bibnamefont
  {Kindermans}}, \bibinfo {author} {\bibfnamefont {H.~E.}\ \bibnamefont
  {Sauceda~Felix}}, \bibinfo {author} {\bibfnamefont {S.}~\bibnamefont
  {Chmiela}}, \bibinfo {author} {\bibfnamefont {A.}~\bibnamefont {Tkatchenko}},
  \ and\ \bibinfo {author} {\bibfnamefont {K.-R.}\ \bibnamefont {M\"{u}ller}},\
  }in\ \href
  {https://proceedings.neurips.cc/paper/2017/file/303ed4c69846ab36c2904d3ba8573050-Paper.pdf}
  {\emph {\bibinfo {booktitle} {Advances in Neural Information Processing
  Systems}}},\ Vol.~\bibinfo {volume} {30},\ \bibinfo {editor} {edited by\
  \bibinfo {editor} {\bibfnamefont {I.}~\bibnamefont {Guyon}}, \bibinfo
  {editor} {\bibfnamefont {U.~V.}\ \bibnamefont {Luxburg}}, \bibinfo {editor}
  {\bibfnamefont {S.}~\bibnamefont {Bengio}}, \bibinfo {editor} {\bibfnamefont
  {H.}~\bibnamefont {Wallach}}, \bibinfo {editor} {\bibfnamefont
  {R.}~\bibnamefont {Fergus}}, \bibinfo {editor} {\bibfnamefont
  {S.}~\bibnamefont {Vishwanathan}}, \ and\ \bibinfo {editor} {\bibfnamefont
  {R.}~\bibnamefont {Garnett}}}\ (\bibinfo  {publisher} {Curran Associates,
  Inc.},\ \bibinfo {year} {2017})\BibitemShut {NoStop}%
\bibitem [{\citenamefont {Musaelian}\ \emph {et~al.}(2023)\citenamefont
  {Musaelian}, \citenamefont {Batzner}, \citenamefont {Johansson},
  \citenamefont {Sun}, \citenamefont {Owen}, \citenamefont {Kornbluth},\ and\
  \citenamefont {Kozinsky}}]{allegro}%
  \BibitemOpen
  \bibfield  {author} {\bibinfo {author} {\bibfnamefont {A.}~\bibnamefont
  {Musaelian}}, \bibinfo {author} {\bibfnamefont {S.}~\bibnamefont {Batzner}},
  \bibinfo {author} {\bibfnamefont {A.}~\bibnamefont {Johansson}}, \bibinfo
  {author} {\bibfnamefont {L.}~\bibnamefont {Sun}}, \bibinfo {author}
  {\bibfnamefont {C.~J.}\ \bibnamefont {Owen}}, \bibinfo {author}
  {\bibfnamefont {M.}~\bibnamefont {Kornbluth}}, \ and\ \bibinfo {author}
  {\bibfnamefont {B.}~\bibnamefont {Kozinsky}},\ }\href {\doibase
  10.1038/s41467-023-36329-y} {\bibfield  {journal} {\bibinfo  {journal}
  {Nature Communications}\ }\textbf {\bibinfo {volume} {14}} (\bibinfo {year}
  {2023}),\ 10.1038/s41467-023-36329-y}\BibitemShut {NoStop}%
\bibitem [{\citenamefont {Batzner}\ \emph {et~al.}(2022)\citenamefont
  {Batzner}, \citenamefont {Musaelian}, \citenamefont {Sun}, \citenamefont
  {Geiger}, \citenamefont {Mailoa}, \citenamefont {Kornbluth}, \citenamefont
  {Molinari}, \citenamefont {Smidt},\ and\ \citenamefont {Kozinsky}}]{nequip}%
  \BibitemOpen
  \bibfield  {author} {\bibinfo {author} {\bibfnamefont {S.}~\bibnamefont
  {Batzner}}, \bibinfo {author} {\bibfnamefont {A.}~\bibnamefont {Musaelian}},
  \bibinfo {author} {\bibfnamefont {L.}~\bibnamefont {Sun}}, \bibinfo {author}
  {\bibfnamefont {M.}~\bibnamefont {Geiger}}, \bibinfo {author} {\bibfnamefont
  {J.~P.}\ \bibnamefont {Mailoa}}, \bibinfo {author} {\bibfnamefont
  {M.}~\bibnamefont {Kornbluth}}, \bibinfo {author} {\bibfnamefont
  {N.}~\bibnamefont {Molinari}}, \bibinfo {author} {\bibfnamefont {T.~E.}\
  \bibnamefont {Smidt}}, \ and\ \bibinfo {author} {\bibfnamefont
  {B.}~\bibnamefont {Kozinsky}},\ }\href {\doibase 10.1038/s41467-022-29939-5}
  {\bibfield  {journal} {\bibinfo  {journal} {Nature Communications}\ }\textbf
  {\bibinfo {volume} {13}} (\bibinfo {year} {2022}),\
  10.1038/s41467-022-29939-5}\BibitemShut {NoStop}%
\bibitem [{\citenamefont {Geiger}\ \emph
  {et~al.}(2022{\natexlab{a}})\citenamefont {Geiger}, \citenamefont {Smidt},
  \citenamefont {Alby}, \citenamefont {Miller}, \citenamefont {Boomsma},
  \citenamefont {Dice}, \citenamefont {Lapchevskyi}, \citenamefont {Weiler},
  \citenamefont {{Michał Tyszkiewicz}}, \citenamefont {Uhrin}, \citenamefont
  {Batzner}, \citenamefont {Madisetti}, \citenamefont {Frellsen}, \citenamefont
  {Jung}, \citenamefont {Sanborn}, \citenamefont {{Jkh}}, \citenamefont
  {{Mingjian Wen}}, \citenamefont {Rackers}, \citenamefont {Rød},\ and\
  \citenamefont {Bailey}}]{nequip_zenodo}%
  \BibitemOpen
  \bibfield  {author} {\bibinfo {author} {\bibfnamefont {M.}~\bibnamefont
  {Geiger}}, \bibinfo {author} {\bibfnamefont {T.}~\bibnamefont {Smidt}},
  \bibinfo {author} {\bibfnamefont {M.}~\bibnamefont {Alby}}, \bibinfo {author}
  {\bibfnamefont {B.~K.}\ \bibnamefont {Miller}}, \bibinfo {author}
  {\bibfnamefont {W.}~\bibnamefont {Boomsma}}, \bibinfo {author} {\bibfnamefont
  {B.}~\bibnamefont {Dice}}, \bibinfo {author} {\bibfnamefont {K.}~\bibnamefont
  {Lapchevskyi}}, \bibinfo {author} {\bibfnamefont {M.}~\bibnamefont {Weiler}},
  \bibinfo {author} {\bibnamefont {{Michał Tyszkiewicz}}}, \bibinfo {author}
  {\bibfnamefont {M.}~\bibnamefont {Uhrin}}, \bibinfo {author} {\bibfnamefont
  {S.}~\bibnamefont {Batzner}}, \bibinfo {author} {\bibfnamefont
  {D.}~\bibnamefont {Madisetti}}, \bibinfo {author} {\bibfnamefont
  {J.}~\bibnamefont {Frellsen}}, \bibinfo {author} {\bibfnamefont
  {N.}~\bibnamefont {Jung}}, \bibinfo {author} {\bibfnamefont {S.}~\bibnamefont
  {Sanborn}}, \bibinfo {author} {\bibnamefont {{Jkh}}}, \bibinfo {author}
  {\bibnamefont {{Mingjian Wen}}}, \bibinfo {author} {\bibfnamefont
  {J.}~\bibnamefont {Rackers}}, \bibinfo {author} {\bibfnamefont
  {M.}~\bibnamefont {Rød}}, \ and\ \bibinfo {author} {\bibfnamefont
  {M.}~\bibnamefont {Bailey}},\ }\href {\doibase 10.5281/ZENODO.7430260}
  {\enquote {\bibinfo {title} {e3nn/e3nn: 2022-12-12},}\ } (\bibinfo {year}
  {2022}{\natexlab{a}})\BibitemShut {NoStop}%
\bibitem [{\citenamefont {Geiger}\ \emph
  {et~al.}(2022{\natexlab{b}})\citenamefont {Geiger}, \citenamefont {Smidt},
  \citenamefont {M.}, \citenamefont {Miller}, \citenamefont {Boomsma},
  \citenamefont {Dice}, \citenamefont {Lapchevskyi}, \citenamefont {Weiler},
  \citenamefont {Tyszkiewicz}, \citenamefont {Batzner}, \citenamefont
  {Madisetti}, \citenamefont {Uhrin}, \citenamefont {Frellsen}, \citenamefont
  {Jung}, \citenamefont {Sanborn}, \citenamefont {Wen}, \citenamefont
  {Rackers}, \citenamefont {Rød},\ and\ \citenamefont {Bailey}}]{e3nn}%
  \BibitemOpen
  \bibfield  {author} {\bibinfo {author} {\bibfnamefont {M.}~\bibnamefont
  {Geiger}}, \bibinfo {author} {\bibfnamefont {T.}~\bibnamefont {Smidt}},
  \bibinfo {author} {\bibfnamefont {A.}~\bibnamefont {M.}}, \bibinfo {author}
  {\bibfnamefont {B.~K.}\ \bibnamefont {Miller}}, \bibinfo {author}
  {\bibfnamefont {W.}~\bibnamefont {Boomsma}}, \bibinfo {author} {\bibfnamefont
  {B.}~\bibnamefont {Dice}}, \bibinfo {author} {\bibfnamefont {K.}~\bibnamefont
  {Lapchevskyi}}, \bibinfo {author} {\bibfnamefont {M.}~\bibnamefont {Weiler}},
  \bibinfo {author} {\bibfnamefont {M.}~\bibnamefont {Tyszkiewicz}}, \bibinfo
  {author} {\bibfnamefont {S.}~\bibnamefont {Batzner}}, \bibinfo {author}
  {\bibfnamefont {D.}~\bibnamefont {Madisetti}}, \bibinfo {author}
  {\bibfnamefont {M.}~\bibnamefont {Uhrin}}, \bibinfo {author} {\bibfnamefont
  {J.}~\bibnamefont {Frellsen}}, \bibinfo {author} {\bibfnamefont
  {N.}~\bibnamefont {Jung}}, \bibinfo {author} {\bibfnamefont {S.}~\bibnamefont
  {Sanborn}}, \bibinfo {author} {\bibfnamefont {M.}~\bibnamefont {Wen}},
  \bibinfo {author} {\bibfnamefont {J.}~\bibnamefont {Rackers}}, \bibinfo
  {author} {\bibfnamefont {M.}~\bibnamefont {Rød}}, \ and\ \bibinfo {author}
  {\bibfnamefont {M.}~\bibnamefont {Bailey}},\ }\href {\doibase
  10.5281/zenodo.6459381} {\enquote {\bibinfo {title} {Euclidean neural
  networks: e3nn},}\ } (\bibinfo {year} {2022}{\natexlab{b}})\BibitemShut
  {NoStop}%
\bibitem [{\citenamefont {Geiger}\ and\ \citenamefont
  {Smidt}(2022)}]{e3nn_paper}%
  \BibitemOpen
  \bibfield  {author} {\bibinfo {author} {\bibfnamefont {M.}~\bibnamefont
  {Geiger}}\ and\ \bibinfo {author} {\bibfnamefont {T.}~\bibnamefont {Smidt}},\
  }\href {\doibase 10.48550/ARXIV.2207.09453} {\enquote {\bibinfo {title}
  {e3nn: Euclidean neural networks},}\ } (\bibinfo {year} {2022})\BibitemShut
  {NoStop}%
\bibitem [{\citenamefont {Townsend}\ \emph {et~al.}(2020)\citenamefont
  {Townsend}, \citenamefont {Micucci}, \citenamefont {Hymel}, \citenamefont
  {Maroulas},\ and\ \citenamefont {Vogiatzis}}]{PI_townsend}%
  \BibitemOpen
  \bibfield  {author} {\bibinfo {author} {\bibfnamefont {J.}~\bibnamefont
  {Townsend}}, \bibinfo {author} {\bibfnamefont {C.~P.}\ \bibnamefont
  {Micucci}}, \bibinfo {author} {\bibfnamefont {J.~H.}\ \bibnamefont {Hymel}},
  \bibinfo {author} {\bibfnamefont {V.}~\bibnamefont {Maroulas}}, \ and\
  \bibinfo {author} {\bibfnamefont {K.~D.}\ \bibnamefont {Vogiatzis}},\ }\href
  {\doibase 10.1038/s41467-020-17035-5} {\bibfield  {journal} {\bibinfo
  {journal} {Nature Communications}\ }\textbf {\bibinfo {volume} {11}},\
  \bibinfo {pages} {3230} (\bibinfo {year} {2020})}\BibitemShut {NoStop}%
\bibitem [{\citenamefont {Bigi}\ \emph {et~al.}(2023)\citenamefont {Bigi},
  \citenamefont {Pozdnyakov},\ and\ \citenamefont {Ceriotti}}]{wigner_kernels}%
  \BibitemOpen
  \bibfield  {author} {\bibinfo {author} {\bibfnamefont {F.}~\bibnamefont
  {Bigi}}, \bibinfo {author} {\bibfnamefont {S.~N.}\ \bibnamefont
  {Pozdnyakov}}, \ and\ \bibinfo {author} {\bibfnamefont {M.}~\bibnamefont
  {Ceriotti}},\ }\href@noop {} {\enquote {\bibinfo {title} {Wigner kernels:
  body-ordered equivariant machine learning without a basis},}\ } (\bibinfo
  {year} {2023}),\ \Eprint {http://arxiv.org/abs/2303.04124} {arXiv:2303.04124
  [physics.chem-ph]} \BibitemShut {NoStop}%
\bibitem [{\citenamefont {Browning}\ \emph {et~al.}(2022)\citenamefont
  {Browning}, \citenamefont {Faber},\ and\ \citenamefont {Anatole~von
  Lilienfeld}}]{browning2022gpu}%
  \BibitemOpen
  \bibfield  {author} {\bibinfo {author} {\bibfnamefont {N.~J.}\ \bibnamefont
  {Browning}}, \bibinfo {author} {\bibfnamefont {F.~A.}\ \bibnamefont {Faber}},
  \ and\ \bibinfo {author} {\bibfnamefont {O.}~\bibnamefont {Anatole~von
  Lilienfeld}},\ }\href@noop {} {\bibfield  {journal} {\bibinfo  {journal} {The
  Journal of Chemical Physics}\ }\textbf {\bibinfo {volume} {157}},\ \bibinfo
  {pages} {214801} (\bibinfo {year} {2022})}\BibitemShut {NoStop}%
\bibitem [{\citenamefont {Atz}\ \emph {et~al.}(2022)\citenamefont {Atz},
  \citenamefont {Isert}, \citenamefont {B{\"o}cker}, \citenamefont
  {Jim{\'e}nez-Luna},\ and\ \citenamefont {Schneider}}]{atz2022delta}%
  \BibitemOpen
  \bibfield  {author} {\bibinfo {author} {\bibfnamefont {K.}~\bibnamefont
  {Atz}}, \bibinfo {author} {\bibfnamefont {C.}~\bibnamefont {Isert}}, \bibinfo
  {author} {\bibfnamefont {M.~N.}\ \bibnamefont {B{\"o}cker}}, \bibinfo
  {author} {\bibfnamefont {J.}~\bibnamefont {Jim{\'e}nez-Luna}}, \ and\
  \bibinfo {author} {\bibfnamefont {G.}~\bibnamefont {Schneider}},\ }\href@noop
  {} {\bibfield  {journal} {\bibinfo  {journal} {Physical Chemistry Chemical
  Physics}\ }\textbf {\bibinfo {volume} {24}},\ \bibinfo {pages} {10775}
  (\bibinfo {year} {2022})}\BibitemShut {NoStop}%
\bibitem [{\citenamefont {Bannwarth}\ \emph {et~al.}(2019)\citenamefont
  {Bannwarth}, \citenamefont {Ehlert},\ and\ \citenamefont {Grimme}}]{gfn2xtb}%
  \BibitemOpen
  \bibfield  {author} {\bibinfo {author} {\bibfnamefont {C.}~\bibnamefont
  {Bannwarth}}, \bibinfo {author} {\bibfnamefont {S.}~\bibnamefont {Ehlert}}, \
  and\ \bibinfo {author} {\bibfnamefont {S.}~\bibnamefont {Grimme}},\ }\href
  {\doibase 10.1021/acs.jctc.8b01176} {\bibfield  {journal} {\bibinfo
  {journal} {Journal of Chemical Theory and Computation}\ }\textbf {\bibinfo
  {volume} {15}},\ \bibinfo {pages} {1652} (\bibinfo {year}
  {2019})}\BibitemShut {NoStop}%
\bibitem [{\citenamefont {Montavon}\ \emph {et~al.}(2013)\citenamefont
  {Montavon}, \citenamefont {Rupp}, \citenamefont {Gobre}, \citenamefont
  {Vazquez-Mayagoitia}, \citenamefont {Hansen}, \citenamefont {Tkatchenko},
  \citenamefont {M{\"u}ller},\ and\ \citenamefont {von Lilienfeld}}]{qm7b}%
  \BibitemOpen
  \bibfield  {author} {\bibinfo {author} {\bibfnamefont {G.}~\bibnamefont
  {Montavon}}, \bibinfo {author} {\bibfnamefont {M.}~\bibnamefont {Rupp}},
  \bibinfo {author} {\bibfnamefont {V.}~\bibnamefont {Gobre}}, \bibinfo
  {author} {\bibfnamefont {A.}~\bibnamefont {Vazquez-Mayagoitia}}, \bibinfo
  {author} {\bibfnamefont {K.}~\bibnamefont {Hansen}}, \bibinfo {author}
  {\bibfnamefont {A.}~\bibnamefont {Tkatchenko}}, \bibinfo {author}
  {\bibfnamefont {K.-R.}\ \bibnamefont {M{\"u}ller}}, \ and\ \bibinfo {author}
  {\bibfnamefont {O.~A.}\ \bibnamefont {von Lilienfeld}},\ }\href
  {http://stacks.iop.org/1367-2630/15/i=9/a=095003} {\bibfield  {journal}
  {\bibinfo  {journal} {New Journal of Physics}\ }\textbf {\bibinfo {volume}
  {15}},\ \bibinfo {pages} {095003} (\bibinfo {year} {2013})}\BibitemShut
  {NoStop}%
\bibitem [{\citenamefont {Blum}\ and\ \citenamefont
  {Reymond}(2009)}]{970m_qm7b}%
  \BibitemOpen
  \bibfield  {author} {\bibinfo {author} {\bibfnamefont {L.~C.}\ \bibnamefont
  {Blum}}\ and\ \bibinfo {author} {\bibfnamefont {J.-L.}\ \bibnamefont
  {Reymond}},\ }\href@noop {} {\bibfield  {journal} {\bibinfo  {journal} {J.
  Am. Chem. Soc.}\ }\textbf {\bibinfo {volume} {131}},\ \bibinfo {pages} {8732}
  (\bibinfo {year} {2009})}\BibitemShut {NoStop}%
\bibitem [{\citenamefont {Christensen}\ and\ \citenamefont {von
  Lilienfeld}(2020)}]{anders_gradients_role}%
  \BibitemOpen
  \bibfield  {author} {\bibinfo {author} {\bibfnamefont {A.~S.}\ \bibnamefont
  {Christensen}}\ and\ \bibinfo {author} {\bibfnamefont {O.~A.}\ \bibnamefont
  {von Lilienfeld}},\ }\href {\doibase 10.1088/2632-2153/abba6f} {\bibfield
  {journal} {\bibinfo  {journal} {Machine Learning: Science and Technology}\
  }\textbf {\bibinfo {volume} {1}},\ \bibinfo {pages} {045018} (\bibinfo {year}
  {2020})}\BibitemShut {NoStop}%
\bibitem [{\citenamefont {Christensen}\ and\ \citenamefont
  {lilienfeld}(2020)}]{rmd17}%
  \BibitemOpen
  \bibfield  {author} {\bibinfo {author} {\bibfnamefont {A.~S.}\ \bibnamefont
  {Christensen}}\ and\ \bibinfo {author} {\bibfnamefont {A.~V.}\ \bibnamefont
  {lilienfeld}},\ }\href {\doibase 10.6084/m9.figshare.12672038.v3} {\
  (\bibinfo {year} {2020}),\ 10.6084/m9.figshare.12672038.v3}\BibitemShut
  {NoStop}%
\bibitem [{\citenamefont {Chmiela}\ \emph {et~al.}(2017)\citenamefont
  {Chmiela}, \citenamefont {Tkatchenko}, \citenamefont {Sauceda}, \citenamefont
  {Poltavsky}, \citenamefont {Sch\"{u}tt},\ and\ \citenamefont
  {M\"{u}ller}}]{gdml}%
  \BibitemOpen
  \bibfield  {author} {\bibinfo {author} {\bibfnamefont {S.}~\bibnamefont
  {Chmiela}}, \bibinfo {author} {\bibfnamefont {A.}~\bibnamefont {Tkatchenko}},
  \bibinfo {author} {\bibfnamefont {H.~E.}\ \bibnamefont {Sauceda}}, \bibinfo
  {author} {\bibfnamefont {I.}~\bibnamefont {Poltavsky}}, \bibinfo {author}
  {\bibfnamefont {K.~T.}\ \bibnamefont {Sch\"{u}tt}}, \ and\ \bibinfo {author}
  {\bibfnamefont {K.-R.}\ \bibnamefont {M\"{u}ller}},\ }\href {\doibase
  10.1126/sciadv.1603015} {\bibfield  {journal} {\bibinfo  {journal} {Science
  Advances}\ }\textbf {\bibinfo {volume} {3}} (\bibinfo {year} {2017}),\
  10.1126/sciadv.1603015}\BibitemShut {NoStop}%
\bibitem [{\citenamefont {Chmiela}\ \emph {et~al.}(2018)\citenamefont
  {Chmiela}, \citenamefont {Sauceda}, \citenamefont {M\"{u}ller},\ and\
  \citenamefont {Tkatchenko}}]{sgdml}%
  \BibitemOpen
  \bibfield  {author} {\bibinfo {author} {\bibfnamefont {S.}~\bibnamefont
  {Chmiela}}, \bibinfo {author} {\bibfnamefont {H.~E.}\ \bibnamefont
  {Sauceda}}, \bibinfo {author} {\bibfnamefont {K.-R.}\ \bibnamefont
  {M\"{u}ller}}, \ and\ \bibinfo {author} {\bibfnamefont {A.}~\bibnamefont
  {Tkatchenko}},\ }\href {\doibase 10.1038/s41467-018-06169-2} {\bibfield
  {journal} {\bibinfo  {journal} {Nature Communications}\ }\textbf {\bibinfo
  {volume} {9}} (\bibinfo {year} {2018}),\
  10.1038/s41467-018-06169-2}\BibitemShut {NoStop}%
\bibitem [{\citenamefont {Christensen}\ \emph {et~al.}(2019)\citenamefont
  {Christensen}, \citenamefont {Faber},\ and\ \citenamefont {von
  Lilienfeld}}]{op_response_anders}%
  \BibitemOpen
  \bibfield  {author} {\bibinfo {author} {\bibfnamefont {A.~S.}\ \bibnamefont
  {Christensen}}, \bibinfo {author} {\bibfnamefont {F.~A.}\ \bibnamefont
  {Faber}}, \ and\ \bibinfo {author} {\bibfnamefont {O.~A.}\ \bibnamefont {von
  Lilienfeld}},\ }\href {\doibase 10.1063/1.5053562} {\bibfield  {journal}
  {\bibinfo  {journal} {The Journal of Chemical Physics}\ }\textbf {\bibinfo
  {volume} {150}},\ \bibinfo {pages} {064105} (\bibinfo {year} {2019})},\
  \Eprint {http://arxiv.org/abs/https://doi.org/10.1063/1.5053562}
  {https://doi.org/10.1063/1.5053562} \BibitemShut {NoStop}%
\end{thebibliography}%
\end{document}

\title{
Supplementary Information: Many-body distribution functionals as compact quantum machine learning
representations} 
\maketitle

\begin{figure*}[htb]
          \centering
          \includegraphics[width=0.8\linewidth]{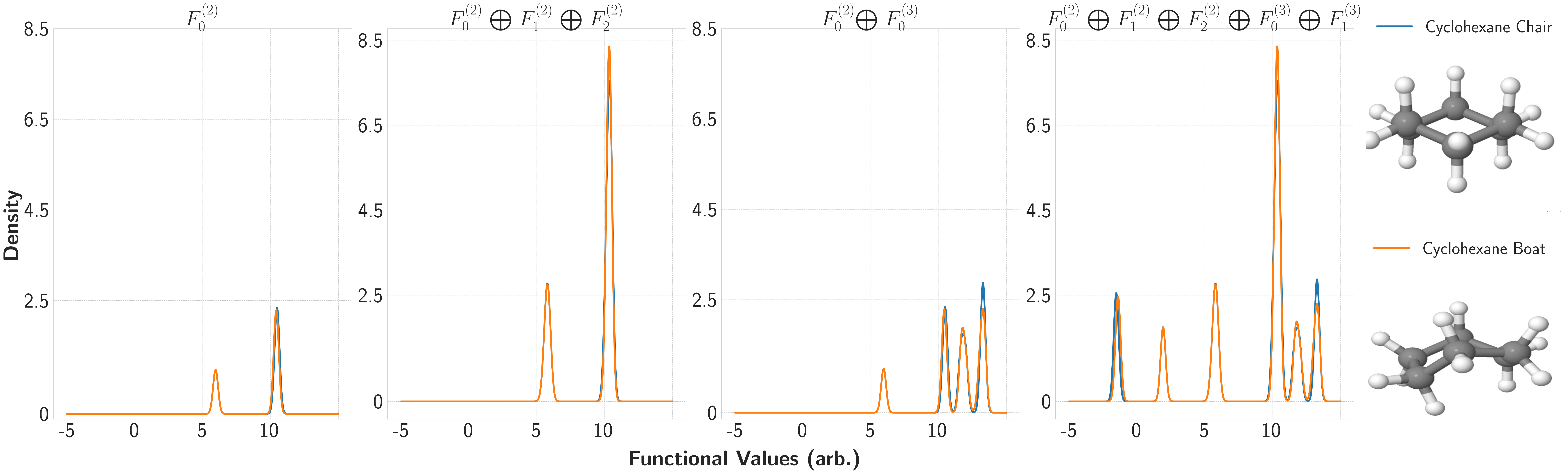}\label{fig:fig3b}
          \caption{DF versions with different number of functionals (Eqs. 11-15) for boat- and chair-conformers of cyclohexane showing effect of inclusion of higher order derivatives and many-body terms on the representation resolution.
          }
 \end{figure*}
 To generate the intermediate geometries we used the Climbing Image-Nudged Elastic Band method (CI-NEB)\cite{ci-neb} as implemented in the Atomic-Simulation-Environment (ASE)\cite{ase} using the ORCA5.0\cite{orca5} calculator.
Reactant and product states were optimized using the L-BFGS algorithm\cite{lbfgs} as implemented in ASE. Subsequently, a CI-NEB calculation was performed using 7 images (9 total geometries) to obtaine a minimal energy path connecting the chair- to the boat-like configuration.
The default convergence criteria ($f_{max}$ = 0.05 eV/Angstrom) and the method PBE0/6-31G* were used in all calulations.

 \begin{figure}[htb]
          \centering           
          \includegraphics[width=\linewidth]{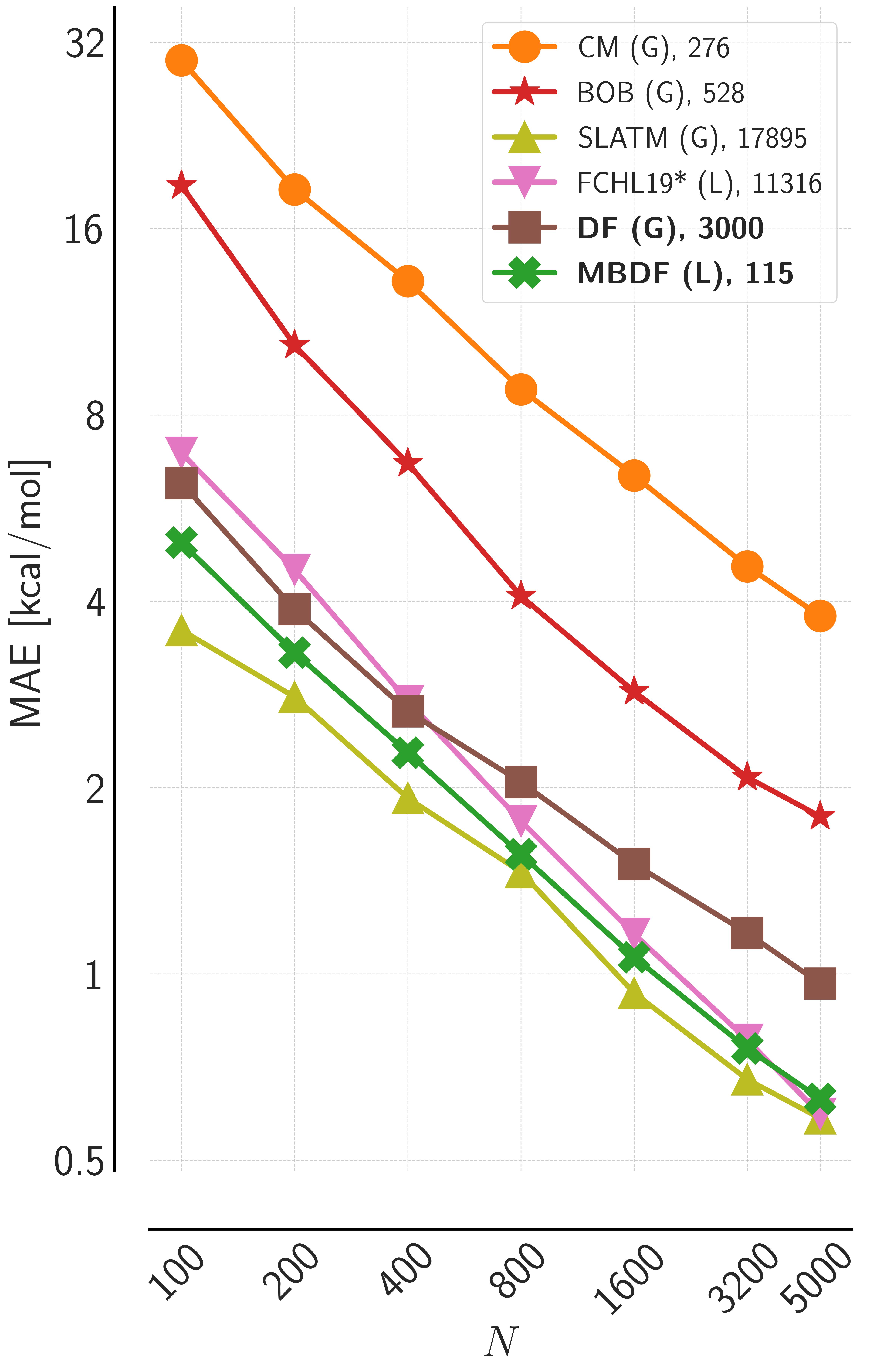}
          \caption{MBDF/DF performance and comparison to CM~\cite{CM}, BOB~\cite{bob}, SLATM~\cite{amons_slatm} and FCHL19~\cite{fchl19} representation based atomization energy estimates using the QM7b ($\sim$7k organic molecules with up to 7 heavy atoms) data set~\cite{qm7b}. 
          Training and testing data drawn at random. 
          Prediction mean absolute errors (MAE) are shown as a function of training set size.
          Numbers in legend denote representation size (feature vector dimensions), G and L denote Global and Local kernels respectively.
          Asterisk denotes representation with reduced hyperparameters used in this work.}
     \label{fig:qm7b}
 \end{figure}

  \begin{figure*}[htb]
          \centering           
          \includegraphics[width=\linewidth]{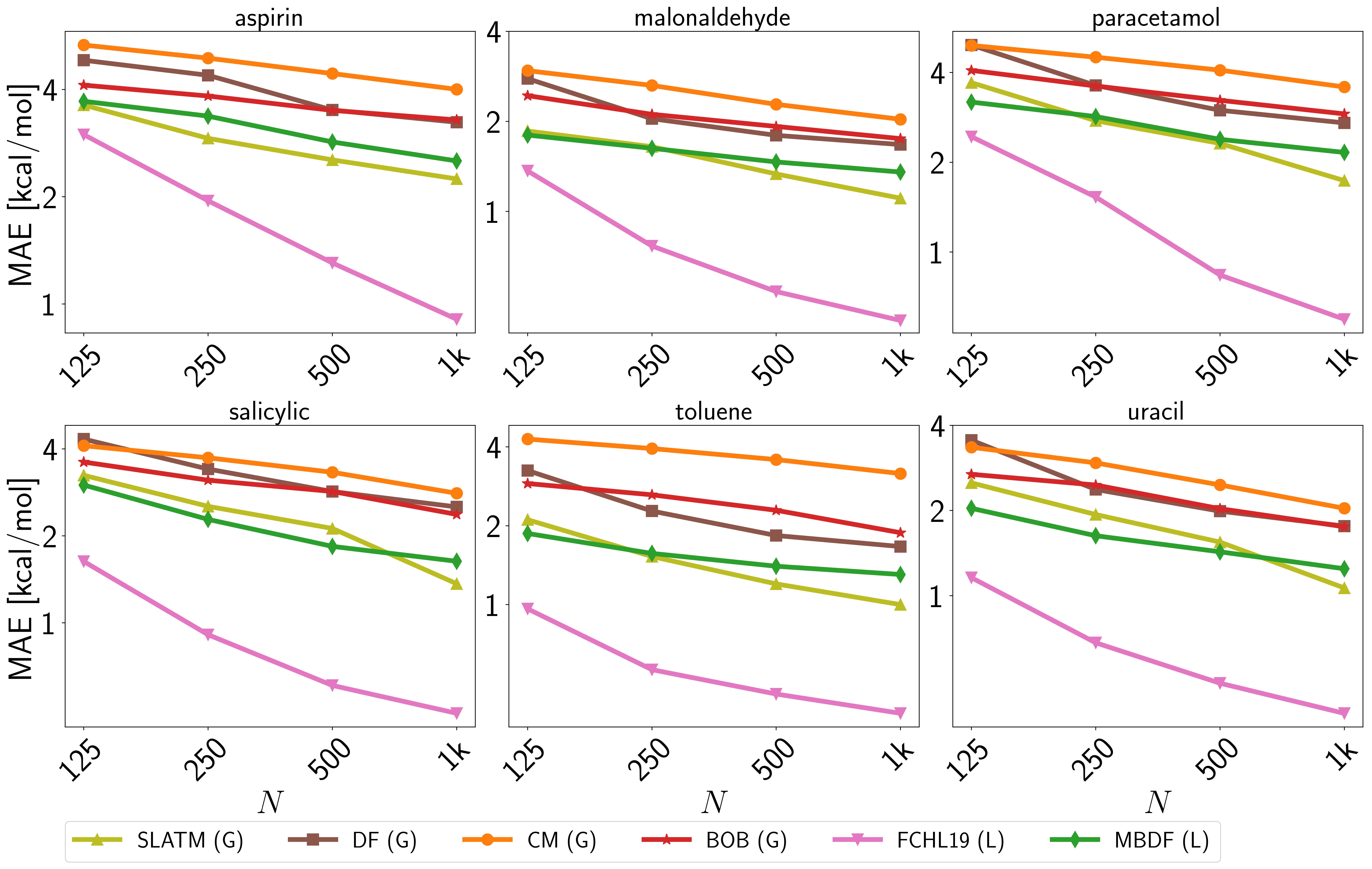}
          \caption{MBDF/DF performance and comparison to CM~\cite{CM}, BOB~\cite{bob}, SLATM~\cite{amons_slatm} and FCHL19~\cite{fchl19} representation based learning curves for energies of a few molecules from the revised MD17 dataset\cite{gdml,sgdml,anders_gradients_role,rmd17}. 
          Training and testing (1k structures) data drawn at random. 
          Prediction mean absolute errors (MAE) are shown as a function of training set size.
          G and L denote Global and Local kernels respectively.}
     \label{fig:md17}
 \end{figure*}

\begin{figure*}[ht!]
          \centering           
          \includegraphics[width=0.99\textwidth]{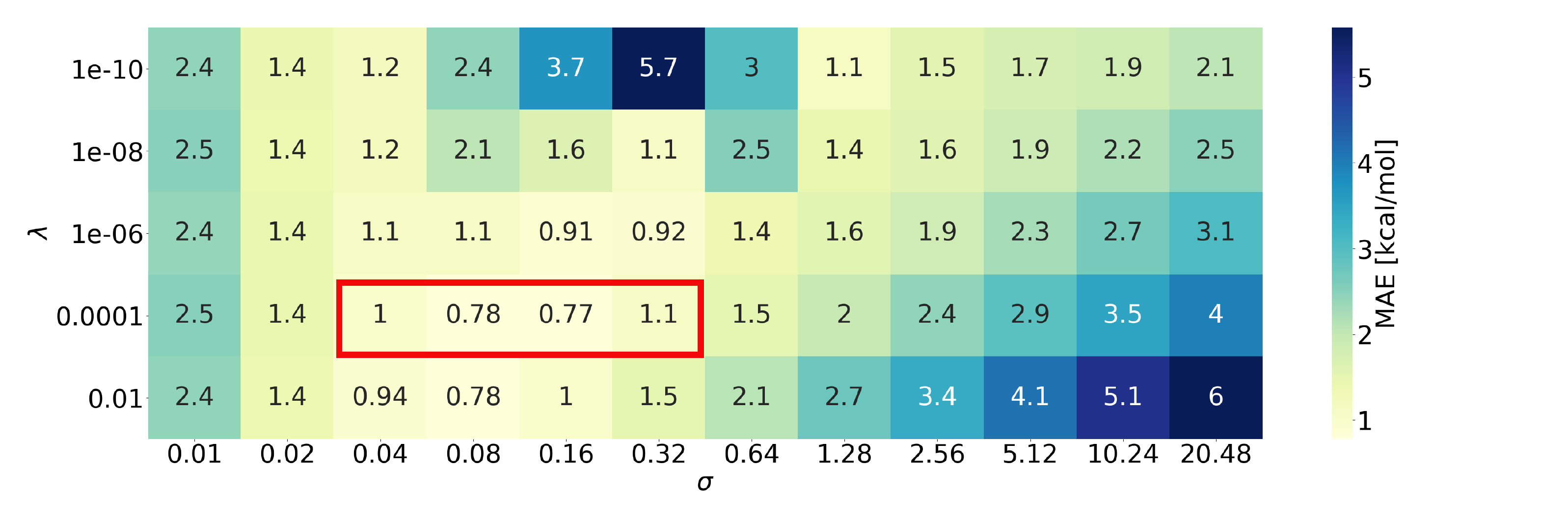}
          \caption{MBDF QML model based hyperparameter scan of local Gaussian kernel length-scale $\sigma$ and noise-level $\lambda$ with Cholesky decomposition for the kernel inversion. Dataset includes 7000 molecules from QM7b with optimal values for length-scale between $\sigma = 0.04$ and $0.32$ at $\lambda = 0.0001$. Train/test-split 6k/1k.}

     \label{SI:fig:hyperparam}
\end{figure*}

\begin{figure*}[ht!]
          \centering           
          \includegraphics[width=0.99\textwidth]{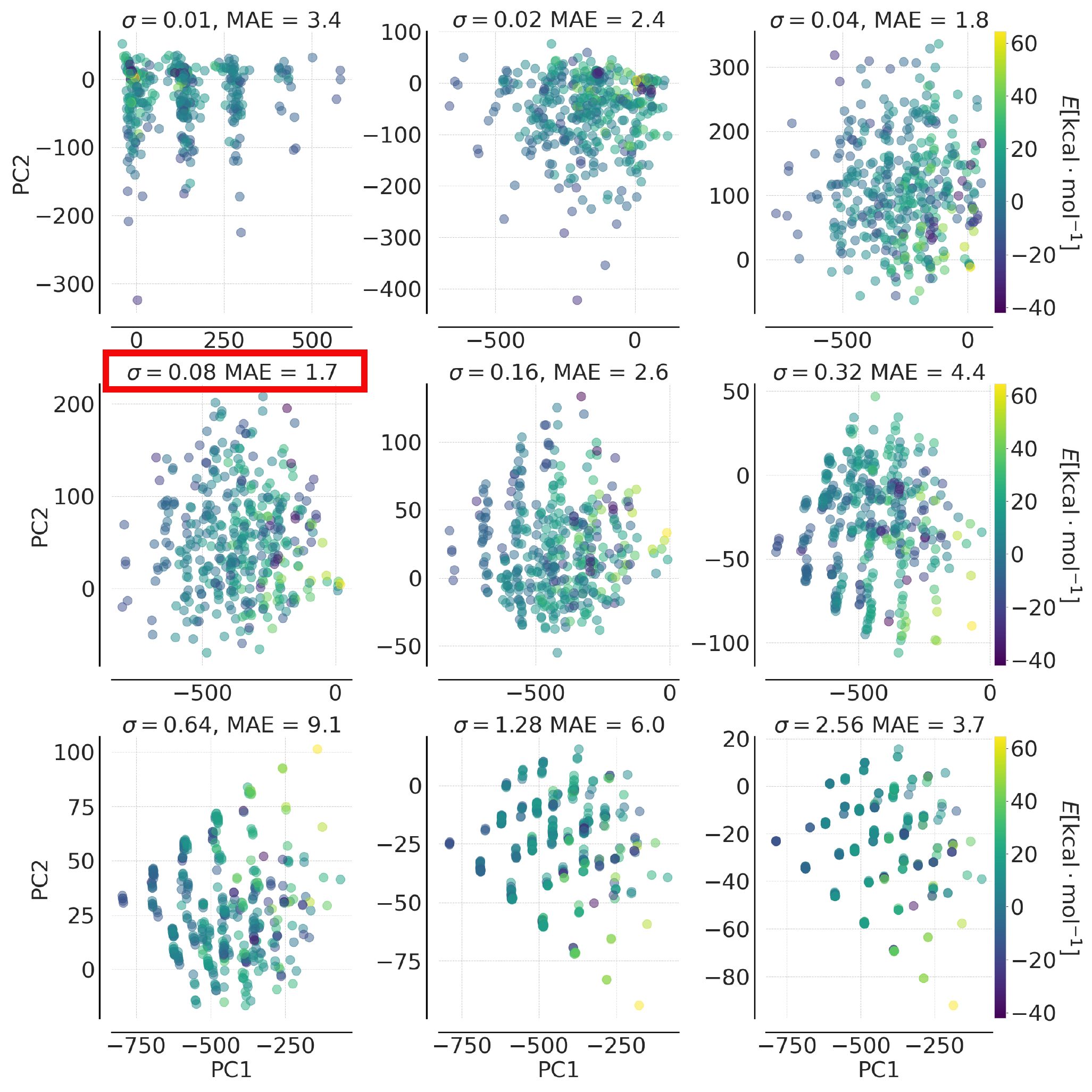}
          \caption{
          Impact of length-scale on kernel principle component analysis (PCA) for the first 400 molecules in QM7b. A scan over different $\sigma$'s using a singular value decomposition (SVD) was performed to obtain training mean absolute errors (MAE) in kcal/mol.  Colour code corresponds to atomization energies. 
          }
     \label{SI:fig:3kernel_pcas}
\end{figure*}

%\begin{figure*}[ht!]
%          \centering           
%          \includegraphics[width=0.99\textwidth]{figs/kernel_pca_50.png}
%          \caption{Zoom into  MBD based kernel (Gaussian) PCA for the first 100 molecules of $\sigma = 0.256$. Sum formulas are shown, colour code corresponds to atomization energy.}
%     \label{SI:fig:single_kernel_pca}
%\end{figure*}

\bibliography{literature}